\documentclass[11pt,a4paper]{article}
\usepackage{amssymb}
\usepackage{jheppub}
\usepackage[T1]{fontenc}
\usepackage{hyperref}

\input{tcilatex}
\affiliation[a]{Lab of High Energy Physics, Faculty of Science, Univ
Mohammed V-Agdal, Rabat, Morocco } \affiliation[b]{Centre of Physics
and Mathematics, CPM, Av Ibn Battota, Rabat, Morocco}
\emailAdd{h-saidi@fsr.ac.ma}

\abstract{\  \  \  \  \  \  \  \  \newline Decomposing the Higgs
potential  $\mathcal{V}_{higgs}$ of next - to - Minimal
Supersymmetric Standard Model (\emph{n-MSSM}) \ as the sum of three
contributions like
$\mathcal{V}_{ch}+\mathcal{V}_{kah}+\mathcal{V}_{expl}$ and assuming
the two following things: $\left( a\right) $ $\mathcal{V}_{higgs}$
dominated by $\mathcal{V}_{ch}$ coming from the chiral sector of
supersymmetry: $\mathcal{V}_{ch}=\left \vert \mathrm{\nu }\right
\vert ^{2}\mathcal{U}$ with $\left \vert \mathrm{\nu }\right \vert $
large and $\frac{\mathrm{r}}{\left \vert \mathrm{\nu }\right \vert
}<<1$; and $\left( b\right) $ replacing the chiral down Higgs
superfield doublet $\left( \boldsymbol{H}_{d}\right) ^{i}$ of
\emph{n-MSSM} by a chiral anti-doublet $\left( \boldsymbol{\Phi
}_{d}\right) _{i}$; we derive the explicit geometry of the Higgs
fields in the ground state $\left \vert \Sigma _{higgs}\right
\rangle $ found to be given by \textrm{two intersecting conifolds}.
We show as well that the property  $\tan \beta _{susy}=1$ living at
singularity $r=0$ is a supersymmetric signal; and deviation away
reads in terms of the Kahler parameter $r$ and the $\vartheta
_{_{W}}$- Weinberg angle as $\tan \beta \simeq
1+\frac{\mathrm{r}}{2\left \vert \mathrm{\nu }\right \vert }\sin
^{2}\vartheta _{_{W}}$. Other related issues are also studied.}
 \arxivnumber{ }

\begin{document}

\title{\vspace{-2cm}%
\rightline{\mbox{\small {HEP preprint-july-2014}} \vspace
{1cm}}\textbf{Geometry of the Ground State of Higgs Fields in Next-to-MSSM}}
\author[]{El Hassan Saidi }
\maketitle

\flushbottom

\section{Introduction}

\label{sec:intro}

The experimental discovery of a Higgs-like particle with a mass
$M_{h}\simeq 126$ $GeV$ is a great success of the $SU_{L}\left(
2\right) \times U_{Y}\left( 1\right) $ standard model (SM) of
electroweak unification
\textrm{\cite{A1,A2}}. After three decades since the obtention of the W$%
^{\pm }$ and Z$^{0}$ gauge particles, the recent observation of a
Higgs-like constitutes another big step towards understanding the
architecture of elementary particles and their interactions around
the electroweak scale. This discovery also opens a window on
exploration of extended candidates of SM among which supersymmetric
theories; in particular the Minimal Supersymmetric Standard Model
(\emph{MSSM}) \textrm{\cite{B1}-\cite{B9}} and
extensions such as the next - to - MSSM extension \textrm{\cite{C1}-\cite%
{C12}} we are interested in here in this study.

\  \  \  \  \newline Motivated by the LHC experimental breakthrough,
and by the basic role played by the Higgs fields that capture data
on ground state of standard model and on masses of its particles, we
focus in this paper on the Higgs sector of the next -to- MSSM in
superspace ( shortly \emph{n-MSSM}) and study the following points:

\begin{description}
\item[$1)$] \  \  \ The effect of putting a chiral anti-doublet $\Phi _{\bar{%
\imath}}$ at \emph{the place} of the chiral superfield doublet $\boldsymbol{H%
}_{d}^{i}$; but the others Higgs superfields $\boldsymbol{H}_{u}^{i}$ and $%
\boldsymbol{S}$ unchanged; that is modifying the quantum numbers of
the Higgs \emph{chiral} superfields of \emph{n-MSSM} like\newline \
\  \  \  \  \  \  \  \  \newline
\begin{equation}
\begin{tabular}{llllllllll}
\hline
&  &  &  &  &  &  &  &  &  \\
\multicolumn{3}{l}{\  \  \  \  \  \  \  \emph{n-MSSM}} &  & \multicolumn{3}{l}{$%
\rightarrow $} & \multicolumn{3}{l}{\  \  \  \  \  \  \  \  \  \
\emph{n-MSSM}$^{\ast
}$} \\
{\small doublet} & : & $\  \left( \boldsymbol{H}_{u}\right) ^{i}$ &
&  &
\multicolumn{1}{||l}{} &  & {\small doublet} & : & $\  \left( \boldsymbol{H}%
_{u}\right) ^{i}$ \\
{\small doublet} & : & $\  \left( \boldsymbol{H}_{d}\right) ^{i}$ &
&  & \multicolumn{1}{||l}{} &  & {\small anti-doublet} & : & $\
\left(
\boldsymbol{\Phi }\right) _{\bar{\imath}}$ \\
{\small singlet} & : & $\  \  \boldsymbol{S}$ &  &  &
\multicolumn{1}{||l}{} &
& {\small singlet} & : & $\  \  \boldsymbol{S}$ \\
&  &  &  &  &  &  &  &  &  \\ \hline
\end{tabular}
\label{TA}
\end{equation}%
\begin{equation*}
\end{equation*}%
This quantum number change of the down Higgs superfield is
manifested at the interaction level between $\boldsymbol{\Phi
}_{\bar{\imath}}$ and the non abelian gauge superfields by a change
of the sign of the $SU_{L}\left( 2\right) $ gauge coupling constant
$g_{su_{2}}$ ( see eqs \ref{319}). It has
been motivated by the opposite hypercharges of $\boldsymbol{H}_{u}^{i}$ and $%
\boldsymbol{H}_{d}^{i}$ under $U_{Y}\left( 1\right) $ symmetry
transformations; and also by (a) and (b) given here below:\newline \
\  \  \  \  \  \  \newline $\left( \mathbf{a}\right) $ the seeking
for a non trivial geometric interpretation of the Higgs ground state
$\left \vert \Sigma \right \rangle $ in terms of intersecting
manifolds; this property is indicated by the
superspace lagrangian density of \emph{n-MSSM}; the 4 chiral superfields $%
\boldsymbol{\Phi }_{i}$ and $\boldsymbol{H}_{u}^{i}$ obey the
typical complex
3D conifold relations:%
\begin{equation*}
\begin{tabular}{llllll}
$\  \  \  \  \  \  \  \  \  \  \  \  \boldsymbol{\Phi
}_{i}\boldsymbol{H}_{u}^{i}$ & $=$
& $\mathrm{\nu }$ & : & \  \  \  \  \  \  \  & $\mathfrak{C}_{\nu }$ \\
$\boldsymbol{\bar{H}}_{ui}\boldsymbol{H}_{u}^{i}-\boldsymbol{\Phi }_{i}%
\boldsymbol{\bar{\Phi}}^{i}$ & $=$ & $\mathrm{r}$ & : &  & $\mathfrak{R}_{r}$%
\end{tabular}%
\end{equation*}%
showing that Higgs fields ground state lives on the intersection of
the threefolds $\mathfrak{C}_{\nu }$ and $\mathfrak{R}_{r}$ given by
a complex
surface $\mathcal{S}$; see eqs (\ref{ces}). The other constraints on $%
\boldsymbol{\Phi }_{i}$ and $\boldsymbol{H}_{u}^{i}$, coming from
the minimization of the scalar potential, give extra data on the
precise location of the Higgs VEVs within the surface $\mathcal{S}$.

\  \  \  \  \  \  \  \  \newline $\left( \mathbf{b}\right) $
exhibiting explicitly the dual role of the component scalar Higgs
fields $\boldsymbol{h}_{u}$ and $\boldsymbol{h}_{d}$; seen that one
Higgs doublet, say $\boldsymbol{h}_{u}$, is sufficient to break
$SU_{L}\left( 2\right) \times U_{Y}\left( 1\right) $ gauge symmetry
down to $U_{em}\left( 1\right) $. This duality, which is understood in \emph{%
n-MSSM}, is precisely given by the relation $\boldsymbol{\Phi }_{i}%
\boldsymbol{H}_{u}^{i}=\mathrm{\nu }$ and remarkably captured by the
extra chiral singlet superfield $\boldsymbol{S}$ of \emph{n-MSSM}.

\item[$2)$] \  \  \ The underlying geometry of the quantum phases of the Higgs
ground state namely the supersymmetric phase $\left \vert \Sigma
_{susy}\right \rangle $ and the non supersymmetric one $\left \vert
\Sigma _{n-susy}\right \rangle $. We first study the structure of
these geometries, given by compact complex curves inside the surface
$\mathcal{S}$, respectively denoted as $\Sigma _{susy}$ and $\Sigma
_{n-susy}$; and then derive the explicit expression of their metrics
as well as specific
properties. In the first case, we show amongst others that the ratio $\frac{%
\upsilon _{h}}{\upsilon _{d}}$ of the Higgs VEVs corresponds exactly
to
\begin{equation*}
\tan \beta _{susy}=1
\end{equation*}%
and in the case of $\Sigma _{n-susy}$, we find that this ratio
deviates from
unity as%
\begin{equation*}
\tan \beta =\frac{\mathrm{r}\sin ^{2}\vartheta _{_{W}}}{2\left \vert \mathrm{%
\nu }\right \vert }+\sqrt{1+\left( \frac{\mathrm{r}\sin ^{2}\vartheta _{_{W}}%
}{2\left \vert \mathrm{\nu }\right \vert }\right) ^{2}}
\end{equation*}%
with $\left \vert \frac{\mathrm{r}}{\mathrm{\nu }}\right \vert <<1$
and \textrm{r} a real number; non necessary positif. We find
moreover that the metric of $\Sigma _{n-susy}$ reads in terms of the
$\vartheta _{_{W}}$- Weinberg angle like
\begin{equation}
ds_{\Sigma
_{\text{n-susy}}}^{2}=\frac{1}{4}\mathcal{R}_{_{H}}^{2}\left(
d\theta ^{2}+4\cos ^{2}\frac{\theta }{2}d\phi ^{2}\right)
\label{S2}
\end{equation}%
\begin{equation*}
\end{equation*}%
with $\theta $ and $\phi $ standing for the usual coordinates of the
real 2-sphere, and the radius $\mathcal{R}_{H}$ given by
\begin{equation*}
\mathcal{R}_{_{H}}^{2}=\frac{2\left \vert \mathrm{\nu }\right \vert ^{2}+%
\mathrm{r}^{2}\sin ^{2}\vartheta _{_{W}}+\mathrm{r}\sqrt{4\left
\vert
\mathrm{\nu }\right \vert ^{2}+\mathrm{r}^{2}\sin ^{4}\vartheta _{_{W}}}}{%
\mathrm{r}^{2}\sin ^{2}\vartheta _{_{W}}+\sqrt{4\left \vert \mathrm{\nu }%
\right \vert ^{2}+\mathrm{r}^{2}\sin ^{4}\vartheta _{_{W}}}}
\end{equation*}%
\begin{equation*}
\end{equation*}%
From this point of view, the real Fayet Iliopoulos (FI) coupling constant $%
\mathrm{r}\sim M^{2}$ may be then thought of as given by the mass
scale where supersymmetry breaks down. In other works, if taking the
high energy limit $\frac{\mathrm{r}}{\left \vert \mathrm{\nu }\right
\vert }=0$, the
squared radius tends to $\left. \mathcal{R}_{_{H}}^{2}\right \vert _{\mathrm{%
r}=0}=\left \vert \mathrm{\nu }\right \vert $ and one is left with
the
supersymmetric phase $\left \vert \Sigma _{\text{susy}}\right \rangle $.%
\newline
\  \  \  \newline
Furthermore, we find that the point $\mathcal{P}_{u_{{\small em}}{\small (1)}%
}$, where the $SU_{L}\left( 2\right) \times U_{Y}\left( 1\right) $
gauge symmetry breaks spontaneously down to $U_{em}\left( 1\right)
$, sits in our geometric description, at the south pole
\begin{equation*}
\theta =\pi
\end{equation*}%
of the real 2-sphere (\ref{S2}) where the metric degenerates and the
"electromagnetic" circle $\mathbb{S}_{em}^{1}$, parameterized by the angle $%
\phi $, shrinks to $\mathcal{P}_{u_{{\small em}}{\small
(1)}}$.\newline \  \  \  \  \newline We also derive the following
formula for the energy density of the non
supersymmetric ground state $\left \vert \Sigma _{\text{n-susy}%
}\right
\rangle $%
\begin{equation*}
\mathcal{E}_{\min }=\mathcal{E}_{0}+\frac{\sin ^{2}\left( 2\vartheta
_{_{W}}\right) }{32}\mathrm{r}^{4}
\end{equation*}%
with the constant $\mathcal{E}_{0}$ related to explicit
supersymmetry breaking.
\end{description}

\  \  \  \newline The presentation of this paper is as follows:
\textrm{In section 2}, we present the key idea of our proposal;
describe the method of approaching the solutions of non linear
coupled equations; and give a summary of main results obtained in
this work. \textrm{In section 3}, we review basic aspects on the
quantum charges of the Higgs fields under the $SU_{L}\left( 2\right)
\times U_{Y}\left( 1\right) $ gauge symmetry and develop our
proposal regarding the $SU_{L}\left( 2\right) $ gauge charge of the
Higgs
doublet $H_{d}$. In our method, we replace the chiral superfield \emph{%
doublet} $H_{d}^{i}$ by an \emph{anti-doublet }$\Phi _{i}$ whose $%
SU_{L}\left( 2\right) \times U_{Y}\left( 1\right) $ quantum charges,
including those of $SU_{L}\left( 2\right) $, are given by the
opposite of those of $H_{u}$. \textrm{In section 4}, we study the
superfield formulation
of n-MSSM; but by using the \emph{anti-doublet }$\Phi _{i}$ at the place of $%
H_{d}^{i}$. \textrm{In section 5}, we consider the case $r=0$; while
letting the complex $\nu $ arbitrary; and study the set of exact
solutions of the auxiliary field's equations of motion; their
intersections as well as the phase of ground state. \textrm{In
section 6}, we switch on the real FI coupling constant $r\neq 0$ and
examine the solution of the equations of motion of the auxiliary
fields. In this case, we show that supersymmetry is broken and we
determine the energy density of the ground state as well as the
deviation of the $\tan \beta $ ratio of the Higgs VEVs. \textrm{In
section 7}, we explore general aspects of explicit supersymmetry
breaking
and in \textrm{section 8}, we give a conclusion and make some comments.%
\textrm{\ Sections 9 and 10 }are respectively devoted to appendices
on
useful tools on next - to -MSSM and the metric of the \textrm{intersecting }%
conifold geometry\textrm{.}

\section{n-MSSM* and summary of results}

In this section, we draw the main lines behind \emph{n-MSSM}* of table (\ref%
{TA}) and give a summary of some of the results obtained in this
study; technical details and other results are exposed in the
forthcoming sections and in the two appendices.

\subsection{General on \emph{n-MSSM }in superspace}

We begin by recalling useful ingredients on supersymmetric gauge
theories in superspace taking as an example the next -to- MSSM
extension of $U_{Y}\left( 1\right) \times SU_{L}\left( 2\right) $
standard model; and focussing on Higgs sector and interactions with
gauge radiation.

\  \  \

\emph{Quantum charges of n-MSSM Higgs}\newline Next -to- MSSM
extension of standard model involves \emph{5} Higgs chiral
superfields: a hyperchargeless iso-singlet $\boldsymbol{S}$ and two
doublets
$\boldsymbol{H}_{u}=\left( H_{u}^{+},H_{u}^{0}\right) $ and $\boldsymbol{H}%
_{d}=\left( H_{d}^{0},H_{d}^{-}\right) $ with opposite hypercharge, $%
y_{u}=-y_{d}=1$; but same charge under $SU_{L}\left( 2\right) $. A
way to
see how these quantum charges are manifested in the lagrangian density $%
\mathcal{L}$ of the model is through the gauge covariant derivatives
of the leading scalar field components $S$, $\left( h_{u}^{i}\right)
$ and $\left(
h_{d}^{i}\right) $ of the chiral superfields $\boldsymbol{S}$, $\boldsymbol{H%
}_{u}$ and $\boldsymbol{H}_{d}$. While $\partial _{\mu }S$ is
un-affected under gauge symmetry transformations, $\left( \partial
_{\mu }h_{u}\right) ^{i}$\ and $\left( \partial _{\mu }h_{d}\right)
^{i}$\ get modified into
gauge covariant derivatives $\left( \nabla _{\mu }h_{u}\right) ^{i}$ and $%
\left( \nabla _{\mu }h_{d}\right) ^{i}$ depending on the bosonic
gauge
fields $B_{\mu }$ and $W_{\mu }^{A}$ as given below%
\begin{equation}
\begin{tabular}{lll}
&  &  \\
$\left( \nabla _{\mu }h_{u}\right) ^{i}$ & $=$ & $\left[ \delta
_{j}^{i}\partial _{\mu }-i\frac{g^{\prime }}{2}\delta _{j}^{i}B_{\mu
}-igW_{\mu }^{A}\left( \frac{\tau _{A}}{2}\right) _{j}^{i}\right]
h_{u}^{j}$
\\
&  &  \\
$\left( \nabla _{\mu }h_{d}\right) ^{i}$ & $=$ & $\left[ \delta
_{j}^{i}\partial _{\mu }+i\frac{g^{\prime }}{2}\delta _{j}^{i}B_{\mu
}-igW_{\mu }^{A}\left( \frac{\tau _{A}}{2}\right) _{j}^{i}\right]
h_{d}^{j}$
\\
&  &
\end{tabular}
\label{dh}
\end{equation}%
These gauge covariant derivatives differ only by the sign of the $%
U_{Y}\left( 1\right) $ gauge coupling constant $g^{\prime }$; this
is because the Higgs fields $h_{u}^{i}$ and $h_{d}^{i}$ have
opposite
hypercharges; but the same $SU_{L}\left( 2\right) $ charge,%
\begin{equation}
\begin{tabular}{lllll}
$\left[ Y,h_{u}^{i}\right] $ & $=+h_{u}^{i}$ & , & $\left[ T^{A},h_{u}^{i}%
\right] $ & $=\frac{1}{2}\left( \tau ^{A}\right) _{j}^{i}h_{u}^{j}$ \\
$\left[ Y,h_{d}^{i}\right] $ & $=-h_{d}^{i}$ & , & $\left[ T^{A},h_{d}^{i}%
\right] $ & $=\frac{1}{2}\left( \tau ^{A}\right) _{j}^{i}h_{d}^{j}$ \\
&  &  &  &
\end{tabular}%
\end{equation}%
where $Y$ and $T^{A}$ are the hermitian generators of the
$U_{Y}\left( 1\right) \times SU_{L}\left( 2\right) $ gauge symmetry.
Later on, we consider as well a gauge covariant derivative
$\mathcal{D}_{\mu }\varphi _{i} $ of the anti-doublet $\varphi _{i}$
where the sign of both gauge
couplings $g^{\prime }$ and $g$ of the $U_{Y}\left( 1\right) $ and $%
SU_{L}\left( 2\right) $ gauge symmetry factors are flipped; see
eq(\ref{FL}) for comparison and to fix the idea.

\

\emph{Scalar potential}\ $\mathcal{V}_{sca}$\  \  \newline In
supersymmetric gauge theories describing the interacting dynamics
between supersymmetric chiral matter and supersymmetric gauge
multiplets, the scalar potential energy density $\mathcal{V}_{sca}$
is positive and is given by the sum of two basic terms
\begin{equation}
\mathcal{V}_{sca}=\mathcal{V}_{ch}+\mathcal{V}_{re}\geq 0
\label{sv}
\end{equation}%
with $\mathcal{V}_{ch}$ and $\mathcal{V}_{re}$ describing
respectively the contribution coming from the auxiliary fields $F$
(depending on chiral superpotential) and the auxiliary fields $D$
(Kahler sector). \newline
In the case of \emph{n-MSSM}, the potential energy densities $\mathcal{V}%
_{ch}$ and $\mathcal{V}_{re}$ read as
\begin{equation}
\begin{tabular}{lll}
$\mathcal{V}_{ch}$ & $=$ & $\bar{F}_{S}F_{S}+\dsum \limits_{i=1}^{2}\bar{F}%
_{ui}F_{u}^{i}+\dsum \limits_{i=1}^{2}\bar{F}_{di}F_{d}^{i}\geq 0$ \\
$\mathcal{V}_{re}$ & $=$ & $\frac{1}{2}\left( D^{\prime }\right) ^{2}+\frac{1%
}{2}\dsum \limits_{A=1}^{3}D_{{\small A}}D^{{\small A}}\geq 0$%
\end{tabular}%
\end{equation}%
The term $\mathcal{V}_{ch}$ is the sum of \emph{5} quadratic
monomials in
one one to one with the number of the \emph{5} Higgs chiral superfields $%
\boldsymbol{S}$, $\boldsymbol{H}_{u}^{i}$ and
$\boldsymbol{H}_{d}^{i}$ and so with the number of bosonic Higgs
fields $S$, $\left( h_{u}^{i}\right) $ and $\left( h_{d}^{i}\right)
$. The potential $\mathcal{V}_{ch}$ depends
moreover on the complex coupling constants $\left \{ \lambda _{x},\bar{%
\lambda}_{x}\right \} $ of the intra superfield Higgs interactions; typically%
\begin{equation}
W=\lambda \boldsymbol{SH}_{u}\boldsymbol{H}_{d}+\nu \boldsymbol{S}+\frac{%
\kappa }{3}\boldsymbol{S}^{3}  \label{sp}
\end{equation}%
Similarly, the term $\mathcal{V}_{re}$ is given by the sum of
\emph{4}
quadratic monomials in one to one with the \emph{4} gauge multiplets $%
\boldsymbol{V}^{\prime }$ and $\boldsymbol{V}^{A}$ of the
$U_{Y}\left( 1\right) \times SU_{L}\left( 2\right) $ gauge symmetry
group. It depends on
the real gauge coupling constants $g$ and $g^{\prime }$; but not on $%
\left \{ \lambda _{x},\bar{\lambda}_{x}\right \} $.

\  \  \  \  \  \newline
To deal with the \emph{total} scalar\textrm{\footnote{%
The total scalar potential $\mathcal{V}$ in \emph{n-MSSM} contains
as well
an extra term $\mathcal{V}_{exl}$ breaking supersymmetry explicitly; $%
\mathcal{V}$ is bounded below and is a gauge invariant quartic
polynom in Higgs fields.}} potential in \emph{n-MSSM}, one first
uses the equations of motion of the auxiliary fields to get the
expressions of the F's and the D's in terms of the bosonic scalar
fields $S$, $h_{u}^{i},$ $h_{d}^{i}$. Generally, these are quadratic
quantities in the Higgs fields; and so the scalar potential
\begin{equation*}
\mathcal{V}=\mathcal{V}_{{\small n}\text{{\small
-}}MSSM}+\mathcal{V}_{exl}
\end{equation*}%
is \emph{a quartic} polynom in $S$, $h_{u}^{i},$ $h_{d}^{i}$ bounded
from below; with sections having generally \emph{3} extrema:
\emph{2} minima and a local extremum.\newline Then one determines
the set of Higgs configurations that minimize the total scalar
potential
\begin{equation*}
\mathcal{V}\left( h_{u},h_{d},S,\text{ }\bar{h}_{u},\bar{h}_{d},\bar{S}%
\right)
\end{equation*}%
to get the ground state $\left \vert \Sigma _{Higgs}\right \rangle $
of the\ Higgs fields. \newline If thinking about $\mathcal{V}$ as
the mapping

\begin{equation*}
\begin{tabular}{lllll}
$\mathcal{V}$ & $:$ & $\  \  \  \  \  \mathbb{C}^{5}$ & $\  \  \  \
\  \  \
\rightarrow $ \  \  \  \  \  \  & $\  \  \  \  \  \  \  \mathbb{R}$ \\
&  & $\left( h_{u},h_{d},S\right) $ &  & $\mathcal{V}\left(
h_{u},h_{d},S\right) $ \\
&  &  &  &
\end{tabular}%
\end{equation*}%
the ground state $\Sigma _{Higgs}$ of the \emph{5} complex Higgs field $%
h_{u},h_{d},S$ lives then inside of $\mathbb{C}^{5}$ and is given by
the kernel of $\mathrm{\delta }_{var}\mathcal{V}$; the variation of
the total scalar potential with respect to the \emph{5} Higgs
fields,
\begin{equation}
\Sigma _{Higgs}=\ker \left( \mathrm{\delta }_{var}\mathcal{V}\right)
\subset \mathbb{C}^{5}  \label{s}
\end{equation}%
This set is completely characterized by the two real gauge couplings
constants $g$ and $g^{\prime }$; as well as by the complex coupling
constants $\left \{ \lambda _{\left \{ x\right \}
},\bar{\lambda}_{\left \{ x\right \} }\right \} $ of the intra Higgs
interactions including those
coming from the explicit supersymmetry breaking terms; \textrm{see eq(\ref%
{3V}) and section 7 as well as appendix A for further details}.

\subsection{The proposed \emph{n-MSSM*} and first results}

Motivated by the determination of the three followings:

\begin{itemize}
\item the topological shape of the Higgs manifold $\Sigma _{Higgs}$
associated with Higgs ground state,

\item the explicit expression of the metric $ds_{\Sigma }^{2}$ of $\Sigma
_{Higgs}$; and

\item the property of $ds_{\Sigma }^{2}$ that singles out the electrically
neutral point $P_{u_{1}^{em}}\in \Sigma _{Higgs}$ where $U_{Y}\left(
1\right) \times SU_{L}\left( 2\right) $ symmetry breaks down to $%
U_{em}\left( 1\right) $,
\end{itemize}

\  \  \  \newline we consider in this study the Higgs sector of
\emph{n-MSSM} on which we impose \emph{two assumptions} that allow
very explicit calculations and permit more insight into the
structure $\Sigma _{Higgs}$.

\  \  \  \newline
$\mathbf{a})$ \emph{assumption I}: $\mathcal{V}_{ch}>>\mathcal{V}_{re}+%
\mathcal{V}_{exl}$\newline the total the scalar potential
$\mathcal{V}$ is dominated by the contribution coming from the
chiral sector of supersymmetry; this property fixes the topological
shape of the manifold where lives $\Sigma _{Higgs}$,

\  \  \  \newline
$\mathbf{b})$ \emph{assumption II}: \emph{n-MSSM} \ $\rightarrow $\  \  \emph{%
n-MSSM}$^{\ast }$\newline the two chiral Higgs doublet superfields
are assumed to be dual superfields;
in the sense they have opposite quantum numbers; i.e: they have opposite $%
U_{Y}\left( 1\right) $ hypercharges and also opposite $SU_{L}\left(
2\right) $ ones. \newline With this some how natural assumption, one
expects that the self dual point to be a critical point of
$ds_{\Sigma }^{2}$ where a phase transition takes place; As we will
see later self dual point corresponds precisely to the
supersymmetric phase.

\  \  \newline The first hypothesis is further explored in sub-sub-
section 2.2.1 and the second one is studied in sub-sub- section
2.2.2.

\subsubsection{More on the condition $\mathcal{V}_{ch}>>\mathcal{V}_{re}+%
\mathcal{V}_{exl}$}

First, notice that from the explicit expression of the total scalar
potential
\begin{eqnarray*}
\mathcal{V} &=&\mathcal{V}_{susy}+\mathcal{V}_{exl} \\
&=&\left( \mathcal{V}_{ch}+\mathcal{V}_{re}\right)
+\mathcal{V}_{exl}
\end{eqnarray*}%
one learns that in order to determine the set $\Sigma _{Higgs}$ of eq(\ref{s}%
) one has to solve the 5 complex equations%
\begin{equation}
\begin{tabular}{lll}
$\frac{\partial \mathcal{V}}{\partial S}$ & $=$ & $0$ \\
$\frac{\partial \mathcal{V}}{\partial h_{u}^{i}}$ & $=$ & $0$ \\
$\frac{\partial \mathcal{V}}{\partial h_{d}^{i}}$ & $=$ & $0$%
\end{tabular}%
\end{equation}%
and look for their intersecting solutions. Clearly this system of
complex equations has solutions; but difficult to figure out
explicitly due to the number of relations; their couplings as well
as the cubic non linearities in the field variables.

\  \  \  \

\emph{1) the condition on scalar potential}\newline In order to get
explicit solutions of the above relations, we make the
hypothesis%
\begin{equation*}
\mathcal{V}_{ch}>>\mathcal{V}_{re}+\mathcal{V}_{exl}
\end{equation*}%
\textrm{which should be thought of as corresponding to a particular
region in the moduli space of coupling constants }of \emph{n-MSSM}.
In fact this region corresponds to high energy limit where lives the
supersymmetric phase of the model. \newline With this condition, one
may think about the total scalar potential of the
model as mainly given by $\mathcal{V}_{ch}$ with a perturbation $\mathrm{%
\delta }_{pert}\mathcal{V}=\mathcal{V}_{pert}$ equal to $\left( \mathcal{V}%
_{re}+\mathcal{V}_{exl}\right) $. Explicitly,
\begin{equation}
\mathcal{V}=\mathcal{V}_{ch}+\mathcal{V}_{pert}
\end{equation}%
with%
\begin{equation}
\mathcal{V}_{pert}=\frac{1}{2}\left( D^{\prime }\right) ^{2}+\frac{1}{2}%
\dsum \limits_{A=1}^{3}D_{{\small A}}D^{{\small
A}}+\mathcal{V}_{exl}
\end{equation}%
Then use the approximation to compute explicitly the ground state
$\left \vert \Sigma \right \rangle $ and its phases by proceeding as
follows:

\begin{itemize}
\item First, compute the minimum of $\mathcal{V}_{ch}=\sum_{I}\bar{F}%
_{I}F^{I}$ obtained by solving the condition%
\begin{equation*}
\delta \mathcal{V}_{ch}=\dsum \limits_{I}\bar{F}_{I}\left( \delta
F^{I}\right) +F^{I}\left( \delta \bar{F}_{I}\right) =0
\end{equation*}%
and remarkably given by the zero energy condition%
\begin{equation}
\mathcal{V}_{ch}=\bar{F}_{ui}F_{u}^{i}+\bar{F}_{di}F_{d}^{i}+\bar{F}%
_{S}F_{S}=0
\end{equation}%
or equivalently%
\begin{equation}
\begin{tabular}{lll}
$\bar{F}_{S}$ & $=$ & $0$ \\
$\bar{F}_{ui}$ & $=$ & $0$ \\
$\bar{F}_{di}$ & $=$ & $0$%
\end{tabular}
\label{3F}
\end{equation}%
The explicit expression of the solutions of these relations are a
priori not difficult to obtain seen that they are quadratic in the
Higgs fields; let us
denote these solutions as%
\begin{equation}
\begin{tabular}{lll}
$\left \langle S\right \rangle ,$ & $\  \  \  \  \  \left \langle
h_{u}^{i}\right
\rangle ,$ & $\  \  \  \  \  \left \langle h_{d}^{i}\right \rangle $%
\end{tabular}%
\end{equation}%
and refer to them collectively as $\xi $; they parameterize a submanifold $%
\mathfrak{C}_{\nu }$ in $\mathbb{C}^{5}$.

\item Then, put these solutions back into the total scalar potential $%
\mathcal{V}$, we get the following hermitian function depending on
the $\xi $ moduli
\begin{equation}
\begin{tabular}{lll}
$\mathcal{V}\left( \xi \right) $ & $=$ & $\mathcal{V}_{ch}\left( \xi
\right) +\mathcal{V}_{re}\left( \xi \right) +\mathcal{V}_{exl}\left(
\xi \right) $
\\
&  &  \\
& $=$ & $\mathcal{V}_{re}\left( \xi \right) +\mathcal{V}_{exl}\left(
\xi
\right) $%
\end{tabular}%
\end{equation}%
because $\mathcal{V}_{ch}\left( \xi \right) =0$.

\item Next, compute the minimum of
\begin{equation*}
\mathcal{V}_{pert}=\mathcal{V}_{re}\left( \xi \right) +\mathcal{V}%
_{exl}\left( \xi \right)
\end{equation*}%
to end with the values of the Higgs fields minimizing the potential
energy
density; and which we denote like,%
\begin{equation*}
\xi _{\min }=\left \{ \left \langle S\right \rangle _{\min },\left
\langle h_{u}^{i}\right \rangle _{\min },\  \left \langle
h_{d}^{i}\right \rangle _{\min }\right \}
\end{equation*}%
The set of these minima gives the ground state $\left \vert \Sigma
\right \rangle $ of the Higgs fields.\newline \  \  \  \  \  \  \  \
\  \  \  \  \  \newline
Clearly, the geometry of the Higgs ground state $\Sigma _{\nu }$ depends on $%
\mathcal{V}_{ch}\left( \xi \right) $, $\mathcal{V}_{re}\left( \xi
\right) $ and $\mathcal{V}_{exl}\left( \xi \right) $; but in order
to get more insight into the role of auxiliary fields $F$ and $D$ in
\emph{n-MSSM}, we first consider the case where $\mathcal{V}_{exl}$
is switched off; and turn later to study the effect of switching on
$\mathcal{V}_{exl}$; for details see \textrm{section 7}.
\end{itemize}

\  \  \  \  \

\emph{2) switching off} $\mathcal{V}_{exl}\left( \xi \right) :\mathcal{V}%
_{pert}=\mathcal{V}_{re}\left( \xi \right) $\newline
From an abstract point of view; if thinking about $\mathcal{V}_{ch}$ and $%
\mathcal{V}_{re}$ as the two maps%
\begin{equation}
\begin{tabular}{lllll}
$\mathcal{V}_{ch}$ & $:$ & $\  \  \mathbb{C}^{5}$ & $\  \
\rightarrow $ & $\  \
\mathbb{R}_{+}$ \\
$\mathcal{V}_{re}$ & $:$ & $\  \  \ker \mathcal{V}_{ch}$ & $\  \
\rightarrow $
& $\  \  \mathbb{R}_{+}$%
\end{tabular}%
\end{equation}%
then, according to whether $\ker \mathcal{V}_{re}$ is the empty set $%
\varnothing $ or not, we distinguish two phases: a supersymmetric
phase and a non supersymmetric one as follows

\begin{eqnarray*}
&&%
\begin{tabular}{llllll}
\hline \hline
&  &  &  &  &  \\
\  \  \ {\small phase} &  & \multicolumn{4}{l}{\  \  \  \  \  \  \
\  \ {\small \  \  \
\  \  \  \ ground state }$\Sigma $} \\
&  &  &  &  &  \\
$\ker \mathcal{V}_{re}\neq \varnothing $ & {\small :} &  & $\Sigma
_{susy}$
& $=$ & $\ker \mathcal{V}_{ch}\dbigcap \ker \mathcal{V}_{re}$ \\
&  &  &  &  &  \\
$\ker \mathcal{V}_{re}=\varnothing $ & : &  & $\Sigma _{n-susy}$ & $=$ & $%
\ker \mathcal{V}_{ch}\dbigcap \ker \left( \mathrm{\delta }\mathcal{V}%
_{re}\right) $ \\
&  &  &  &  &  \\ \hline \hline
\end{tabular}
\\
&&
\end{eqnarray*}

Let us illustrate these two phases on \emph{n-MSSM}; and give two
comments regarding their physical implications.

\  \  \  \

\emph{3) illustration}\newline As an illustration of these phases,
consider the equations of motion of the
auxiliary fields in \emph{n-MSSM}; that is the \emph{5} complex equations (%
\ref{3F}) of the F- auxiliary fields
\begin{equation}
\begin{tabular}{lll}
$Sh_{u}^{i}$ & $=$ & $0$ \\
$Sh_{d}^{i}$ & $=$ & $0$ \\
$\varepsilon _{ij}h_{u}^{i}h_{d}^{j}+\kappa S^{2}-\mathrm{\mu }$ & $=$ & $0$%
\end{tabular}
\label{S2H}
\end{equation}%
with $\kappa $ and $\mathrm{\mu }$ the two complex numbers in the
superpotential W given by (\ref{sp}); and the \emph{4} hermitian
equations
of the D- auxiliary fields%
\begin{eqnarray}
\sum_{i=1}^{2}\left(
\bar{h}_{ui}h_{u}^{i}-\bar{h}_{di}h_{d}^{i}\right) &=&0
\notag \\
\sum_{i=1}^{2}\left( \tau ^{A}\right) _{j}^{i}\left( \bar{h}_{ui}h_{u}^{j}+%
\bar{h}_{di}h_{d}^{j}\right) &=&0  \label{2D}
\end{eqnarray}%
Clearly for the case $\mathrm{\mu }=0$, all these equations are solved by $%
S=h_{u}^{i}=h_{d}^{i}=0$; and so the origin of $\mathbb{C}^{5}$
belongs to the supersymmetric phase $\Sigma _{susy}$. \newline
In general, eq(\ref{S2H}) has two sets of solutions $\Omega _{1}$ and $%
\Omega _{2}$ as given hereafter

\begin{equation}
\begin{tabular}{lll}
$\Omega _{1}$ & $=$ & $\left \{ \left( S,h_{u}^{i},h_{d}^{i}\right)
\in
\mathbb{C}^{5}\text{ \  \ }|\text{ \  \ }S=\pm \sqrt{\frac{\mathrm{\mu }}{%
\kappa }}\text{ \ },\  \ h_{u}^{i}=h_{d}^{i}=0\right \} $ \\
&  &  \\
$\Omega _{2}$ & $=$ & $\left \{ \left( S,h_{u}^{i},h_{d}^{i}\right)
\in \mathbb{C}^{5}\text{ \  \ }|\text{ \  \ }S=0\text{ \ },\  \
\varepsilon
_{ij}h_{u}^{i}h_{d}^{j}=\mathrm{\mu }\right \} $%
\end{tabular}%
\end{equation}%
\begin{equation*}
\end{equation*}%
Higgs configurations in the set $\Omega _{1}$ are not physically
interesting
since they lead to vanishing VEVs for the Higgs doublets; i.e: $%
h_{u}^{i}=h_{d}^{i}=0$. However the configurations in the set
$\Omega _{2}$ have non zero VEVs for the Higgs doublets
\begin{equation}
\varepsilon _{ij}h_{u}^{i}h_{d}^{j}=\mathrm{\mu }  \label{mh}
\end{equation}%
they parameterize a complex \emph{3D} geometry inside
$\mathbb{C}^{5}$ which is nothing but a complex \emph{3D} deformed
conifold with complex deformation $\mathrm{\mu }$. \newline Seen
that the Higgs doublets $h_{u}^{i}$ and $h_{d}^{i}$ can never take
zero values for $\mathrm{\mu }\neq 0$; supersymmetry is
spontaneously broken since the constraints (\ref{2D}) cannot be
fulfilled as they require the vanishing of the VEVs of the doublets.
This feature can be checked directly
by computing the scalar energy which is positive definite $\mathcal{V}%
_{re}>0 $; or equivalently by trying to solve the equations of
motion of the
auxiliary D- fields (\ref{2D}). Indeed, for the singular limit $\mathrm{\mu }%
=0$ for instance, equation (\ref{mh}) is remarkably solved by the
non zero
quantities%
\begin{equation}
h_{u}^{i}=a\zeta ^{i},\qquad h_{d}^{i}=b\zeta ^{i}  \label{er}
\end{equation}%
thanks to the property $\varepsilon _{ij}\zeta ^{i}\zeta ^{j}=0$.
\newline In this solution, $a$ and $b$ are arbitrary complex numbers
whose absolute
values are the usual Higgs VEVs $\upsilon _{u}$ and $\upsilon _{d}$ of \emph{%
n-MSSM}; and $\zeta ^{i}$ is complex doublet that parameterize the
real 3-sphere,
\begin{equation*}
\bar{\zeta}_{i}\zeta ^{i}=1,\qquad \varepsilon _{ij}\zeta ^{i}\zeta
^{j}=0
\end{equation*}%
and leading to%
\begin{equation*}
\begin{tabular}{lll}
$\bar{h}_{ui}h_{u}^{i}$ & $=\left \vert a\right \vert ^{2}$ &
$=\upsilon
_{u}^{2}$ \\
$\bar{h}_{di}h_{d}^{i}$ & $=\left \vert b\right \vert ^{2}$ &
$=\upsilon
_{d}^{2}$ \\
$\tan \beta $ & $=\frac{\upsilon _{u}}{\upsilon _{d}}$ &
\end{tabular}%
\end{equation*}%
However putting the solution (\ref{er}) back into the constraint relations (%
\ref{2D}) coming from the auxiliary fields D, one obtains two
constraint
relations on the numbers $a,$ $b$ as given below%
\begin{equation*}
\begin{tabular}{lll}
$\left( \left \vert a\right \vert ^{2}-\left \vert b\right \vert
^{2}\right)
$ & $=$ & $0$ \\
$\left( \left \vert a\right \vert ^{2}+\left \vert b\right \vert
^{2}\right)
\left( \dsum \limits_{i=1}^{2}\left( \tau ^{A}\varepsilon \right) _{ij}\bar{%
\zeta}^{(i}\zeta ^{j)}\right) $ & $=$ & $0$%
\end{tabular}%
\end{equation*}%
The first condition is very remarkable as it leads to
\begin{equation*}
\left \vert a\right \vert =\left \vert b\right \vert \qquad
\Rightarrow \qquad \tan \beta =\frac{\left \vert a\right \vert
}{\left \vert b\right \vert }=1
\end{equation*}%
it may be interpreted as a necessary condition for supersymmetry
seen that is follows from the vanishing condition of an auxiliary
field. The second
constraint requires however $a=b=0$; and leads to the undesired $%
h_{u}^{i}=h_{d}^{i}=0$.

\  \

\emph{4)} \emph{two comments}\newline We give two comments: one on
the phases of the ground state $\Sigma $ of the Higgs fields; and
the other on the expression of the ratio $\tan \beta $ of the Higgs
VEVs.

\begin{itemize}
\item In the singular limit $\mathrm{\mu }=0$, the solution of both the
auxiliary fields F-type and D- type is given by the point
\begin{equation*}
S=h_{u}^{i}=h_{d}^{i}=0
\end{equation*}%
and the Higgs ground state $\left \vert \Sigma \right \rangle $
corresponds to a supersymmetric phase living at the origin of
$\mathbb{C}^{5}$. \newline For $\mathrm{\mu }\neq 0$, supersymmetry
is spontaneously broken for
\begin{equation*}
S=0;\qquad \varepsilon _{ij}h_{u}^{i}h_{d}^{j}=\mathrm{\mu }
\end{equation*}%
this breaking is roughly coming from the Kahler sector. Indeed, for
non zero
VEVs for the Higgs doublets satisfying (\ref{mh}), the configuration%
\begin{equation}
h_{u}^{i}=a\zeta ^{i},\qquad h_{d}^{i}=b\zeta ^{i}+c\varepsilon ^{ik}\bar{%
\zeta}_{k}  \label{ab}
\end{equation}%
solves the 5 complex equations (\ref{S2H}) of the chiral sector for
\begin{equation*}
ac=\mathrm{\nu }
\end{equation*}%
as well as the hermitian singlet relation $D^{\prime }=\bar{h}_{ui}h_{u}^{i}-%
\bar{h}_{di}h_{d}^{i}=0$ by taking
\begin{equation*}
\left \vert a\right \vert ^{2}=\left \vert b\right \vert ^{2}+\left
\vert c\right \vert ^{2}
\end{equation*}%
but not the isotriplet constraint relation%
\begin{equation}
D^{A}=\sum_{i=1}^{2}\left( \tau ^{A}\right) _{j}^{i}\left( \bar{h}%
_{ui}h_{u}^{j}+\bar{h}_{di}h_{d}^{j}\right) =0  \label{DA}
\end{equation}%
With this reasoning, one may think about $D^{A}$ as the responsible
of the breaking of supersymmetry.

\item from above comment, a natural question arises: why eqs(\ref{ab}) solve
the $U_{Y}\left( 1\right) $ constraint $D^{\prime }=0$ given by
first relation of (\ref{2D}) and why they do not solve the
$SU_{L}\left( 2\right) $
constraint $D^{A}=0$. If eqs(\ref{ab}) could also solve the constraint $%
D^{A}=0$, then one disposes of an interesting information on
supersymmetry since in this view, the value
\begin{equation}
\tan \beta =1  \label{tan}
\end{equation}%
may be interpreted as a \emph{supersymmetric signal}.\newline
In what follows, we explore further this issue; the price to pay for having (%
\ref{tan}) without changing the number of degrees of freedom is
remarkably
low as it is given by just modifying the quantum numbers of $\boldsymbol{H}%
_{d}^{i}$ as done below.
\end{itemize}

\subsubsection{An anti-doublet $\Phi _{i}$ \emph{at place} of the doublet%
\emph{\ }$\boldsymbol{H}_{d}^{i}$}

If keeping the singlet $\boldsymbol{S}$ and the doublet
$\boldsymbol{H}_{u}$
as in \emph{n-MSSM}; but replacing the\ chiral doublet $\boldsymbol{H}%
_{d}^{i}$ by a chiral anti-doublet $\left( \Phi _{i}\right) $, i.e:%
\begin{equation*}
\boldsymbol{H}_{d}^{i}\qquad \rightarrow \qquad \Phi _{i}
\end{equation*}%
one can have the property (\ref{tan}). This replacement, which is
allowed by $SL\left( 2,C\right) $ representation theory analysis to
be developed in
next section, is manifested by the \emph{change of the sign} of the $%
SU_{L}\left( 2\right) $ gauge coupling constant $g$ as shown on the
gauge
covariant derivatives,%
\begin{equation}
\begin{tabular}{lll}
&  &  \\
$\left( \nabla _{\mu }h\right) ^{i}$ & $=$ & $\left[ \delta
_{j}^{i}\partial _{\mu }-i\frac{g^{\prime }}{2}\delta _{j}^{i}B_{\mu
}-igW_{\mu }^{A}\left(
\frac{\tau _{A}}{2}\right) _{j}^{i}\right] h^{i}$ \\
&  &  \\
$\left( \mathcal{D}_{\mu }\varphi \right) _{i}$ & $=$ & $\left[
\delta _{i}^{j}\partial _{\mu }+i\frac{g^{\prime }}{2}\delta
_{i}^{j}B_{\mu }+igW_{\mu }^{A}\left( \frac{\tau _{A}}{2}\right)
_{i}^{j}\right] \varphi
_{j}$ \\
&  &
\end{tabular}
\label{FL}
\end{equation}%
these gauge covariant derivatives should be compared with the
corresponding \emph{n-MSSM} ones given by eqs(\ref{dh}).\newline
With this change, the new equations of motion of the F-auxiliary
fields read
as%
\begin{equation}
\begin{tabular}{lll}
$Sh_{u}^{i}$ & $=$ & $0$ \\
$S\varphi _{i}$ & $=$ & $0$ \\
$\varphi _{i}h_{u}^{i}+\kappa S^{2}$ & $=$ & $\mathrm{\nu }$%
\end{tabular}
\label{ni}
\end{equation}%
and those of the auxiliary fields D get modified like%
\begin{equation}
\begin{tabular}{lll}
$\left( \bar{h}_{i}h^{i}-\varphi _{i}\bar{\varphi}^{i}\right) $ & $=$ & $%
\mathrm{r}$ \\
$\left( \tau ^{A}\right) _{j}^{i}\left( \bar{h}_{i}h^{j}-\varphi _{i}\bar{%
\varphi}^{j}\right) $ & $=$ & $0$%
\end{tabular}
\label{in}
\end{equation}%
where the complex $\mathrm{\nu }$ and the real $\mathrm{r}$ are
Fayet Iliopoulos coupling constants that appear in the superspace
lagrangian density in the usual manner namely
\begin{equation}
\mathrm{\nu }\int d^{2}\theta \boldsymbol{S}+\mathrm{\bar{\nu}}\int d^{2}%
\bar{\theta}\boldsymbol{S}^{\dagger }+\mathrm{r}\int d^{2}\theta \boldsymbol{%
V}_{u_{_{1}}}  \label{nr}
\end{equation}%
Here $\left \vert \mathrm{\nu }\right \vert $ is assumed large with
respect to $r$.

\subsection{Summary of other results}

The physically interesting solutions of the constraint relations (\ref{ni}-%
\ref{in}) correspond to zero VEV for the isosinglet ($S=0$) and non
zero VEVs for $h^{i}$ and $\varphi _{i}$. Depending on the values of
the Kahler parameter $r$, we distinguish two phases of the ground
state $\Sigma $:
\newline
$\left( i\right) $ a supersymmetric ground state with $r=0$; and
\newline $\left( ii\right) $ a non supersymmetric one for $r\neq 0$.

\subsubsection{Supersymmetric phase $r=0$}

In this phase, we find that the VEVs of the Higgs fields $h^{i}$ and $%
\varphi _{i}$ solving the equations of motion of all auxiliary fields (\ref%
{ni}-\ref{in}) are given by

\begin{eqnarray}
h^{i} &=&\sqrt{\left \vert \mathrm{\nu }\right \vert }\text{ }\left(
\begin{array}{c}
\cos \frac{\mathrm{\theta }}{2}e^{\frac{i}{2}\left( \psi +\phi \right) } \\
\\
\sin \frac{\mathrm{\theta }}{2}e^{\frac{i}{2}\left( \psi -\phi \right) }%
\end{array}%
\right)  \notag \\
&&  \label{lso} \\
\varphi _{i} &=&\sqrt{\left \vert \mathrm{\nu }\right \vert }\text{
}\left(
\begin{array}{c}
\cos \frac{\mathrm{\theta }}{2}e^{-\frac{i}{2}\left( \psi +\phi \right) }%
\text{ } \\
\\
\sin \frac{\mathrm{\theta }}{2}e^{-\frac{i}{2}\left( \psi -\phi \right) }%
\end{array}%
\right)  \notag
\end{eqnarray}%
\begin{equation*}
\end{equation*}%
where the complex $\mathrm{\nu }$ is as in (\ref{ni}) and the real $\mathrm{%
\theta },$ $\psi $ and $\phi $ are the usual angles\textrm{\footnote{%
not confuse the angle $\mathrm{\theta }$ with the Grassman variable
denoted by the same letter. }} of the real 3-sphere
$\mathbb{S}^{3}.$\newline The VEVs in the supersymmetric ground
state $\Sigma _{\text{susy}}$ obey the
property%
\begin{equation*}
\bar{h}_{i}h^{i}=\varphi _{i}\bar{\varphi}^{i}=\left \vert \mathrm{\nu }%
\right \vert
\end{equation*}%
and lead to%
\begin{equation*}
\tan \beta _{susy}=1
\end{equation*}%
The metric of the Higgs ground state $\Sigma _{\text{susy}}$ is
induced from
the metric of the complex space $\mathbb{C}^{4}$ namely%
\begin{equation}
ds^{2}=d\bar{h}_{i}dh^{i}+d\varphi _{i}d\bar{\varphi}^{i}
\end{equation}%
By substituting (\ref{lso}) into this relation, we end with the
following
expression%
\begin{equation}
ds_{_{\Sigma _{\text{susy}}}}^{2}=\frac{1}{2}\left \vert \mathrm{\nu
}\right
\vert \left( d\mathrm{\theta }^{2}+d\psi ^{2}+d\phi ^{2}+2\cos \mathrm{%
\theta }d\psi d\phi \right)
\end{equation}%
Moreover, seen that $ds^{2}$ is invariant by $U\left( 1\right) $
phase changes of the Higgs fields
\begin{equation}
h^{i\prime }=e^{i\alpha }h^{i},\qquad \varphi _{i}^{\prime
}=e^{-i\alpha }\varphi _{i}  \label{sy}
\end{equation}%
one can use this symmetry to gauge away a real degree of freedom by setting $%
\psi -\phi =0$; this leads to
\begin{eqnarray}
h^{i} &=&\sqrt{\left \vert \mathrm{\nu }\right \vert }\text{ }\left(
\begin{array}{c}
\cos \frac{\mathrm{\theta }}{2}e^{i\phi } \\
\\
\sin \frac{\mathrm{\theta }}{2}%
\end{array}%
\right)  \notag \\
&& \\
\varphi _{i} &=&\sqrt{\left \vert \mathrm{\nu }\right \vert }\text{
}\left(
\begin{array}{c}
\cos \frac{\mathrm{\theta }}{2}e^{-i\phi }\text{ } \\
\\
\sin \frac{\mathrm{\theta }}{2}%
\end{array}%
\right)  \notag
\end{eqnarray}%
and
\begin{equation}
ds_{_{\Sigma _{susy}}}^{2}=\frac{1}{2}\left \vert \mathrm{\nu
}\right \vert \left( d\mathrm{\theta }^{2}+4\cos
^{2}\frac{\mathrm{\theta }}{2}d\phi ^{2}\right)  \label{osl}
\end{equation}%
\begin{equation*}
\end{equation*}%
which is nothing but the metric of a real 2-sphere $\mathbb{S}_{susy}^{2}$.%
\newline
Notice moreover the following properties:

\begin{description}
\item[$\mathbf{1)}$] \  \  \  \emph{geometry of ground state}: $\Sigma _{\text{%
susy}}\sim CP^{1}\sim S^{2}$\  \newline The ground state $\Sigma
_{susy}$ of the Higgs fields is given by a real 2-sphere with radius
\begin{equation*}
R_{\Sigma _{susy}}=\sqrt{\left \vert \mathrm{\nu }\right \vert }
\end{equation*}%
and metric $ds_{_{\Sigma _{susy}}}^{2}$ as above. So the absolute of
the complex Fayet Iliopoulos coupling constant $\mathrm{\nu }$ is,
up to the factor $4\pi $, the area of $\Sigma _{\text{susy}}$.
\newline In the limit $\mathrm{\nu }\rightarrow 0$, the 2-sphere
shrinks to the origin of $\mathbb{C}^{2}\subset \mathbb{C}^{4}$; the
metric
\begin{equation*}
\mathcal{G}=\frac{\left \vert \mathrm{\nu }\right \vert }{2}\left(
\begin{array}{cc}
1 & 0 \\
0 & 4\cos ^{2}\frac{\mathrm{\theta }}{2}%
\end{array}%
\right)
\end{equation*}%
of the ground state $ds_{_{\Sigma _{susy}}}^{2}$ becomes singular;
but the value of the ratio $\tan \beta _{susy}$ of the VEVs is
unaffected,
\begin{equation*}
\tan \beta _{susy}=1
\end{equation*}%
Actually the limit $\mathrm{\nu }\rightarrow 0$ is forbidden in our
approximation since we have assumed $\mathcal{V}_{ch}$ large with
respect to the other contributions to the total scalar potential
$\mathcal{V}$.

\item[$\mathbf{2)}$] \  \  \  \emph{self dual point}\newline
In the phase $r=0$, the doublet $h^{i}$\ and anti-doublet $\varphi
_{i}$\
parameterize same sphere since we have$\ $%
\begin{equation*}
\mathbb{S}_{h}^{2}\subset \mathbb{S}_{\varphi }^{2}\qquad \emph{and\  \qquad }%
\mathbb{S}_{\varphi }^{2}\subset \mathbb{S}_{h}^{2}
\end{equation*}%
This property is manifested in various ways; for example through the
solution $h^{i}$ and $\varphi _{i}$ of the constraint relations
which happen to be related like
\begin{equation*}
\varphi _{i}=\overline{\left( h^{i}\right) }
\end{equation*}%
This means that $\varphi _{i}$ is basically playing the role of the
Higgs anti-doublet $\bar{h}_{i}$ of standard model. We also have the
two following remarkable features:

\begin{itemize}
\item the Higgs configurations (\ref{lso}) describe a degenerate geometric
situation where the real 2-spheres $\mathbb{S}_{h}^{2}$ and $\mathbb{S}%
_{\varphi }^{2}$ with defining eqs
\begin{eqnarray}
\mathbb{S}_{h}^{2} &:&\left \vert \bar{h}_{1}\right \vert ^{2}+\left
\vert
\bar{h}_{2}\right \vert ^{2}=\varrho _{h}^{2}  \notag \\
&&  \label{SS} \\
\mathbb{S}_{\varphi }^{2} &:&\left \vert \varphi _{1}\right \vert
^{2}+\left
\vert \varphi _{2}\right \vert ^{2}=\varrho _{\varphi }^{2}  \notag \\
&&  \notag
\end{eqnarray}%
overlap and merge into a single 2-sphere $\mathbb{S}_{susy}^{2}$
with area
\begin{equation*}
4\pi \varrho _{h}^{2}=4\pi \varrho _{\varphi }^{2}=4\pi \left \vert \mathrm{%
\nu }\right \vert
\end{equation*}

\item the ratio $\tan \beta _{susy}$ of the VEVs of Higgs fields $h^{i}$ and
$\varphi _{i}$ captures precisely the merging property of $\mathbb{S}%
_{h}^{2} $ and $\mathbb{S}_{\varphi }^{2}$ into a unique sphere $\mathbb{S}%
_{susy}^{2} $. Deviation of $\tan \beta $ from its supersymmetric value $%
\tan \beta _{susy}=1$ leads to a splitting of the two spheres and
then to supersymmetry breaking.
\end{itemize}
\end{description}

\subsubsection{Non supersymmetric phase $r\neq 0$}

In this case, we find that the previous solution (\ref{lso}), giving
the VEVs of the Higgs fields $h^{i}$ and $\varphi _{i}$ for $r=0,$
gets modified
as follows%
\begin{equation}
h^{i}=\varrho \text{ }\left(
\begin{array}{c}
\cos \frac{\mathrm{\theta }}{2}e^{\frac{i}{2}\left( \psi +\phi \right) } \\
\\
\sin \frac{\mathrm{\theta }}{2}e^{\frac{i}{2}\left( \psi -\phi \right) }%
\end{array}%
\right)  \label{hr}
\end{equation}%
and%
\begin{equation}
\varphi _{i}=\frac{\mathrm{\nu }}{\varrho }\text{ }\left(
\begin{array}{c}
\cos \frac{\mathrm{\theta }}{2}e^{-\frac{i}{2}\left( \psi +\phi \right) }%
\text{ } \\
\\
\sin \frac{\mathrm{\theta }}{2}e^{-\frac{i}{2}\left( \psi -\phi \right) }%
\end{array}%
\right)  \label{fr}
\end{equation}%
with%
\begin{equation*}
\end{equation*}%
\begin{equation*}
\varrho ^{2}=\frac{g^{\prime 2}\mathrm{r}+\sqrt{g^{\prime 4}\mathrm{r}%
^{2}+4\left( g^{2}+g^{\prime 2}\right) ^{2}\mathrm{\nu
\bar{\nu}}}}{2\left( g^{2}+g^{\prime 2}\right) }
\end{equation*}%
\begin{equation*}
\end{equation*}%
By taking the limit $r=0$, one recovers exactly the previous
supersymmetric solutions.\newline Notice moreover the following
properties:

\begin{itemize}
\item \emph{Lifting degeneracy of the 2-spheres }$\mathbb{S}_{h}^{2}$ and $%
\mathbb{S}_{\varphi }^{2}$\newline
By switching on of the FI coupling constant r, the previous 2-spheres $%
\mathbb{S}_{h}^{2}$ and $\mathbb{S}_{\varphi }^{2}$ are no longer
merged;
they dissociate and become%
\begin{eqnarray*}
\mathbb{\tilde{S}}_{h}^{2} &:&\left \vert \bar{h}_{1}\right \vert
^{2}+\left
\vert \bar{h}_{2}\right \vert ^{2}=\varrho ^{2} \\
&& \\
\mathbb{\tilde{S}}_{\varphi }^{2} &:&\left \vert \varphi _{1}\right
\vert ^{2}+\left \vert \varphi _{2}\right \vert ^{2}=\frac{\left
\vert \mathrm{\nu
}\right \vert ^{2}}{\varrho ^{2}} \\
&&
\end{eqnarray*}%
with $\varrho $ as in (\ref{rn}). The respective areas of these
2-spheres are given by
\begin{equation*}
A_{h}=4\pi \varrho ^{2}\qquad ,\qquad A_{\varphi }=4\pi \frac{\left
\vert \mathrm{\nu }\right \vert ^{2}}{\varrho ^{2}}
\end{equation*}%
and they become equal ($A_{h}=A_{\varphi }$) exactly for the value
$\varrho ^{2}=\left \vert \mathrm{\nu }\right \vert $ where live
supersymmetry; see fig \ref{su}\ for illustration.
\begin{figure}[tbph]
\begin{center}
\hspace{0cm} \includegraphics[width=8cm]{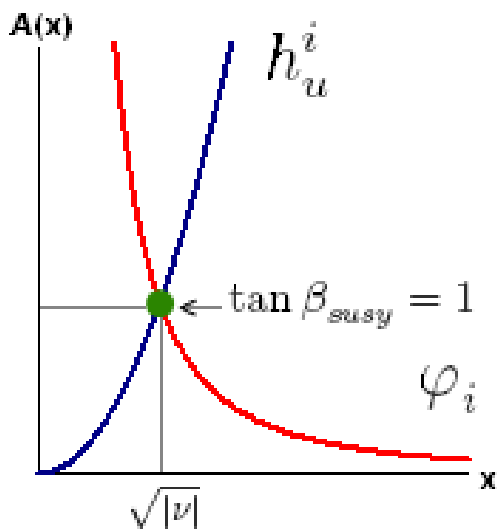}
\end{center}
\par
\vspace{-1cm}
\caption{Variation of the areas A$\left( x\right) $ of the spheres $\mathbb{S%
}_{h}^{2}$ and $\mathbb{S}_{\protect \varphi }^{2}$ in terms of their radius $%
x$. In blue A$_{h}$ and in red A$_{\protect \varphi }$; at
intersecting point $\tan \protect \beta _{susy}=1$ and supersymmetry
is restored.} \label{su}
\end{figure}

\item \emph{Vacuum energy }$\mathcal{E}_{\min }$\newline
Being a non supersymmetric ground state, the energy density $\mathcal{E}%
_{\min }$ of the state $\left \vert \tilde{\Sigma}_{nsusy}\right
\rangle $
is non zero and is given by%
\begin{equation*}
\mathcal{E}_{\min }=\frac{g^{\prime 2}g^{2}}{8\left( g^{2}+g^{\prime
2}\right) }\mathrm{r}^{2}
\end{equation*}%
where $g$ and $g^{\prime }$ are the gauge coupling constants of the $%
SU_{L}\left( 2\right) \times U_{Y}\left( 1\right) $ gauge symmetry
group.

By using the well known standard model relations
\begin{equation*}
\sin \vartheta _{_{W}}=\frac{g^{\prime }}{\sqrt{g^{2}+g^{\prime
2}}},\qquad \cos \vartheta _{_{W}}=\frac{g}{\sqrt{g^{2}+g^{\prime
2}}}
\end{equation*}%
expressing the gauge coupling constants $g$ and $g^{\prime }$ in
terms of
the Weinberg angle $\vartheta _{_{W}}$, we then have%
\begin{equation*}
\mathcal{E}_{\min }=\frac{\sin ^{2}\left( 2\vartheta _{_{W}}\right) }{32}%
\mathrm{r}^{2}
\end{equation*}%
and%
\begin{equation*}
\end{equation*}%
\begin{equation}
\varrho ^{2}=\frac{\mathrm{r}\sin ^{2}\vartheta _{_{W}}}{2}+\left
\vert \mathrm{\nu }\right \vert \sqrt{1+\frac{\mathrm{r}^{2}\sin
^{4}\vartheta _{_{W}}}{4\left \vert \mathrm{\nu }\right \vert ^{2}}}
\label{rn}
\end{equation}

\item \  \emph{Deviation of} $\tan \beta $ \newline
Using the relation between $\varphi _{i}$ and $\bar{h}_{i}$ that
reads in present case like
\begin{equation*}
\varphi _{i}=\frac{\mathrm{\nu }}{\varrho ^{2}}\bar{h}_{i},\qquad
\varphi
_{i}\bar{\varphi}^{i}=\frac{\left \vert \mathrm{\nu }\right \vert ^{2}}{%
\varrho ^{2}}
\end{equation*}%
one learns that the ratio of the Higgs VEVs gets modified and leads
to a deviation with respect to $\tan \beta _{susy}=1$. This
deviation $\tan \beta =\frac{\varrho ^{2}}{\left \vert \mathrm{\nu
}\right \vert }$ and is expressed in terms of the Weinberg angle of
standard model of electroweak
interactions like%
\begin{equation}
\tan \beta =\frac{\mathrm{r}\sin ^{2}\vartheta _{_{W}}}{2\left \vert \mathrm{%
\nu }\right \vert }+\sqrt{1+\left( \frac{\mathrm{r}\sin ^{2}\vartheta _{_{W}}%
}{2\left \vert \mathrm{\nu }\right \vert }\right) ^{2}}
\label{tanb}
\end{equation}%
From this formula, we also learn that the dimensionless parameter
\begin{equation*}
\frac{\mathrm{r}}{2\left \vert \mathrm{\nu }\right \vert }
\end{equation*}%
captures data on supersymmetry breaking. For a small supersymmetry
breaking
regime, we have $\frac{\mathrm{r}}{2\left \vert \mathrm{\nu }\right \vert }%
<<1$; so we can expand the above $\tan \beta $ relation; we get up
to first order
\begin{equation}
\tan \beta \simeq 1+\frac{\mathrm{r}\sin ^{2}\vartheta
_{_{W}}}{2\left \vert \mathrm{\nu }\right \vert }+\frac{1}{2}\left(
\frac{\mathrm{r}\sin ^{2}\vartheta _{_{W}}}{2\left \vert \mathrm{\nu
}\right \vert }\right) ^{2}
\end{equation}%
For a large supersymmetry breaking regime corresponding to $\frac{\mathrm{r}%
}{\left \vert \mathrm{\nu }\right \vert }>>1$; we have%
\begin{equation}
\tan \beta \simeq \frac{\mathrm{r}\sin ^{2}\vartheta _{_{W}}}{\left
\vert \mathrm{\nu }\right \vert }
\end{equation}%
but this limit is beyond the used approximation $\mathcal{V}_{ch}>>\mathcal{V%
}_{re}$.

\item \emph{metric of ground state} $\Sigma _{n\text{-}susy}$\newline
In the case $r\neq 0$, the metric of the ground state of the Higgs
fields reads as
\begin{equation}
ds_{\Sigma _{n-susy}}^{2}=\frac{1}{4}\left( \varrho ^{2}+\frac{\left
\vert \mathrm{\nu }\right \vert ^{2}}{\varrho ^{2}}\right) \left[
d\theta ^{2}+d\psi ^{2}+d\phi ^{2}+2\cos \theta d\psi d\phi \left.
\begin{array}{c}
\\
\end{array}%
\right. \right]
\end{equation}%
\begin{equation*}
\end{equation*}%
where the angles $\theta ,$ $\psi $ and $\phi $ are as in eqs(\ref{hr}-\ref%
{fr}) and $\varrho ^{2}$ like in eq(\ref{rn}). We also have the
useful
relation%
\begin{equation}
\varrho ^{2}+\frac{\left \vert \mathrm{\nu }\right \vert ^{2}}{\varrho ^{2}}=%
\mathrm{r}+\frac{2\left \vert \mathrm{\nu }\right \vert
^{2}}{\mathrm{r}\sin
^{2}\vartheta _{_{W}}+\sqrt{4\left \vert \mathrm{\nu }\right \vert ^{2}+%
\mathrm{r}^{2}\sin ^{4}\vartheta _{_{W}}}}  \label{R2}
\end{equation}%
\begin{equation*}
\end{equation*}%
Moreover, because of the symmetry (\ref{sy}), the Higgs fields can
be written like\newline
\begin{equation}
h^{i}=\varrho \text{ }\left(
\begin{array}{c}
\cos \frac{\theta }{2}e^{i\phi } \\
\\
\sin \frac{\theta }{2}%
\end{array}%
\right)
\end{equation}%
and%
\begin{equation}
\varphi _{i}=\frac{\mathrm{\nu }}{\varrho }\text{ }\left(
\begin{array}{c}
\cos \frac{\theta }{2}e^{-i\phi }\text{ } \\
\\
\sin \frac{\theta }{2}%
\end{array}%
\right)
\end{equation}%
The previous metric becomes%
\begin{equation*}
ds_{\Sigma _{n-susy}}^{2}=\frac{\varrho ^{4}+\left \vert \mathrm{\nu
}\right \vert ^{2}}{4\varrho ^{2}}\left[ d\theta ^{2}+4\left( \cos
^{2}\frac{\theta }{2}\right) d\phi ^{2}\right]
\end{equation*}%
with no dependence in $\psi $. \newline From the knowledge of the
electric charges of the components fields of the doublet
$h^{i}=\left( h^{+},h^{0}\right) $ and the anti-doublet $\varphi
_{i}=\left( \varphi ^{-},\varphi ^{0}\right) $, it follows that the angle $%
\phi $ parameterizes precisely the electromagnetic circle $\mathbb{S}%
_{em}^{1}$ with rotation generator $Q_{em}$ given by%
\begin{equation*}
Q_{em}=\frac{\partial }{i\partial \phi }
\end{equation*}

\item \emph{the electromagnetic point }$P_{u_{1}^{em}}$\newline
If setting the angular variable $\theta $ to a constant; say $\theta
=\chi $ with $d\chi =0$; the above metric reduces to the
$\mathbb{S}_{em}^{1}$ metric
\begin{equation*}
ds_{\Sigma _{n-susy}}^{2}|_{\theta =\chi }=\left( \frac{\varrho
^{4}+\left
\vert \mathrm{\nu }\right \vert ^{2}}{\varrho ^{2}}\cos ^{2}\frac{\chi }{2}%
\right) d\phi ^{2}
\end{equation*}%
\begin{equation*}
\end{equation*}%
which vanishes precisely at the south point of the real 2-sphere
\begin{equation*}
\chi =\pi \text{ \ },\text{ \ }\func{mod}\pi
\end{equation*}%
where lives $P_{u_{1}^{em}}$. There the Higgs fields $h^{i}$ and
$\varphi
_{i}$ take the values%
\begin{equation}
\begin{tabular}{lll}
$h^{i}=\varrho $ $\left(
\begin{array}{c}
0 \\
1%
\end{array}%
\right) $ & , & $\varphi _{i}=\frac{\mathrm{\nu }}{\varrho }\text{
}\left(
\begin{array}{c}
0\text{ } \\
1%
\end{array}%
\right) $%
\end{tabular}%
\end{equation}%
and have no electric charges.
\end{itemize}

\section{Higgs vacua and intersecting conifolds}

In this section, we develop the idea where the down Higgs superfield
doublet $\left( \boldsymbol{H}_{d}\right) ^{i}$ of \emph{n-MSSM} is
\emph{replaced} by a chiral superfield \emph{anti-doublet} $\left(
\Phi _{i}\right) $. First, we study the group theoretical set up of
the proposal, then we give its superfield formulation, and after we
describe its link with intersecting conifold geometries.

\subsection{Higgs doublet and anti-doublet in \emph{n-MSSM*}}

Here, we focus on the gauge and Higgs sectors of the next - to -
MSSM and study the basic property that allows to replace the role of
doublet $\left(
\boldsymbol{H}_{d}\right) ^{i}$ by the anti-doublet $\boldsymbol{\Phi }_{%
\bar{\imath}}$. \newline
To fix the ideas, think about the chiral superfields $\left( \boldsymbol{H}%
_{d}\right) ^{i}$ and $\boldsymbol{\Phi }_{\bar{\imath}}$ as
transforming into the two fundamental representations of $SL\left(
2,C\right) $ as exhibited below
\begin{equation}
\begin{tabular}{lllll}
&  &  &  &  \\
\  \ {\small chiral superfield} & : \  \  \  \  \  \  & $\
\boldsymbol{H}_{d}^{i}$
& $\  \  \  \rightarrow $ \  \  \  & $\  \boldsymbol{\Phi }_{\bar{\imath}}$ \\
${\small SL(2)}$ {\small representation \  \  \  \  \ } & : & $(\frac{1}{2}%
,0)_{-}$ &  & $(0,\frac{1}{2})_{-}$ \\
&  &  &  &
\end{tabular}%
\end{equation}%
with the sub-index $\left( -\right) $ referring to the hypercharge.
As described in previous section, this replacement leads to
important consequences; in particular to a change into the two
following:

\begin{itemize}
\item the sign of the $SU_{L}\left( 2\right) $ gauge coupling constant $g$\
in the gauge covariant derivative $\left( \mathcal{D}_{\mu }\varphi
\right) _{\bar{\imath}}$ compared to $\left( \mathcal{D}_{\mu
}h_{d}\right) ^{i}$; and

\item the equation of motion of the auxiliary fields $D^{\prime }$ and $%
D^{A} $; since eqs(\ref{2D}) and (\ref{in}) differ as well by a
sign.
\end{itemize}

\subsubsection{Group theory basis of the proposal}

To justify the replacement of the Higgs doublet
$\boldsymbol{H}_{d}^{i}$ by the anti-doublet $\boldsymbol{\Phi
}_{i}$, it is interesting to begin by recalling rapidly some basic
tools and useful properties in the study of supersymmetric gauge
theories in superspace with focus on \emph{n-MSSM}.

\  \  \  \

\textbf{1)} \emph{Basic ingredients}\newline
First recall that, being a particular supersymmetric gauge theory, \emph{%
n-MSSM} has \emph{29} superfields, among which the \emph{4} gauge
supersymmetric multiplets; and \emph{5} Higgs chiral multiplets
carrying
different charges under the non abelian gauge symmetry%
\begin{equation*}
U_{Y}\left( 1\right) \times SU_{L}\left( 2\right)
\end{equation*}%
In superspace, these supersymmetric Higgs multiplets are described by \emph{5%
} chiral superfields belonging to different representations of the
gauge symmetry; these are:

\begin{itemize}
\item the hyperchargeless chiral iso-singlet $\boldsymbol{S}$ having no
direct interactions with the gauge superfields; and,

\item the two chiral superfield iso-doublets: the up-Higgs $\boldsymbol{H}%
_{u}$ and the down-Higgs $\boldsymbol{H}_{d}$ having opposite
hypercharges; but same quantum charge under $SU_{L}\left( 2\right)
$. \newline
These two doublets couple to the $U_{Y}\left( 1\right) $ gauge multiplet $%
\boldsymbol{V}^{\prime }$ in opposite ways; but with the same manner to the $%
SU_{L}\left( 2\right) $ gauge multiplet $\boldsymbol{V}^{A}$.
\end{itemize}

\  \  \  \newline Recall also that because of superspace chirality
condition of the Higgs
superfields namely%
\begin{equation*}
\mathcal{\bar{D}}_{\dot{a}}\boldsymbol{S}=0\boldsymbol{,\qquad }\mathcal{%
\bar{D}}_{\dot{a}}\boldsymbol{H}_{u}=0,\qquad \mathcal{\bar{D}}_{\dot{a}}%
\boldsymbol{H}_{d}=0
\end{equation*}%
with $\mathcal{\bar{D}}_{\dot{a}}$ the usual superspace covariant
derivative \textrm{\cite{W}}, the gauge symmetry transformations of
the Higgs chiral superfields require going beyond the unitary
$U_{Y}\left( 1\right) \times
SU_{L}\left( 2\right) $ to its complex extension%
\begin{equation*}
\mathbb{C}_{Y}^{\ast }\times SL\left( 2,C\right) \text{ \ }{\Large \supset }%
\text{ \ }U_{Y}\left( 1\right) \times SU_{L}\left( 2\right)
\end{equation*}%
containing the unitary gauge symmetry as a subgroup. This complex
extension
has a larger number of representations; in particular the two inequivalent $%
SL\left( 2,C\right) $ fundamental iso-spinors namely%
\begin{equation*}
\begin{tabular}{lllll}
$(\frac{1}{2},0)$ &  & and &  & $(0,\frac{1}{2})$%
\end{tabular}%
\end{equation*}%
So chiral superfields in \emph{n-MSSM} can transform either in the
fundamental representation $(\frac{1}{2},0)$ or in the anti-fundamental $(0,%
\frac{1}{2})$ one. To fix the ideas, the quantum numbers of chiral
and antichiral superfields under SL$\left( 2,C\right) $ may in
general be as
follows%
\begin{equation*}
\begin{tabular}{lllll}
&  & chiral &  & antichiral \\
$(\frac{1}{2},0)$ & : & $\  \boldsymbol{\Psi }^{i}$ &  & $\  \boldsymbol{\Psi }%
_{\bar{\imath}}^{\dagger }$ \\
$(0,\frac{1}{2})$ & : & $\  \boldsymbol{\Phi }_{\bar{\imath}}$ &  &
$\
\boldsymbol{\Phi }^{\dagger i}$%
\end{tabular}%
\end{equation*}%
Gauge invariance of the Kahler potential $\mathcal{K}\left(
\boldsymbol{\Psi },\boldsymbol{\Psi }^{\dagger }\right) $ of the
superspace lagrangian density of \emph{n-MSSM} is ensured by the
implementation of the exponential
of the gauge superfield multiplet $\boldsymbol{V}$ that transforms in the $(%
\frac{1}{2},\frac{1}{2})$ representation of $SL\left( 2,C\right) $
subject
to a hermiticity condition; and through which the two fundamental $(\frac{1}{%
2},0)$ and $(0,\frac{1}{2})$ couple. In superfield language, the
typical
gauge invariant coupling is given as usual by%
\begin{equation*}
\begin{tabular}{lll}
&  &  \\
$\mathcal{K}\left( \boldsymbol{\Psi },\boldsymbol{\Psi }^{\dagger
}\right) $ & $\sim $ & $\boldsymbol{\Psi }^{\dagger }\times
e^{-\boldsymbol{V}}\times
\boldsymbol{\Psi }$ \\
&  &  \\
& $\equiv $ & $\left[ \boldsymbol{\Psi }^{\dagger }.\mathcal{U}^{\dagger }%
\right] \times \left[ \left( \mathcal{U}^{\dagger }\right) ^{-1}.\text{ }e^{-%
\boldsymbol{V}}.\left( \mathcal{U}\right) ^{-1}\right] \times \left[
\mathcal{U}.\boldsymbol{\Psi }\right] $%
\end{tabular}%
\end{equation*}

\  \  \  \newline where the chiral superfield matrix $\mathcal{U}$
stands for an arbitrary representation group element of
$\mathbb{C}_{Y}^{\ast }\times SL\left(
2,C\right) $ with the properties%
\begin{equation*}
\begin{tabular}{lll}
$\left( \mathcal{U}\right) ^{-1}$ & $\neq $ & $\left(
\mathcal{U}\right)
^{\dagger }$ \\
$\mathcal{\bar{D}}_{\dot{a}}\mathcal{U}$ & $=$ & $0$%
\end{tabular}%
\end{equation*}%
From representation group theory view, the superfield's coupling $%
\boldsymbol{\Psi }^{\dagger }$ $e^{-\boldsymbol{V}}$
$\boldsymbol{\Psi }$
may roughly be thought of as follows%
\begin{eqnarray*}
Tr\left[ \text{ \ }\left( \frac{1}{2},0\right) \text{ }\otimes \text{ }%
\left( \frac{1}{2},\frac{1}{2}\right) ^{\otimes n}\text{ }\otimes \text{ }%
\left( 0,\frac{1}{2}\right) \text{ \ }\right] ,\qquad n &\geq &0 \\
&&
\end{eqnarray*}%
where we have used with $e^{-\boldsymbol{V}}=\sum \frac{1}{n!}\left( -%
\boldsymbol{V}\right) ^{n}$. The correspondence between the leading complex $%
SL\left( 2,C\right) $ representations and the unitary $SU_{L}\left(
2\right)
$ ones is as collected below%
\begin{eqnarray}
&&%
\begin{tabular}{llll}
&  &  &  \\
$\  \  \  \  \ SL\left( 2,C\right) $ & \  \  \  \  \  \  \  \  \  \
& \multicolumn{1}{|l}{\  \  \  \  \  \  \  \  \  \ } & $\
SU_{L}\left( 2\right) $ \  \  \
\  \  \  \  \  \\
$\  \  \  \  \  \  \ (0,0)$ &  & \multicolumn{1}{|l}{} & $\  \  \ 1$ \\
$\  \  \  \  \  \  \ (\frac{1}{2},0)$ &  & \multicolumn{1}{|l}{} & $\  \  \ 2$ \\
$\  \  \  \  \  \  \ (0,\frac{1}{2})$ &  & \multicolumn{1}{|l}{} &
$\  \  \  \bar{2}$
\\
$\  \  \  \  \  \  \ (\frac{1}{2},\frac{1}{2})$ &  &
\multicolumn{1}{|l}{} & $\
2\otimes \bar{2}=1\oplus 3$%
\end{tabular}
\label{bet} \\
&&  \notag
\end{eqnarray}%
from which we learn that $(\frac{1}{2},\frac{1}{2})$ reduces to the
sum of two $SU_{L}\left( 2\right) $ representations.

\  \  \  \

\textbf{2)} \emph{Useful properties}\newline
We give two useful features on the relation between the representations of $%
SL\left( 2,C\right) $ and those of $SU_{L}\left( 2\right) $. These
features can be also learnt from the correspondences of table
(\ref{bet}).

\begin{itemize}
\item besides that it can be subject to \emph{a reality condition}
\begin{equation*}
\boldsymbol{V}^{\dagger }=\boldsymbol{V}
\end{equation*}%
the complex 4- dimensional $SL\left( 2,C\right) $ representation $(\frac{1}{2%
},\frac{1}{2})$, when reduced to $SU_{L}\left( 2\right) $, has the
remarkable property of containing a complex $SU_{L}\left( 2\right) $
iso-singlet as exhibited here below
\begin{equation}
(\frac{1}{2},\frac{1}{2})\text{ \  \  \ }\rightarrow \text{ \  \  \
}1\oplus 3 \label{re}
\end{equation}%
This feature teaches us that complex \emph{4} representation $(\frac{1}{2},%
\frac{1}{2})$ as well as its hermitian version contains an
$SU_{L}\left( 2\right) $ singlet; a desired feature for models
building in elementary particles.

\item the above reduction (\ref{re}) down to $SU_{L}\left( 2\right) $
representation is fact a particular one among two other cousin ones;
namely

\begin{equation}
\begin{tabular}{lllll}
\hline \hline
&  &  &  &  \\
\multicolumn{3}{l}{\  \  \  \  \  \  \  \  \  \  \  \  \  \  \  \  \  \ $SL\left( 2,C\right) $%
} & $\  \  \  \ :$ & $\  \  \ SU_{L}\left( 2\right) $ \  \  \  \  \  \  \  \\
&  &  &  &  \\
$(\frac{1}{2},0)\otimes (\frac{1}{2},0)$ & $=$ & $(0,0)\oplus (1,0)$
\  \  \
& $\  \  \  \  \rightarrow $ \  \  \  \  \  \  \  & $\  \  \  \  \ 1\oplus 3$ \\
$(0,\frac{1}{2})\otimes (0,\frac{1}{2})$ & $=$ & $(0,0)\oplus (0,1)$
& $\  \
\  \  \rightarrow $ & $\  \  \  \  \  \bar{1}\oplus \bar{3}$ \\
$(\frac{1}{2},0)\otimes (0,\frac{1}{2})$ & $=$ & $\  \  \  \ (\frac{1}{2},\frac{%
1}{2})$ & $\  \  \  \  \rightarrow $ & $\  \  \  \  \ 1\oplus 3$ \\
&  &  &  &  \\ \hline \hline &  &  &  &
\end{tabular}
\label{pro}
\end{equation}%
The two first relations cannot be real; they are necessary complex.
But all of the 3 relations lead to complex $SU_{L}\left( 2\right) $
iso-singlets as exhibited on (\ref{pro}). \newline
\end{itemize}

\  \  \

\textbf{3)} \emph{Higgs superfields vs }$\mathbb{C}_{Y}^{\ast
}\times SL\left( 2,C\right) $\newline From the view of the complex
$\mathbb{C}_{Y}^{\ast }\times SL\left( 2,C\right) $ group
representations, the quantum numbers of the Higgs chiral superfield
doublets $\boldsymbol{H}_{u}$ and $\boldsymbol{H}_{d}$; as well
as their adjoint conjugates $\boldsymbol{H}_{u}^{\dagger }$ and $\boldsymbol{%
H}_{d}^{\dagger }$\ are as follows
\begin{eqnarray}
&&%
\begin{tabular}{llllll}
\hline
&  & \multicolumn{2}{l}{\emph{n-MSSM}} &  &  \\
\multicolumn{2}{l}{\  \  \  \  \ {\small chiral superfields }} & \
\  \  \  \  \  & \multicolumn{1}{||l}{\  \  \  \  \ } &
\multicolumn{2}{l}{\  \ {\small antichiral
superfields }} \\
&  &  & \multicolumn{1}{||l}{} &  &  \\
$\  \  \  \  \  \boldsymbol{H}_{u}$ & $\  \  \  \
\boldsymbol{H}_{d}$ &  & \multicolumn{1}{||l}{} & $\  \  \  \  \  \
\boldsymbol{H}_{u}^{\dagger }$ & $\  \
\  \  \  \  \  \boldsymbol{H}_{d}^{\dagger }$ \\
$\  \ (\frac{1}{2},0)_{+}$ \  & $\  \  \ (\frac{1}{2},0)_{-}$ &  &
\multicolumn{1}{||l}{} & \ $\  \  \ (0,\frac{1}{2})_{-}$ \  & $\  \  \  \  \ (0,%
\frac{1}{2})_{+}$ \  \  \  \  \  \  \\
&  &  &  &  &  \\ \hline
\end{tabular}
\label{ll} \\
&&  \notag
\end{eqnarray}%
where the $\pm q$ charges appearing as sub-indices of the $SL\left(
2,C\right) $ representation $\left( j,j^{\prime }\right) _{\pm q}$
refer to the $\mathbb{C}_{Y}^{\ast }$ action:
\begin{equation*}
\mathbb{C}_{Y}^{\ast }:%
\begin{tabular}{lll}
$\boldsymbol{H}_{x}$ & $\text{\ }\rightarrow \text{ \ }$ & $e^{\Lambda _{0}%
\frac{Y}{2}}\boldsymbol{H}_{x}$ \\
$\boldsymbol{H}_{x}^{\boldsymbol{\dagger }}$ & $\text{\ }\rightarrow
\text{
\ }$ & $\boldsymbol{H}_{x}^{\boldsymbol{\dagger }}e^{\Lambda _{0}^{%
\boldsymbol{\dagger }}\frac{Y}{2}}$%
\end{tabular}%
\end{equation*}%
with $\Lambda _{0}$ a chiral superfield
($\mathcal{\bar{D}}_{\dot{a}}\Lambda
_{0}=0$) and $\Lambda _{0}^{\boldsymbol{\dagger }}$ anti-chiral. $\mathbb{C}%
_{Y}^{\ast }$ is the complex extension of the $U_{Y}\left( 1\right)
$
hypercharge symmetry. Because of their opposite hypercharges and eq(\ref{pro}%
), we also have

\begin{eqnarray}
&&%
\begin{tabular}{l||l}
\  \  \ {\small chiral superfields } & \  \  \  \ {\small antichiral
superfields }
\\
$\  \  \  \  \  \  \boldsymbol{H}_{u}\boldsymbol{H}_{d}$ & $\  \  \
\  \  \  \  \  \  \  \
\  \boldsymbol{H}_{u}^{\dagger }\  \boldsymbol{H}_{d}^{\dagger }$ \\
$(0,0)_{_{0}}\oplus (1,0)_{_{0}}$ \  \  \  \  \  \  \  \  & \ $\  \
\  \  \ (0,0)_{_{0}}\oplus (0,1)_{_{0}}$ \  \  \
\end{tabular}
\\
&&  \notag
\end{eqnarray}%
showing that the composite chiral superfield
\begin{equation*}
\boldsymbol{H}_{u}\boldsymbol{H}_{d}
\end{equation*}%
and its monomials are good candidate for the chiral sector of
supersymmetry; $\boldsymbol{H}_{u}\boldsymbol{H}_{d}$ can be coupled
to the iso-singlet chiral superfield $\boldsymbol{S}$; but also to a
hyperchargeless isotriplet chiral superfield $\vec{\Delta}$;
\textrm{see appendix}
\begin{equation*}
\begin{tabular}{lllll}
$\boldsymbol{H}_{u}S\boldsymbol{H}_{d}$ &  &  & $\vec{\Delta}.\left(
\boldsymbol{H}_{u}\vec{T}\boldsymbol{H}_{d}\right) $ &
\end{tabular}%
\end{equation*}

\  \  \  \  \

\textbf{4)} up Higgs $\boldsymbol{H}_{u}$ \emph{as a doublet and
down Higgs} $\boldsymbol{\Phi }$ \emph{as an anti-doublet}\newline
In our proposal, the $\boldsymbol{H}_{d}$ of \emph{n-MSSM} is
replaced by the anti-doublet chiral superfield $\boldsymbol{\Phi }$
with quantum numbers under $SL\left( 2,C\right) $ as follows
\begin{eqnarray}
&&%
\begin{tabular}{llllll}
\hline \hline
&  & \multicolumn{2}{l}{\emph{n-MSSM*}} &  &  \\
\multicolumn{2}{l}{\  \  \  \  \  \ {\small chiral superfields \  \
\  \ }} & \  \  \ \  \  \  & \multicolumn{1}{||l}{\  \  \  \  \ } &
\multicolumn{2}{l}{\  \  \  \
{\small antichiral superfields }} \\
$\  \  \  \  \  \  \  \boldsymbol{H}_{u}^{i}$ & $\  \  \  \  \  \boldsymbol{\Phi }_{\bar{%
\imath}}$ &  & \multicolumn{1}{||l}{} & $\  \  \  \  \  \boldsymbol{H}_{u\bar{%
\imath}}^{\dagger }$ & $\  \  \  \  \  \boldsymbol{\Phi }^{i\dagger }$ \\
$\  \  \ (\frac{1}{2},0)_{+}$ \  & $\  \ (0,\frac{1}{2})_{-}$ \  &
&
\multicolumn{1}{||l}{} & \ $\  \  \ (0,\frac{1}{2})_{-}$ \  & $\  \  \ (\frac{1}{%
2},0)_{+}$ \\
&  &  &  &  &  \\ \hline \hline
\end{tabular}
\label{tab} \\
&&  \notag
\end{eqnarray}%
with gauge transformations like
\begin{equation}
\begin{tabular}{lllll}
$\boldsymbol{H}_{u}$ & $\  \  \rightarrow $ \  &
$\boldsymbol{H}_{u}^{\prime }$
& $=$ & $\mathcal{U}_{sl_{2}}.\boldsymbol{H}_{u}$ \\
$\boldsymbol{\Phi }$ & $\  \  \rightarrow $ \  & $\boldsymbol{\Phi
}^{\prime }$
& $=$ & $\boldsymbol{\Phi }.\left( \mathcal{U}_{sl_{2}}\right) ^{-1}$%
\end{tabular}%
\end{equation}%
With these quantum number assignments, the Higgs superfields $\boldsymbol{H}%
_{u}$ and $\boldsymbol{\Phi }$ have opposite charges under both
factors of the gauge symmetry. \newline For the interesting case of
the chiral sector of superspace lagrangian density, compatibility
between chirality and $U_{Y}\left( 1\right) \times SU_{L}\left(
2\right) $ gauge invariance is explicitly exhibited in the
following table%
\begin{eqnarray}
&&%
\begin{tabular}{llll}
\hline \hline
&  & \  \  &  \\
{\small symmetry group} &  & \multicolumn{1}{|l}{\  \  \  \ {\small
chiral }} &
\multicolumn{1}{|l}{\  \  \  \  \  \  \ {\small antichiral }} \\
{\small composite} &  & \multicolumn{1}{|l}{$\  \  \  \  \boldsymbol{H}_{u}%
\boldsymbol{\Phi }_{d}$} & \multicolumn{1}{|l}{$\  \  \  \  \  \boldsymbol{H}%
_{u}^{\dagger }\  \boldsymbol{\Phi }_{d}^{\dagger }$} \\
${\small SL(2)\times C}_{{\small Y}}^{{\small \ast }}$ &  &
\multicolumn{1}{|l}{\  \  \ $(\frac{1}{2},\frac{1}{2})_{_{0}}$} &
\multicolumn{1}{|l}{\ $\  \  \  \  \  \ (\frac{1}{2},\frac{1}{2})_{_{0}}$} \\
${\small SU}_{L}{\small (2)\times U(1)}$ &  & \multicolumn{1}{|l}{$\
\  \ 1_{_{0}}\oplus 3_{_{0}}$ \  \  \  \  \  \  \ } &
\multicolumn{1}{|l}{$\  \  \  \  \  \
\  \bar{1}_{_{0}}\oplus \bar{3}_{_{0}}$ \  \  \  \  \  \  \  \ } \\
&  &  &  \\ \hline \hline
\end{tabular}
\label{hfh} \\
&&  \notag
\end{eqnarray}%
From this table we learn that as far as $U_{Y}\left( 1\right) \times
SU_{L}\left( 2\right) $ gauge symmetry is concerned, the charge
assignments as in (\ref{tab}) gives another possibility in looking
for supersymmetric extensions of standard model (SM) of electroweak
theory. The study of this extension is one of the objectives of the
present study.

\subsubsection{Gauge transformations}

The supersymmetric extension of the standard model requires at least
\emph{4} Higgs supersymmetric chiral multiplets; these are given by
the usual chiral superfield doublets $\boldsymbol{H}_{u}$ and
$\boldsymbol{H}_{d}$. In next - to - MSSM, one has, in addition to
the two above superfield doublets, the iso-singlet superfield
$\boldsymbol{S}$.

\  \  \  \  \newline In our proposal baptized as \emph{n-MSSM*},
instead of $\boldsymbol{H}_{u}$ and $\boldsymbol{H}_{d}$, we have
$\boldsymbol{H}_{u}$ and $\boldsymbol{\Phi }$ with gauge symmetry
charges as in (\ref{tab}-\ref{hfh}). Therefore, the gauge
transformations of the \emph{5} Higgs chiral superfields are as
follows:

\begin{itemize}
\item the chiral superfields $\boldsymbol{S}$ and $\boldsymbol{H}_{u}\equiv
\left( \boldsymbol{H}^{i}\right) $ are exactly as in \emph{n-MSSM}.
Under generic $SL\left( 2,C\right) $ transformations preserving
chirality, these \emph{3} superfields transform as usual like
\begin{equation}
\begin{tabular}{lllll}
$\boldsymbol{S}$ & $\  \rightarrow $ \  & $\boldsymbol{S}^{\prime }$ & $=$ & $%
\boldsymbol{S}$ \\
$\boldsymbol{H}_{u}$ & $\  \rightarrow $ & $\boldsymbol{H}_{u}^{\prime }$ & $%
= $ & $\mathcal{U}_{_{sl_{2}}}\times \boldsymbol{H}_{u}$%
\end{tabular}
\label{so}
\end{equation}%
with gauge transformation
\begin{equation}
\mathcal{U}_{_{sl_{2}}}=e^{+\Lambda _{A}T^{A}}\text{ \  \  \ },\qquad \mathcal{%
\bar{D}}_{\dot{a}}\text{ }\mathcal{U}_{_{sl_{2}}}=0
\end{equation}%
and the gauge parameter $\Lambda =\Lambda _{A}T^{A}$ is a chiral
superfield valued in the Lie algebra of $SU_{L}\left( 2\right) $
gauge symmetry. Infinitesimally, we have
\begin{equation}
\begin{tabular}{lll}
&  &  \\
$\delta _{_{sl_{{\small 2}}}}\boldsymbol{S}$ & $=$ & $0$ \\
$\delta _{_{sl_{{\small 2}}}}\boldsymbol{H}^{i}$ & $=$ & $\Lambda
_{A}\left(
\frac{\tau ^{A}}{2}\right) _{k}^{i}\boldsymbol{H}^{k}$ \\
&  &
\end{tabular}%
\end{equation}

\item the \emph{2} chiral superfields making $\boldsymbol{\Phi }_{i}$ behave
as the components of an \emph{anti-doublet} of $SL\left( 2,C\right)
$. By anti-doublet, we mean that under non abelian gauge
transformations, the
chiral superfields $\boldsymbol{\Phi }$ transform like%
\begin{equation}
\begin{tabular}{lll}
$\boldsymbol{\Phi }$ & $\rightarrow $ & $\boldsymbol{\Phi }^{\prime }=%
\boldsymbol{\Phi }\times \left( \mathcal{U}_{_{sl_{2}}}\right) ^{-1}$%
\end{tabular}
\label{sa}
\end{equation}%
with
\begin{eqnarray}
\mathcal{U}_{_{sl_{2}}}^{-1} &=&e^{-\Lambda _{A}T^{A}}\text{ \  \  \
},\qquad
\mathcal{\bar{D}}_{\dot{a}}\left( \mathcal{U}_{_{sl_{2}}}^{-1}\right) =0 \\
&&  \notag
\end{eqnarray}%
standing for the inverse of $\mathcal{U}_{_{sl_{2}}}$. Infinitesimally%
\begin{equation}
\begin{tabular}{lll}
$\delta _{_{sl_{{\small 2}}}}\Phi _{i}$ & $=$ & $-\Lambda _{A}\Phi
_{k}\left( \frac{\tau ^{A}}{2}\right) _{i}^{k}$%
\end{tabular}%
\end{equation}%
Notice that like in eq(\ref{so}), the gauge transformation
(\ref{sa}) preserves as well the chirality property. Notice also
that the electric charges of the $\boldsymbol{H}_{u}$ and
$\boldsymbol{H}_{d}$ components are as
\begin{eqnarray}
\left( \boldsymbol{H}_{u}\right) ^{i} &=&\left(
\begin{array}{c}
H_{u}^{+} \\
H_{u}^{0}%
\end{array}%
\right) \qquad ,\qquad \left( \boldsymbol{H}_{u}^{\dagger }\right)
_{i}=\left( \bar{H}_{u}^{-},\bar{H}_{u}^{0}\right)  \notag \\
\left( \boldsymbol{H}_{d}\right) ^{i} &=&\left(
\begin{array}{c}
H_{d}^{0} \\
H_{d}^{-}%
\end{array}%
\right) \qquad ,\qquad \left( \boldsymbol{H}_{d}^{\dagger }\right)
_{i}=\left( \bar{H}_{d}^{0},\bar{H}_{d}^{+}\right)
\end{eqnarray}%
while for the anti-doublet $\boldsymbol{\Phi }$ they are given by
\begin{equation}
\left( \boldsymbol{\Phi }\right) _{i}=\left( \Phi ^{-},\Phi
^{0}\right) \qquad ,\qquad \left( \boldsymbol{\Phi }^{\dagger
}\right) ^{i}=\left(
\begin{array}{c}
\bar{\Phi}^{+} \\
\bar{\Phi}^{0}%
\end{array}%
\right)
\end{equation}
\end{itemize}

\  \  \  \  \  \newline The opposite choice of the quantum numbers
for $\boldsymbol{\Phi }$, with respect to the standard
$\boldsymbol{H}_{d}$ of \emph{n-MSSM}, affects the
sign of the $SU_{L}\left( 2\right) $ gauge coupling constants $\boldsymbol{%
\Phi }_{i}$ and $\boldsymbol{\bar{\Phi}}^{i}$ as shown on the
following Kahler superfield potentials
\begin{equation}
\begin{tabular}{lll}
&  &  \\
$\mathcal{K}\left( \boldsymbol{H}_{u},\boldsymbol{H}_{u}^{\dagger
}\right) $
& $\  \sim $ \  & $\boldsymbol{H}_{u}^{\dagger }\times e^{-g\boldsymbol{V}%
_{su_{2}}}\times \boldsymbol{H}_{u}$ \\
&  &  \\
& $\  \equiv $ \  & $\boldsymbol{H}_{u}^{\dagger
}\mathcal{U}^{\dagger
}\times \mathcal{U}^{\dagger -1}$ $e^{-g\boldsymbol{V}_{su_{2}}}$ $\mathcal{U%
}^{-1}\times \mathcal{U}\boldsymbol{H}_{u}$ \\
&  &  \\
&  &  \\
$\mathcal{K}\left( \boldsymbol{\Phi },\boldsymbol{\Phi }^{\dagger
}\right) $ & $\  \sim $ & $\boldsymbol{\Phi }\times
e^{+g\boldsymbol{V}_{su_{2}}}\times
\boldsymbol{\Phi }^{\dagger }$ \\
&  &  \\
& $\  \equiv $ & $\boldsymbol{\Phi }\mathcal{U}^{-1}\times \mathcal{U}$ $e^{+g%
\boldsymbol{V}_{su_{2}}}$ $\mathcal{U}^{\dagger }\times
\mathcal{U}^{\dagger
-1}\boldsymbol{\Phi }^{\dagger }$ \\
&  &
\end{tabular}%
\end{equation}%
with $\boldsymbol{V}_{su_{2}}$ the hermitian $SU\left( 2\right) $
gauge superfields. Explicitly, by expanding $e^{\pm
g\boldsymbol{V}_{su_{2}}}$, we have for the first order in the gauge
coupling constant $g$ the following
tri-superfield couplings%
\begin{equation}
\begin{tabular}{lll}
&  &  \\
$-g\text{ }\bar{H}_{i}\times \left( \boldsymbol{V}_{su_{2}}\right)
_{k}^{i}\times H^{k}$ & $\qquad ,\qquad $ & $+g\text{ }\Phi
_{i}\times
\left( \boldsymbol{V}_{su_{2}}\right) _{k}^{i}\times \bar{\Phi}^{k}$ \\
&  &
\end{tabular}
\label{319}
\end{equation}%
Using the chiral superfield doublet $H^{i}$ and the chiral anti-doublet $%
\Phi _{i}$, one can build the iso-scalar
\begin{equation}
\Phi _{i}H^{i}\text{ \  \ }\rightarrow \text{ \  \ }\Phi
_{i}^{\prime }H^{i\prime }=\Phi _{i}H^{i}
\end{equation}%
that preserves chirality and has no hypercharge.\newline In addition
to above features, the exotic gauge transformation (\ref{sa}) has
been also motivated by looking for a link between supersymmetry on
ground state of Higgs fields and intersecting complex 3D conifold
singularities. More details are given below.

\subsection{Link with conifold geometry}

We start by recalling that in the study of complex \emph{3D}
conifold geometries; one distinguishes two remarkable kinds of
conifolds: resolved and deformed \textrm{\cite{X1}-\cite{XF}}. These
geometries are in one to one correspondence with the Kahler and
chiral sectors of a particular class
of supersymmetric gauge theories, such as the Higgs sector of \emph{n-MSSM* }%
we are interested in this study. Generally, by using \emph{4}
complex coordinates, the singular version of these complex
threefolds are respectively defined the following hypersurfaces
\begin{equation}
\begin{tabular}{lllll}
&  &  &  &  \\
{\small Kahler} & : \  \  \  \  \  \  \  & $\left \vert
\boldsymbol{z}_{1}\right \vert ^{2}+\left \vert
\boldsymbol{z}_{2}\right \vert ^{2}-\left \vert
\boldsymbol{z}_{3}\right \vert ^{2}-\left \vert
\boldsymbol{z}_{4}\right
\vert ^{2}$ & $=$ & $0$ \\
&  &  &  &  \\
{\small Chiral} & : & $\  \  \  \  \  \  \  \  \  \  \  \  \  \  \  \  \  \boldsymbol{w}_{1}%
\boldsymbol{w}_{4}-\boldsymbol{w}_{2}\boldsymbol{w}_{3}$ & $=$ & $0$ \\
&  &  &  &
\end{tabular}
\label{ces}
\end{equation}%
In our analysis, the \emph{4} complex variables $\boldsymbol{z}_{I}$
are the same as the $\boldsymbol{w}_{I}$'s and should thought of as
given by the leading scalar fields $\left( h^{i}\right) $ and
$\left( \varphi _{i}\right) $ of the chiral superfield doublet
$\left( H_{u}^{i}\right) $ and the chiral anti-doublet $\left( \Phi
_{i}\right) $.

\subsubsection{Auxiliary field eqs of motion and conifold}

In \emph{n-MSSM*}, there are $5=1+2+2$ complex auxiliary fields type
$F$; and $4=1+3$ hermitian auxiliary fields type $D$. These fields,
which scales as mass$^{2}$, carry different $U_{Y}\left( 1\right)
\times SU_{L}\left( 2\right) $ charges and are denoted in our
proposal respectively as exhibited
on following table%
\begin{equation}
\begin{tabular}{lllllll}
\hline \hline
&  &  &  &  &  &  \\
{\small superfields} & : \  \  \  \  & $\boldsymbol{S}$ &
$\boldsymbol{H}^{i}$
& $\boldsymbol{\Phi }_{i}$ & $\boldsymbol{V}_{0}^{\prime }$ & $\boldsymbol{V}%
^{A}$ \\
{\small bosonic fields} & : & $S,$ & $h^{i},$ & $\varphi _{i},$ &
$B_{\mu }$
& $W_{\mu }^{A}$ \\
{\small fermionic fields} & : & $\tilde{S}_{\alpha },$ &
$\tilde{h}_{\alpha
}^{i},$ & $\tilde{\varphi}_{\alpha i},$ & $\tilde{\lambda}_{\alpha }$ & $%
\tilde{\lambda}_{\alpha }^{A}$ \\
{\small auxiliary fields} & : & $F_{S},$ & $F^{i},$ & $G_{i},$ &
$D^{\prime
} $ & $D^{A}$ \\
{\small quantum charges} & : & $1_{0}$ & $2_{+}$ & $\bar{2}_{-}$ &
$1_{0}$ &
$3_{0}$ \\
&  &  &  &  &  &  \\ \hline \hline
\end{tabular}%
\end{equation}%
It happens that the equations of motion of the two hyperchargeless
iso-singlets auxiliary fields $F_{S}$ and $D^{\prime }$ have much to
do with the equations (\ref{ces}) of the conifold singularities.
Indeed, by using component Higgs fields on ground state ($h^{i}$ and
$\varphi _{i}$ constant
fields), we will show that the equation of motion of the hermitian $%
D^{\prime }$ and the complex $F_{S}$ can brought to the following forms%
\begin{equation}
\bar{h}_{i}h^{i}-\varphi _{i}\bar{\varphi}^{i}=\mathrm{r}
\label{fcc}
\end{equation}%
and
\begin{equation}
\varphi _{i}h^{i}=\mathrm{\nu }  \label{ccc}
\end{equation}%
The complex number $\mathrm{\nu }$ and the real $\mathrm{r}$ are
Fayet Iliopoulos coupling constants appearing in (\ref{nr}); and
respectively
interpreted as complex and Kahler deformations of conifold singularity.%
\newline
In next section, we show that the general form of the equations of
motion of the full set of auxiliary fields of \emph{n-MSSM} are
given the vanishing
condition of the following relations%
\begin{equation}
\begin{tabular}{lll}
&  &  \\
$\bar{F}_{S}$ & $=$ & $\kappa S^{2}+\lambda \left( h^{i}\varepsilon
_{ij}\varphi ^{j}-\mathrm{\nu }\right) $ \\
$D^{\prime }$ & $=$ & $\frac{g^{\prime }}{2}\left[
\bar{h}_{i}h^{i}-\varphi
_{i}\bar{\varphi}^{i}-\mathrm{r}\right] $%
\end{tabular}
\label{cfc}
\end{equation}%
and
\begin{equation}
\begin{tabular}{lll}
$\bar{F}_{i}$ & $=$ & $+\lambda S\varphi _{i}$ \\
$\bar{G}^{i}$ & $=$ & $+\lambda Sh^{i}$ \\
$D^{A}$ & $=$ & $\frac{g}{2}\left( \tau ^{A}\right) _{j}^{i}\left[ \bar{h}%
_{i}h^{j}-\varphi _{i}\bar{\varphi}^{j}\right] $ \\
&  &
\end{tabular}%
\end{equation}%
For $S=0$, eqs(\ref{cfc}) reduce exactly to (\ref{fcc}-\ref{ccc}).

\subsubsection{Solutions of auxiliary field eqs: Anticipation}

The general solution of the auxiliary fields equations of motion
depends on the complex parameter $\mathrm{\nu }$ and the real
$\mathrm{r}$. In the case $\mathrm{r}=0$ and $\mathrm{\nu }$ an
arbitrary complex number, the field equations $F^{i}=G_{i}=0$ are
trivially solved by $S=0$ and one is left with the following
\begin{equation}
\begin{tabular}{lll}
$\varphi _{i}h^{i}$ & $=$ & $\mathrm{\nu }$ \\
$\bar{h}_{i}h^{i}-\varphi _{i}\bar{\varphi}^{i}$ & $=$ & $\mathrm{0}$ \\
$\  \  \  \  \  \  \ D^{A}$ & $=$ & $0$ \\
&  &
\end{tabular}
\label{nu}
\end{equation}%
We will show later on that the solution of the two first relations
are given
by, see also \textrm{eqs(\ref{los}) for details},%
\begin{equation}
\begin{tabular}{lll}
$h^{i}$ & $=$ & $\left( \mathrm{\nu \bar{\nu}}\right) ^{\frac{1}{4}}\mathrm{f%
}^{i}$ \\
$\varphi _{i}$ & $=$ & $\frac{\mathrm{\nu }}{\left( \mathrm{\nu \bar{\nu}}%
\right) ^{1/4}}\mathrm{\bar{f}}_{i}$%
\end{tabular}%
\end{equation}%
with%
\begin{equation}
\begin{tabular}{lllll}
&  &  &  &  \\
$\mathrm{f}^{i}$ & $=\left(
\begin{array}{c}
e^{\frac{i}{2}\left( \psi +\phi \right) }\text{ }\cos \frac{\theta
}{2}\text{
} \\
\\
e^{\frac{i}{2}\left( \psi -\phi \right) }\text{ }\sin \frac{\theta
}{2}\text{
}%
\end{array}%
\right) $ & , & $\mathrm{\bar{f}}_{i}$ & $=\left(
\begin{array}{c}
e^{-\frac{i}{2}\left( \psi +\phi \right) }\text{ }\cos \frac{\theta }{2}%
\text{ } \\
\\
e^{-\frac{i}{2}\left( \psi -\phi \right) }\text{ }\sin \frac{\theta }{2}%
\end{array}%
\right) $ \\
&  &  &  &
\end{tabular}
\label{arr}
\end{equation}%
We find that the iso-triplet equation $D^{A}=0$ is also exactly solved by (%
\ref{arr}); thanks to the replacement of the doublet
$\boldsymbol{H}_{d}$ by the anti-doublet $\boldsymbol{\Phi
}$.\newline
These Higgs configurations parameterize a real 2-sphere $\mathbb{S}%
^{2}=SU\left( 2\right) /U\left( 1\right) $, preserve supersymmetry
and leads to the VEVs ratio
\begin{equation}
\tan \beta _{susy}=1
\end{equation}%
We find as well that by switching on the hermitian FI coupling constant $%
\mathrm{r}\neq 0$, supersymmetry gets broken; and one ends with the
following deviation of the VEVs ratio%
\begin{equation}
\begin{tabular}{lll}
&  &  \\
$\tan \beta $ & $=$ & $\frac{g^{\prime 2}\mathrm{r}}{2\left(
g^{2}+g^{\prime
2}\right) \sqrt{\mathrm{\nu \bar{\nu}}}}+\sqrt{1+\left( \frac{g^{\prime 2}%
\mathrm{r}}{2\left( g^{2}+g^{\prime 2}\right) \sqrt{\mathrm{\nu \bar{\nu}}}}%
\right) ^{2}}$ \\
&  &
\end{tabular}%
\end{equation}%
where $g$ and $g^{\prime }$ are the gauge coupling constants of $%
SU_{L}\left( 2\right) \times U_{Y}\left( 1\right) $ symmetry. For%
\begin{equation*}
\xi =\frac{g^{\prime 2}\mathrm{r}}{2\left( g^{2}+g^{\prime 2}\right) \sqrt{%
\mathrm{\nu \bar{\nu}}}}=\frac{\mathrm{r}\sin ^{2}\vartheta
_{_{W}}}{2\left \vert \mathrm{\nu }\right \vert }<<1
\end{equation*}%
we have%
\begin{equation*}
\tan \beta \simeq 1+\xi +\frac{1}{2}\xi ^{2}
\end{equation*}

\  \  \  \  \  \newline To distinguish the slightly modified
\emph{n-MSSM*} we are interested in here from the usual next - to -
MSSM, we shall refer to it below as the \emph{conifold model}; its
content in superfields and the interacting dynamics of its Higgs
multiplets are described with some useful details in next section.

\section{n-MSSM$^{\ast }$ as another extension of SM}

\begin{figure}[tbph]
\begin{center}
\hspace{0cm} \includegraphics[width=10cm]{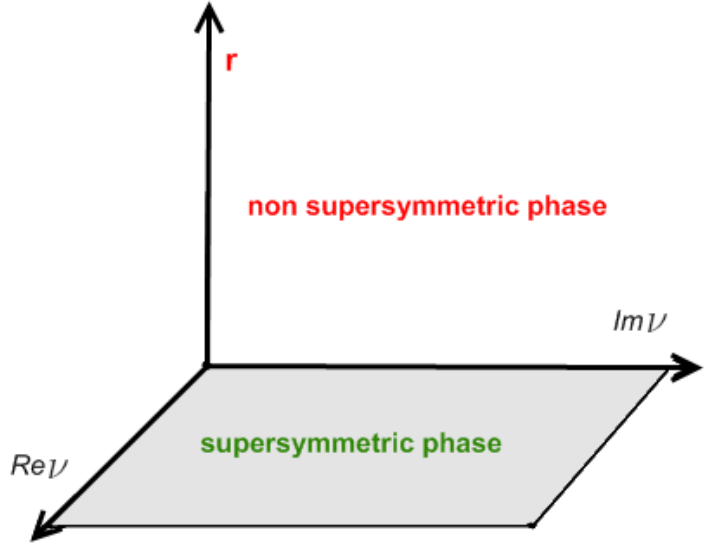}
\end{center}
\par
\vspace{-1cm} \caption{phases of Higgs ground state: $\left(
i\right) $ $r=0$ supersymmetric phase; and $\left( ii\right) $
$r\neq 0$ the non supersymmetric one.} \label{PS}
\end{figure}

\emph{n-MSSM}$^{\ast }$ is a \emph{4D} supersymmetric field model
given by \emph{n-MSSM}; but\emph{\ }with\emph{\ the doublet
}$\boldsymbol{H}_{d}$ replaced by the \emph{anti-doublet}
$\boldsymbol{\Phi }$; it has two phases
characterized by the Fayet-Iliopoulous coupling constant $\left( \mathrm{r},%
\mathrm{\nu }\right) $ as shown on fig \ref{PS}: $\left( i\right) $
a supersymmetric phase given by $r=0$ but an arbitrary complex
parameter $\nu $ as in (\ref{nu}); and $\left( ii\right) $ a non
supersymmetric phase associated with the switching on of the
hermitian $r\neq 0$.\ A comment on explicit breaking of
supersymmetry will be given in \textrm{section 7}.

\subsection{Superfield content and lagrangian density}

We first describe the underlying quantum numbers of the superfield's
content of \emph{n-MSSM*} conifold model; then we study the
structure of the gauge invariant superspace lagrangian density.

\subsubsection{Superfield spectrum}

By ignoring lepton's and quark's sectors of \emph{n-MSSM*}, as they
are not
directly involved in the determination of the Higgs ground state $%
\left \vert \Sigma _{higgs}\right \rangle $, the superfield content
of the conifold model may be then restricted to \emph{9}
supersymmetric multiplets described by \emph{9} superfields,
\emph{4} hermitian and \emph{5} chiral; these are:

\begin{itemize}
\item \emph{the }$U_{Y}\left( 1\right) \times SU_{L}\left( 2\right) $\emph{\
gauge multiplets}\newline
They are given by the usual \emph{4} hermitian superfields $\boldsymbol{V}%
_{u_{1}}\left( x,\theta ,\bar{\theta}\right) $ and $\boldsymbol{V}%
_{su_{2}}\left( x,\theta ,\bar{\theta}\right) $ valued in the Lie
algebra of the $U_{Y}\left( 1\right) \times SU_{L}\left( 2\right) $
gauge symmetry
\begin{equation}
\begin{tabular}{lll}
$\boldsymbol{V}_{u_{1}}$ & $=$ & $\boldsymbol{V}_{0}\frac{Y}{2}$ \\
$\boldsymbol{V}_{su_{2}}$ & $=$ & $\boldsymbol{V}_{A}T^{A}$%
\end{tabular}%
\end{equation}%
with $\frac{Y}{2}$ the hermitian generator of the $U_{Y}\left(
1\right) $
and $T^{A}$ the 3 generators of $SU_{L}\left( 2\right) $. The $\boldsymbol{V}%
^{\prime }$ has no hypercharge and behaves as an iso-singlet under $%
SU_{L}\left( 2\right) $. The $\boldsymbol{V}_{A}$'s form an
iso-triplet and have no hypercharge as well.\newline The $\theta
$-expansion of these superfields are as usual; for the example of
$V_{0}\left( x,\theta ,\bar{\theta}\right) $ solving the reality
condition $V_{0}^{\dagger }=V_{0}$ reads as in general like
\textrm{\cite{W}}

\begin{equation}
\begin{tabular}{lll}
$V_{0}$ & $=$ & $\upsilon +i\theta .\varsigma -i\bar{\theta}.\bar{\varsigma}+%
\frac{i}{2}\theta ^{2}C-\frac{i}{2}\bar{\theta}^{2}\bar{C}\ -\theta
\sigma
^{\mu }\bar{\theta}B_{\mu }$ \\
&  & $+i\theta ^{2}\bar{\theta}.\left( \bar{\lambda}^{\prime }+\frac{i}{2}%
\bar{\sigma}^{\mu }\partial _{\mu }\varsigma \right) -i\bar{\theta}%
^{2}\theta .\left( \lambda ^{\prime }+\frac{i}{2}\sigma ^{\mu
}\partial
_{\mu }\bar{\varsigma}\right) $ \\
&  & $+\frac{1}{2}\theta ^{2}\bar{\theta}^{2}\left( D^{\prime }+\frac{1}{2}%
\square \upsilon \right) $ \\
&  &
\end{tabular}%
\end{equation}%
where, the extra fields $\upsilon $, $\varsigma _{a}$ and $C$ are
pure gauge degrees of freedom. Obviously $B_{\mu }$ is the vector
gauge field, $\lambda _{a}^{\prime }$ the gaugino and $D^{\prime }$
the auxiliary field. \newline
A similar expansion can be written down for the three $\boldsymbol{V}_{A}$%
's; their bosonic gauge fields are denoted by $W_{\mu }^{A}$ and the
\emph{3} auxiliary field ones as $D_{A}$.

\item \emph{5} \emph{chiral superfields}\newline
They are given by the chiral superfield singlet $\boldsymbol{S}$;
the chiral
superfield doublet $\boldsymbol{H}^{i}$ and\ the anti-doublet $\boldsymbol{%
\Phi }_{i}$ with quantum numbers under $U_{Y}\left( 1\right) \times
SU_{L}\left( 2\right) $ as follows

\begin{equation}
\begin{tabular}{l||l}
{\small chiral superfields \  \  \ } & $\  \ {\small SU}_{{\small L}}{\small %
(2)\times U}_{{\small Y}}{\small (1)}$ \\
$\  \  \  \  \  \  \ H^{i}$ & $\  \  \  \  \  \  \  \  \  \ 2_{+1}$ \\
$\  \  \  \  \  \  \  \Phi _{\bar{\imath}}$ & $\  \  \  \  \  \  \  \  \  \  \bar{2}_{-1}$ \\
$\  \  \  \  \  \  \  \boldsymbol{S}$ & $\  \  \  \  \  \  \  \  \  \ 1_{0}$%
\end{tabular}%
\end{equation}

\  \  \  \  \newline The iso-singlet $\boldsymbol{S}$ captures the
extension of the Higgs sector of the minimal supersymmetric standard
model, MSSM; it plays an important
role in solving the auxiliary field equations of motion $F^{i}=0$ and $%
G_{i}=0$. \  \  \  \newline Notice also that $\boldsymbol{H}^{i}$
and $\boldsymbol{\Phi }_{i}$ have opposite quantum number under
$U_{Y}\left( 1\right) \times SU_{L}\left( 2\right) $; this is as
well an important feature in deriving non trivial solutions for the
auxiliary field's equations of motion.\newline The $\theta $-
expansions of these chiral superfields are given by
\begin{equation}
\begin{tabular}{lll}
&  &  \\
$\boldsymbol{H}^{i}\left( y,\theta \right) $ & $=$ & $h^{i}\left( y\right) +%
\sqrt{2}\theta .\tilde{h}^{i}+\theta ^{2}F^{i}\left( y\right) $ \\
$\boldsymbol{\Phi }_{\bar{\imath}}\left( y,\theta \right) $ & $=$ &
$\varphi
_{\bar{\imath}}\left( y\right) +\sqrt{2}\theta .\tilde{\varphi}_{\bar{\imath}%
}+\theta ^{2}G_{\bar{\imath}}\left( y\right) $ \\
$\boldsymbol{S}\left( y,\theta \right) $ & $=$ & $S\left( y\right) +\sqrt{2}%
\theta .\tilde{S}+\theta ^{2}F_{S}\left( y\right) $ \\
&  &
\end{tabular}%
\end{equation}%
with $y^{\mu }=x^{\mu }+i\theta \sigma ^{\mu }\bar{\theta}$, and $\tilde{h}%
^{i},$ $\tilde{\varphi}_{\bar{\imath}}$, $\tilde{S}$ designating the
fermionic superpartners. For antichiral superfields, we have
$\bar{y}^{\mu }=x^{\mu }-i\theta \sigma ^{\mu }\bar{\theta}$.
\newline The $U_{Y}\left( 1\right) $ hypercharges of these
superfields are same as in
\emph{n-MSSM*}; while the charges of $\boldsymbol{H}^{i}$ and $\boldsymbol{%
\Phi }_{i}$ under $SU_{L}\left( 2\right) $ are opposite; and are as in eqs(%
\ref{so}-\ref{sa}). Explicitly, we have

\begin{equation}
\begin{tabular}{lllllllll}
$\boldsymbol{H}^{i}$ & $\  \  \rightarrow $ \  \  \  & $\mathcal{U}_{\bar{j}%
}^{i} $ $\boldsymbol{H}^{j}$ &  & , &  & $\boldsymbol{\Phi
}_{\bar{\imath}}$ & $\  \  \rightarrow $ \  \  \  &
$\boldsymbol{\Phi }_{\bar{j}}$ $\left(
\mathcal{U}^{{\small -1}}\right) _{\bar{\imath}}^{j}$ \\
&  &  &  &  &  &  &  &  \\
$\boldsymbol{H}_{\bar{\imath}}^{\dagger }$ & $\  \  \rightarrow $ \  \  \  & $%
\boldsymbol{H}_{\bar{j}}^{\dagger }$ $\mathcal{U}_{i}^{\dagger j}$ &
& , & & $\boldsymbol{\Phi }^{\dagger i}$ & $\  \  \rightarrow $ \  \
\  & $\left(
\mathcal{U}^{{\small -1}}\right) _{\bar{j}}^{\dagger i}$ $\boldsymbol{\Phi }%
^{\dagger j}$%
\end{tabular}
\label{uh}
\end{equation}%
with%
\begin{eqnarray}
\mathcal{U} &=&\ e^{+g\Lambda _{A}T^{A}}\qquad ,\qquad \  \  \mathcal{U}%
^{-1}=\ e^{-g\Lambda _{A}T^{A}}  \notag \\
\mathcal{U}^{\dagger } &=&\  \ e^{+g\Lambda _{A}^{\dagger
}T^{A}}\qquad ,\qquad \mathcal{U}^{\dagger -1}=\  \ e^{-g\Lambda
_{A}^{\dagger }T^{A}}
\label{g} \\
&&  \notag
\end{eqnarray}%
where the \emph{3} gauge super-parameters $\Lambda _{A}$ are chiral
superfields and $\Lambda _{A}^{\dagger }$ the corresponding antichiral ones.%
\newline
From eqs(\ref{g}), we learn that one may go from the gauge matrix $\mathcal{U%
}$ to its inverse $\mathcal{U}^{-1}$ and vice versa just by
performing the sign change
\begin{equation}
gT^{A}\quad \rightarrow \quad \left( -g\right) T^{A}
\end{equation}
\end{itemize}

\subsubsection{Lagrangian density of \emph{n-MSSM*}}

In $\mathcal{N}=1$ superspace, the lagrangian density describing
supersymmetric interacting dynamics of the gauge and Higgs
superfields of
the conifold model (n-MSSM* ) is given by%
\begin{equation}
\boldsymbol{L}_{conif}=\boldsymbol{L}_{gauge}+\boldsymbol{L}_{higgs}
\label{tl}
\end{equation}%
The pure gauge term $\boldsymbol{L}_{gauge}$ is as usual%
\begin{equation}
\begin{tabular}{lll}
$\boldsymbol{L}_{gauge}$ & $=$ & $\int d^{2}\theta Tr\left( \frac{1}{16g^{2}}%
\mathcal{W}^{\alpha }\mathcal{W}_{\alpha }\right) +hc$ \\
&  &  \\
&  & $+\int d^{2}\theta \left( \frac{1}{4}\mathcal{W}^{\prime \alpha }%
\mathcal{W}_{\alpha }^{\prime }\right) +hc$%
\end{tabular}
\label{kt}
\end{equation}%
and the gauge covariant term $\boldsymbol{L}_{higgs}$ describing the
Higgs
sector coupled to gauge multiplets reads as follows%
\begin{equation}
\begin{tabular}{lll}
&  &  \\
$\boldsymbol{L}_{higgs}$ & $=$ & $\dint d^{4}\theta $ $\boldsymbol{S}%
^{\dagger }\boldsymbol{S+}\dint d^{4}\theta $
$\boldsymbol{H}_{i}^{\dagger }\left( e^{-g\boldsymbol{V-g}^{\prime
}\boldsymbol{V}^{\prime }}\right)
_{j}^{i}\boldsymbol{H}^{j}$ \\
&  &  \\
&  & $+\dint d^{4}\theta $ $\boldsymbol{\Phi }_{i}$ $\left( \frac{1}{{\Large %
e}^{-g\boldsymbol{V-g}^{\prime }\boldsymbol{V}^{\prime }}}\right) _{j}^{i}$ $%
\boldsymbol{\Phi }^{\dagger j}$ \\
&  &  \\
&  & $-\dint d^{2}\theta $ $\left( \lambda \boldsymbol{S}\left( \Phi _{i}%
\boldsymbol{H}^{i}\right) +\frac{\kappa
}{3}\boldsymbol{S}^{3}+hc\right) $
\\
&  &  \\
&  & $+\left( \frac{g^{\prime }}{2}\mathrm{r}\dint d^{4}\theta
V\right) +\left( \lambda \mathrm{\bar{\nu}}\dint d^{2}\theta
\boldsymbol{S}+hc\right)
$ \\
&  &
\end{tabular}
\label{lt}
\end{equation}%
with
\begin{eqnarray}
\left( \frac{1}{e^{-g\boldsymbol{V-g}^{\prime }\boldsymbol{V}^{\prime }}}%
\right) &\equiv &e^{+g\boldsymbol{V+g}^{\prime }\boldsymbol{V}^{\prime }} \\
&&  \notag
\end{eqnarray}%
where $g^{\prime }$, $g$ are the $U_{Y}\left( 1\right) \times
SU_{L}\left(
2\right) $ gauge coupling constants and where the complex $\lambda $ and $%
\kappa $ are coupling constants of tri-superfield's in the chiral
superpotential.

\  \newline In above expression, we have also added two kinds of
Fayet-Iliopoulos (FI) terms, one with a real $\mathrm{r}$ coupling
parameter and the other with
complex $\mathrm{\nu }$ one; the scaled convention $\frac{g^{\prime }}{2}%
\mathrm{r}$ and $\lambda \mathrm{\bar{\nu}}$ are for later use. The
existence of the \emph{3} FI couplings $r,$ $\nu ,$ $\bar{\nu}$ is
because of a hidden $\mathcal{N}=2$ supersymmetry property of the
superfield spectrum. The superfields $\boldsymbol{S}$ and
$\boldsymbol{V}_{0}$ combine indeed into a $\mathcal{N}=2$
supersymmetric $U_{Y}\left( 1\right) $ gauge multiplet which is
known to have 3 FI coupling constants.

\  \  \  \  \  \newline Notice that the lagrangian density
$\boldsymbol{L}_{higgs}$ is manifestly invariant under the gauge
symmetry transformations. For gauge changes under
the non abelian factor, the chiral superfields $\boldsymbol{H}^{i}$ and $%
\boldsymbol{\Phi }_{i}$ transform as in (\ref{uh}) and the gauge
superfields
like%
\begin{eqnarray}
e^{-g\boldsymbol{V}}\text{ \  \ } &\rightarrow &\text{ \  \ }\left( \mathcal{U}%
^{{\small -1}}\right) ^{\dagger }\times \left(
e^{-g\boldsymbol{V}}\right)
\times \left( \mathcal{U}^{{\small -1}}\right)  \notag \\
&& \\
e^{+g\boldsymbol{V}}\text{ \  \ } &\rightarrow &\text{ \  \
}\mathcal{U}\times
e^{+g\boldsymbol{V}}\times \left( \mathcal{U}^{\dagger }\right)  \notag \\
&&  \notag
\end{eqnarray}%
with $\mathcal{U}$ and $\mathcal{U}^{-1}$ as in eqs(\ref{g}) and $e^{-g%
\boldsymbol{V}}e^{+g\boldsymbol{V}}=I$ due to $\left[ \boldsymbol{V},%
\boldsymbol{V}\right] =0$. Explicitly, these superfield
transformations read
as%
\begin{eqnarray}
\left( e^{-g\boldsymbol{V}}\right) _{j}^{i}\text{ \  \ }
&\rightarrow &\text{
\  \ }\left( \mathcal{U}^{{\small -1}}\right) _{k}^{\dagger i}\left( e^{-g%
\boldsymbol{V}}\right) _{l}^{k}\left( \mathcal{U}^{{\small
-1}}\right)
_{j}^{l}  \notag \\
&& \\
\left( e^{+g\boldsymbol{V}}\right) _{j}^{i}\text{ \  \ }
&\rightarrow &\text{ \  \ }\mathcal{U}_{k}^{i}\left(
e^{+g\boldsymbol{V}}\right) _{l}^{k}\left(
\mathcal{U}^{\dagger }\right) _{j}^{l}  \notag \\
&&  \notag
\end{eqnarray}%
with lower indices denoting matrix columns and upper ones the rows.

\  \  \

\emph{Used notations: illustrating examples} \newline To fix the
idea on the used notations, we illustrate the above matrix products
by considering a useful example. First take the $2\times 2$ matrices
$\mathcal{U}$ and its inverse $\mathcal{U}^{{\small -1}}$ with
respective complex entries $\left( \mathcal{U}\right) _{\bar{k}}^{l}$ and $%
\left( \mathcal{U}^{{\small -1}}\right) _{l}^{\bar{k}}$ like
\begin{equation}
\mathcal{U}=\left(
\begin{array}{cc}
a & b \\
c & d%
\end{array}%
\right) ,\quad \mathcal{U}^{{\small -1}}=\left(
\begin{array}{cc}
d & -b \\
-c & a%
\end{array}%
\right) \text{, \  \ }ad-bc=1
\end{equation}%
from which we learn%
\begin{equation}
\left(
\begin{array}{cc}
d & -b \\
-c & a%
\end{array}%
\right) ^{T}=\left(
\begin{array}{cc}
0 & -1 \\
1 & 0%
\end{array}%
\right) \left(
\begin{array}{cc}
a & b \\
c & d%
\end{array}%
\right) \left(
\begin{array}{cc}
0 & 1 \\
-1 & 0%
\end{array}%
\right)  \label{qe}
\end{equation}%
showing that $\left( \mathcal{U}^{{\small -1}}\right) ^{T}$ and
$\mathcal{U}$ are related by $\varepsilon _{ij}$ and its transpose.
This feature can be also obtained by using the expression of the
determinant of the $\mathcal{U}$
matrix namely $\left( \mathcal{U}_{1}^{1}\right) \left( \mathcal{U}%
_{2}^{2}\right) $ $-$ $\left( \mathcal{U}_{1}^{2}\right) \left( \mathcal{U}%
_{2}^{1}\right) $ $=$ $1$ that reads by help of the $\varepsilon $-
antisymmetric tensors like
\begin{equation}
\varepsilon _{ij}\left( \mathcal{U}_{k}^{i}\right) \left( \mathcal{U}%
_{l}^{j}\right) =\varepsilon _{kl}  \label{be}
\end{equation}%
or equivalently as $\varepsilon ^{mk}\mathcal{U}_{k}^{i}\varepsilon _{ij}%
\mathcal{U}_{l}^{j}=\delta _{l}^{m}$. From these relations, one
deduces the
link between $\mathcal{U}^{-1}$ and $\mathcal{U}$\ which is nothing but eq(%
\ref{qe}),%
\begin{equation}
\left( \mathcal{U}^{-1}\right) _{j}^{m}=\varepsilon ^{mk}\text{ }\mathcal{U}%
_{k}^{i}\text{ }\varepsilon _{ij}\quad ,\quad \varepsilon
_{km}\left( \mathcal{U}^{-1}\right)
_{j}^{m}=\mathcal{U}_{k}^{i}\varepsilon _{ij} \label{ce}
\end{equation}%
These relations can be as well derived by equating the gauge
transformation
of both sides of the equality $\boldsymbol{\Phi }_{i}\boldsymbol{H}%
^{i}=\varepsilon _{il}\boldsymbol{\Phi }^{l}\boldsymbol{H}^{i}$; we
obtain
\begin{equation*}
\boldsymbol{\Phi }_{j}\left( \mathcal{U}^{-1}\right) _{i}^{j}\left( \mathcal{%
U}_{k}^{i}\right) \boldsymbol{H}^{k}=\varepsilon _{il}\left( \mathcal{U}%
_{k}^{i}\right) \left( \mathcal{U}_{m}^{l}\right) \varepsilon ^{mj}\Phi _{j}%
\boldsymbol{H}^{k}
\end{equation*}%
leading to (\ref{be}-\ref{ce}).

\  \  \

\emph{n-MSSM and conifold model}\  \  \  \  \newline Compared to
usual expression in \emph{n-MSSM}, the superspace lagrangian density
(\ref{lt}) of the conifold model \emph{n-MSSM*} has some special
features; in particular the two following ones.

\begin{itemize}
\item up-Higgs free super-propagators $\left \langle \boldsymbol{H}%
_{i}^{\dagger }\boldsymbol{H}^{j}\right \rangle $ are same as those in \emph{%
n-MSSM}; but $\left \langle \boldsymbol{\Phi }_{i}\boldsymbol{\Phi
^{j\dagger }}\right \rangle $ come with a minus sign compared to $%
\left
\langle \boldsymbol{H}_{di}^{\dagger }\boldsymbol{H}%
_{d}^{j}\right \rangle $; this feature is due to the $SU\left(
2\right) $ representation group property
\begin{equation*}
\boldsymbol{\Phi }_{i}\boldsymbol{\Phi ^{i\dagger
}=}-\boldsymbol{\Phi _{i}^{\dagger }\Phi }^{i}
\end{equation*}

\item by expanding the exponentials in (\ref{lt}), the hermitian
tri-superfield's couplings involving a gauge superfield are given
by,
\begin{equation}
\begin{tabular}{lll}
$-g\left( \boldsymbol{T}_{A}\right) _{j}^{i}\boldsymbol{V}^{A}\boldsymbol{H}%
_{i}^{\dagger }\boldsymbol{H}^{j}$ & , \  \  \  \  \  \  \  \  & $\boldsymbol{-}%
\frac{\boldsymbol{g}\prime }{2}\boldsymbol{V_{0}H}_{i}^{\dagger }\boldsymbol{%
H}^{i}$ \\
&  &  \\
$+g\left( \boldsymbol{T}_{A}\right) _{j}^{i}\boldsymbol{V}^{A}\boldsymbol{%
\Phi }_{\boldsymbol{i}}\boldsymbol{\Phi }^{\dagger j}$ & , \  \  \
\  \  \  \  \
& $\boldsymbol{+}\frac{\boldsymbol{g}\prime }{2}\boldsymbol{V_{0}\Phi }_{%
\boldsymbol{i}}\boldsymbol{\Phi }^{\dagger i}$ \\
&  &
\end{tabular}
\label{fh}
\end{equation}%
these interactions intervene in the structure of the equations of
motion of the auxiliary fields D. As explicitly exhibited, the
vertices with Higgs superfield doublet $\boldsymbol{H}^{i}$ involve
the gauge coupling constants $\left( -g\right) $ and $\left(
-g^{\prime }\right) $ while those with anti-doublet
$\boldsymbol{\Phi }_{\boldsymbol{i}}$ involves $\left( +g\right) $
and $\left( +g^{\prime }\right) $.
\end{itemize}

\  \  \  \newline Recall that in \emph{n-MSSM}, the analogue of
eqs(\ref{fh}) are directly read from the Kahler part of the gauge
covariant superspace density namely
\begin{equation}
\dint d^{4}\theta \boldsymbol{H}_{u}^{\dagger }\left[ e^{-g\boldsymbol{V}%
_{A}T^{A}\boldsymbol{-g}^{\prime
}\boldsymbol{V}_{0}\frac{Y}{2}}\right] \boldsymbol{H}_{u}+\dint
d^{4}\theta \boldsymbol{H}_{d}^{\dagger }\left[
e^{-g\boldsymbol{V}_{A}T^{A}\boldsymbol{-g}^{\prime }\boldsymbol{V}_{0}\frac{%
Y}{2}}\right] \boldsymbol{H}_{d}
\end{equation}%
By expansion of the exponentials, one obtains the tri-superfield's
interactions; these are:

\begin{itemize}
\item the tri-superfield's interactions involving the $U_{Y}\left( 1\right) $
gauge superfield $\boldsymbol{V_{0}}$,%
\begin{eqnarray}
&&\boldsymbol{-}\frac{\boldsymbol{g}\prime }{2}\boldsymbol{H}_{u}^{\dagger }%
\boldsymbol{V_{0}H}_{u}  \notag \\
&&\boldsymbol{+}\frac{\boldsymbol{g}\prime }{2}\boldsymbol{H}_{d}^{\dagger }%
\boldsymbol{V_{0}H}_{d}
\end{eqnarray}
which are as in (\ref{fh}). Because of the opposite hypercharges,
they come also in a pair with opposite sign; and

\item the tri-superfield's interactions involving the $SU_{L}\left( 2\right)
$ gauge superfields $\boldsymbol{V_{A}}$%
\begin{eqnarray}
&&\boldsymbol{-}g\boldsymbol{V_{A}H}_{u}^{\dagger
}T^{A}\boldsymbol{H}_{u}
\notag \\
&&\boldsymbol{-gV_{A}H}_{d}^{\dagger }T^{A}\boldsymbol{H}_{d}
\end{eqnarray}%
have the same sign of the $SU_{L}\left( 2\right) $ gauge coupling
constant g contrary to (\ref{fh}).
\end{itemize}

\subsection{Supersymmetric scalar potential}

First, we give the component field lagrangian; then we study the
scalar potential of the conifold model.

\subsubsection{Component field lagrangian density}

The integration of the superspace lagrangian density
(\ref{tl}-\ref{lt}) with respect to the Grassmann variables $\theta
$ and $\bar{\theta}$ gives the following component field expression

\begin{eqnarray}
\boldsymbol{L} &=&-\frac{1}{4}B_{\mu \nu }^{A}B_{A}^{\mu \nu }-\frac{1}{4}%
W_{\mu \nu }^{A}W_{A}^{\mu \nu }\text{ \ }+  \notag \\
&& \\
&&\left[ \left( \nabla _{\mu }h\right) ^{i}\right] ^{\dagger }\left(
\nabla _{\mu }h\right) ^{i}+\left( \mathcal{D}_{\mu }\varphi \right)
_{i}\left[
\left( \mathcal{D}_{\mu }\varphi \right) _{i}\right] ^{\dagger }-\mathcal{V}%
_{scalar}  \notag \\
&&  \notag
\end{eqnarray}%
where, for simplicity, we have dropped out the fermionic
contributions.
\newline
The operators $\nabla _{\mu }$ and $\mathcal{D}_{\mu }$ are
$U_{Y}\left( 1\right) \times SU_{L}\left( 2\right) $ gauge covariant
derivatives; they are not completely independent as they are
associated with two
representations of same gauge symmetry. The use of different notations $%
\nabla _{\mu }$ and $\mathcal{D}_{\mu }$ is to\textrm{\ exhibit the
opposite
signs of the gauge coupling constants of }the two Higgs doublets with the $%
W_{\mu }^{A}$\ bosons seen that the doublet $h^{i}$ and anti-doublet $%
\varphi _{i}$ have opposite quantum numbers under $SU_{L}\left(
2\right) $
as well. Thus, we have%
\begin{equation}
\begin{tabular}{lll}
&  &  \\
$\left( \nabla _{\mu }h\right) ^{i}$ & $=$ & $\left[ \delta
_{j}^{i}\partial _{\mu }-i\frac{g^{\prime }}{2}\delta _{j}^{i}B_{\mu
}-igW_{\mu }^{A}\left(
\frac{\tau _{A}}{2}\right) _{j}^{i}\right] h^{i}$ \\
&  &  \\
$\left( \mathcal{D}_{\mu }\varphi \right) _{i}$ & $=$ & $\left[
\delta _{i}^{j}\partial _{\mu }+i\frac{g^{\prime }}{2}\delta
_{i}^{j}B_{\mu }+igW_{\mu }^{A}\left( \frac{\tau _{A}}{2}\right)
_{i}^{j}\right] \varphi
_{j}$ \\
&  &
\end{tabular}%
\end{equation}%
showing that one may go from $\left( \nabla _{\mu }h\right) ^{i}$ to
$\left(
\mathcal{D}_{\mu }\varphi \right) _{i}$ and vice versa by interchanging $%
h^{i}$ and anti-doublet $\varphi _{i}$ ( implicitly $g^{\prime }$ $%
\leftrightarrow $ $-g^{\prime }$ ); but also changing the sign of the $%
SU_{L}\left( 2\right) $ gauge coupling constants
\begin{equation}
g\text{ \  \ }\leftrightarrow \text{ }-g
\end{equation}%
The tensors $B_{\mu \nu }^{A}$, $W_{\mu \nu }^{A}$ are respectively
the gauge field strengths of the $SU_{L}\left( 2\right) $ and
$U_{Y}\left(
1\right) $ gauge bosons; they are as follows%
\begin{equation}
\begin{tabular}{lll}
$W_{\mu \nu }^{A}$ & $=$ & $\partial _{\mu }W_{\nu }^{A}-\partial
_{\nu
}W_{\mu }^{A}-igf_{BC}^{A}W_{\mu }^{B}W_{\nu }^{C}$ \\
&  &  \\
$B_{\mu \nu }$ & $=$ & $\partial _{\mu }B_{\nu }-\partial _{\nu
}B_{\mu }$
\\
&  &
\end{tabular}%
\end{equation}%
with $f_{BC}^{A}$ the structure constants of $SU_{L}\left( 2\right)
$; and the three $2\times 2$ matrices $\tau _{A}$ the usual Pauli
matrices whose entries are represented, in our convention, like
\begin{equation}
\begin{tabular}{lllll}
$\left( \tau ^{1}\right) _{i}^{j}=\left(
\begin{array}{cc}
0 & 1 \\
1 & 0%
\end{array}%
\right) $ & , & $\left( \tau ^{2}\right) _{i}^{j}=\left(
\begin{array}{cc}
0 & -i \\
i & 0%
\end{array}%
\right) ,$ &  & $\left( \tau ^{3}\right) _{i}^{j}=\left(
\begin{array}{cc}
1 & 0 \\
0 & -1%
\end{array}%
\right) $ \\
&  &  &  &
\end{tabular}%
\end{equation}%
The coupling of $W_{\mu }^{A}$\ bosons to the Higgs field's current $%
J_{A}^{\mu }$ can be written as%
\begin{equation}
-gW_{\mu }^{A}J_{A}^{\mu }=-gW_{\mu }^{A}\left( J_{A\left( h\right)
}^{\mu }-J_{A\left( \varphi \right) }^{\mu }\right)
\end{equation}%
with $J_{A\left( h\right) }^{\mu }$ and $J_{A\left( \varphi \right)
}^{\mu }$
given by%
\begin{equation}
\begin{tabular}{lll}
$J_{A\left( h\right) }^{\mu }$ & $=$ & $\frac{i}{2}\left[ \bar{h}%
_{k}\partial ^{\mu }h^{j}-\left( \partial ^{\mu }\bar{h}_{k}\right) h^{j}%
\right] \left( \tau _{A}\right) _{j}^{k}$ \\
&  &  \\
$J_{A\left( \varphi \right) }^{\mu }$ & $=$ & $\frac{i}{2}\left[ \bar{\varphi%
}^{j}\left( \partial ^{\mu }\varphi _{k}\right) -\left( \partial ^{\mu }\bar{%
\varphi}^{j}\right) \varphi _{k}\right] \left( \tau _{A}\right) _{j}^{k}$%
\end{tabular}%
\end{equation}%
Notice that the above gauge covariant derivatives can be put into
the condensed form
\begin{equation}
\begin{tabular}{lll}
$\nabla _{\mu }\boldsymbol{h}$ & $=$ & $\left( \partial _{\mu
}-i\Upsilon
_{\mu }\right) \boldsymbol{h}$ \\
$\mathcal{D}_{\mu }\boldsymbol{\varphi }$ & $=$ & $\left( \partial
_{\mu
}+i\Upsilon _{\mu }\right) \boldsymbol{\varphi }$%
\end{tabular}%
\end{equation}%
with%
\begin{equation}
\Upsilon _{\mu }=\frac{g^{\prime }}{2}B_{\mu }+gW_{\mu }^{A}\frac{\tau _{A}}{%
2}
\end{equation}%
By using the expressions of the Pauli matrices, we explicitly have%
\begin{equation}
\begin{tabular}{lll}
$\Upsilon _{\mu }$ & $=$ & $\left(
\begin{array}{cc}
\frac{g^{\prime }}{2}B_{\mu }+\frac{g}{2}W_{\mu }^{3} &
\frac{g}{2}\left(
W_{\mu }^{1}-iW_{\mu }^{2}\right)  \\
\frac{g}{2}\left( W_{\mu }^{1}+iW_{\mu }^{2}\right)  & \frac{g^{\prime }}{2}%
B_{\mu }-\frac{g}{2}W_{\mu }^{3}%
\end{array}%
\right) $ \\
&  &  \\
& $=$ & $\frac{g}{2}\left(
\begin{array}{cc}
\frac{\cos 2\vartheta _{_{W}}}{\cos \vartheta _{_{W}}}Z_{\mu }+\sin
\vartheta _{_{W}}A_{\mu } & \sqrt{2}W_{\mu }^{-} \\
\sqrt{2}W_{\mu }^{+} & \frac{-1}{\cos \vartheta _{_{W}}}Z_{\mu }%
\end{array}%
\right) $%
\end{tabular}%
\end{equation}%
with%
\begin{equation*}
\cos \vartheta _{_{W}}=\frac{g}{\sqrt{g^{2}+g^{\prime 2}}},\qquad
\sin \vartheta _{_{W}}=\frac{g^{\prime }}{\sqrt{g^{2}+g^{\prime 2}}}
\end{equation*}%
We also have for the quadratic terms in the vector gauge fields and
the
Higgs fields%
\begin{equation}
L_{m}=\bar{h}_{i}\left( \Upsilon _{\mu }\Upsilon ^{\mu }\right)
_{j}^{i}h^{j}+\varphi _{i}\left( \Upsilon _{\mu }\Upsilon ^{\mu
}\right) _{j}^{i}\bar{\varphi}^{j}  \label{lm}
\end{equation}%
where%
\begin{equation}
\Upsilon _{\mu }\Upsilon ^{\mu }=\frac{g^{2}}{4}\Theta ,\qquad
\Theta _{i}^{j}=\left(
\begin{array}{cc}
\Theta _{1}^{1} & \Theta _{1}^{2} \\
\Theta _{2}^{1} & \Theta _{2}^{2}%
\end{array}%
\right)
\end{equation}%
with%
\begin{equation}
\Theta _{2}^{2}=2W^{+\mu }W_{\mu }^{-}+\frac{1}{\cos ^{2}\vartheta _{_{W}}}%
Z^{\mu }Z_{\mu }
\end{equation}%
and\ analogous relations for the others $\Theta $'s. \

\  \  \

\emph{Anticipation: Masses of the gauge particles}\newline To see
how the conifold geometry of the ground state is involved in the
masses of the gauge particles, let us compute the masses of the
$W_{\mu }^{\pm }$ and $Z_{\mu }$ by using results presented in the
summary given in previous section and which will be derived later
on. \newline Using the expression (\ref{hr}-\ref{fr}) giving the
values of the Higgs
fields $h^{i}$ and $\varphi _{i}$ in the ground state namely%
\begin{equation}
\begin{tabular}{lll}
$h^{i}$ & $=\varrho f^{i}$ & $=\varrho \text{ }\left(
\begin{array}{c}
\cos \frac{\mathrm{\theta }}{2}e^{\frac{i}{2}\left( \psi +\phi \right) } \\
\sin \frac{\mathrm{\theta }}{2}e^{\frac{i}{2}\left( \psi -\phi \right) }%
\end{array}%
\right) $ \\
&  &  \\
$\varphi _{i}$ & $=\frac{\mathrm{\nu }}{\varrho }\bar{f}_{i}$ & $=\frac{%
\mathrm{\nu }}{\varrho }\text{ }\left(
\begin{array}{c}
\cos \frac{\mathrm{\theta }}{2}e^{-\frac{i}{2}\left( \psi +\phi \right) } \\
\sin \frac{\mathrm{\theta }}{2}e^{-\frac{i}{2}\left( \psi -\phi \right) }%
\end{array}%
\right) $ \\
&  &
\end{tabular}
\label{fg}
\end{equation}%
with%
\begin{equation}
\varrho ^{2}+\frac{\left \vert \mathrm{\nu }\right \vert ^{2}}{\varrho ^{2}}=%
\mathrm{r}+\frac{2\left \vert \mathrm{\nu }\right \vert
^{2}}{\mathrm{r}\sin
^{2}\vartheta _{_{W}}+\sqrt{4\left \vert \mathrm{\nu }\right \vert ^{2}+%
\mathrm{r}^{2}\sin ^{4}\vartheta _{_{W}}}}  \label{rnr}
\end{equation}%
then eq(\ref{lm}) takes then the form%
\begin{equation*}
L_{m}=\frac{g^{2}}{4}\left( \varrho ^{2}+\frac{\left \vert \mathrm{\nu }%
\right \vert ^{2}}{\varrho ^{2}}\right) \text{ }\bar{f}_{i}\Theta
_{j}^{i}f^{j}
\end{equation*}%
For the case where the angles are set to $\mathrm{\theta }=\pi $ and
$\psi
=\phi ,$ the relations (\ref{fg}) reduce to%
\begin{equation*}
\begin{tabular}{lllllll}
$h^{i}$ & $=$ & $\varrho \text{ }\left(
\begin{array}{c}
0 \\
1%
\end{array}%
\right) $ & , & $\varphi _{i}$ & $=$ & $\frac{\mathrm{\nu }}{\varrho
}\text{ }\left(
\begin{array}{c}
0 \\
1%
\end{array}%
\right) $%
\end{tabular}%
\end{equation*}%
and the $U_{Y}\left( 1\right) \times SU_{L}\left( 2\right) $ gauge
symmetry is broken down to $U_{em}\left( 1\right) $. The masses of
the gauge bosons
are then read from the following relation%
\begin{equation*}
L_{m}=\frac{g^{2}}{2}\left( \varrho ^{2}+\frac{\left \vert \mathrm{\nu }%
\right \vert ^{2}}{\varrho ^{2}}\right) \left( W^{+\mu }W_{\mu }^{-}+\frac{1%
}{2\cos ^{2}\vartheta _{_{W}}}Z^{\mu }Z_{\mu }\right)
\end{equation*}%
from which we read the masses of the gauge fields%
\begin{equation*}
M_{W_{\mu }^{\pm }}=g\sqrt{\frac{\varrho ^{4}+\left \vert \mathrm{\nu }%
\right \vert ^{2}}{2\varrho ^{2}}},\qquad M_{Z_{\mu
}}=\frac{M_{W_{\mu }^{\pm }}}{\cos ^{2}\vartheta _{_{W}}},\qquad
M_{A_{\mu }}=0
\end{equation*}%
These masses, which should be thought of as the ones given by the
standard model namely
\begin{eqnarray*}
M_{W} &=&80.385\pm 0.015GeV/c \\
M_{Z} &=&91.1876\pm 0.0021GeV/c
\end{eqnarray*}%
are proportional to the square root of $\varrho ^{2}+\frac{\left
\vert \mathrm{\nu }\right \vert ^{2}}{\varrho ^{2}}$ and so are,
modulo the used assumption, related to the Kahler \textrm{r} and
complex $\mathrm{\nu }$ parameters of the intersecting conifold
geometries as given by eq(\ref{rnr}).

\subsubsection{Scalar potential}

Like in the usual case of next- to - MSSM, the full scalar potential $%
\mathcal{V}_{higgs}$ of the Higgs field of the \emph{n-MSSM}$^{\ast }$ with $%
\boldsymbol{H}_{d}$ replaced by the anti-doublet $\Phi $ ( conifold
model in
our terminology) is given by%
\begin{equation*}
\mathcal{V}_{higgs}=\mathcal{V}_{susy}+\mathcal{V}_{exl}
\end{equation*}%
with $\mathcal{V}_{susy}$ the supersymmetric component%
\begin{equation*}
\mathcal{V}_{susy}=\mathcal{V}_{ch}+\mathcal{V}_{re}
\end{equation*}%
and $\mathcal{V}_{exl}$ the explicit supersymmetry breaking term.

\paragraph{\textbf{1)}\emph{\ supersymmetric contributions}\newline
}

\emph{a) case of doublet }$\boldsymbol{H}_{u}$ \emph{and
anti-doublet }$\Phi $\newline In the conifold model
\emph{n-MSSM}$^{\ast }$ with supersymmetric lagrangian
density (\ref{lt}), the supersymmetric scalar potential reads like%
\begin{equation}
\begin{tabular}{lll}
$\mathcal{V}_{susy}^{conifold}$ & $=$ & $\left( \bar{F}_{i}F^{i}+G_{i}\bar{G}%
^{i}+\bar{F}_{S}F_{S}\right) +$ \\
&  & $\left( \frac{1}{2}D^{\prime 2}+\frac{1}{2}D_{A}D^{A}\right) $%
\end{tabular}%
\end{equation}%
with the auxiliary fields $F$ and $D$ related to the Higgs fields as follows%
\begin{equation}
\begin{tabular}{lll}
&  &  \\
$\bar{F}_{S}$ & $=$ & $\kappa S^{2}+\lambda \left( h^{i}\varepsilon
_{ij}\varphi ^{j}-\mathrm{\nu }\right) $ \\
$\bar{F}_{i}$ & $=$ & $+\lambda S\varphi _{i}$ \\
$\bar{G}^{i}$ & $=$ & $+\lambda Sh^{i}$ \\
&  &  \\
$D^{\prime }$ & $=$ & $\frac{g^{\prime }}{2}\left[
\bar{h}_{i}h^{i}-\varphi
_{i}\bar{\varphi}^{i}-\mathrm{r}\right] $ \\
$D^{A}$ & $=$ & $\frac{g}{2}\left( \tau ^{A}\right) _{j}^{i}\left[ \bar{h}%
_{i}h^{j}-\varphi _{i}\bar{\varphi}^{j}\right] $ \\
&  &
\end{tabular}
\label{es}
\end{equation}%
For later use, notice that the equation of the auxiliary field
$D^{A}$ can
be also expressed into a symmetric manner like%
\begin{equation*}
D^{A}=\frac{g}{4}\left( \varepsilon \tau ^{A}\right) _{ij}\left[ \bar{h}%
^{(i}h^{j)}-\varphi ^{(i}\bar{\varphi}^{j)}\right]
\end{equation*}%
Eqs(\ref{es}) capture basic data on the physical properties of the
Higgs fields $S$, $h^{i}$ and $\varphi _{i}$ in the ground state.
Their vanishing condition involve:

\begin{itemize}
\item \emph{14} hermitian coupled relations:

\begin{itemize}
\item \emph{5} of them complex holomorphic eqs given by for the auxiliary
fields F,

\item \emph{4} hermitian ones for the auxiliary fields D;
\end{itemize}

\item \emph{9} coupling constant moduli

\begin{itemize}
\item \emph{3} complex constants$\ $namely $\lambda ,$ $\kappa ,$ $\nu $

\item \emph{3} real ones$\ g,$ $g^{\prime },$ $r$
\end{itemize}
\end{itemize}

\  \  \newline The determination of the exact solutions of
eqs(\ref{es}) is of a major importance; this allows to get more
insight into the explicit expression of the Higgs VEVs
\begin{equation*}
\left \langle S\right \rangle ,\qquad \left \langle h^{i}\right
\rangle ,\qquad \left \langle \varphi _{i}\right \rangle
\end{equation*}
and the relations between them.

\emph{b) comparison with n-MSSM eqs}\newline
In \emph{n-MSSM }based on the usual two Higgs superfield doublets $%
\boldsymbol{H}_{u}$ and $\boldsymbol{H}_{d}$, the scalar potential
reads as
\begin{equation}
\begin{tabular}{lll}
$\mathcal{V}_{susy}^{n-mssm}$ & $=$ & $\left( \bar{F}_{u}\right)
_{i}\left(
F_{u}\right) ^{i}+\left( \bar{F}_{d}\right) _{i}\left( F_{d}\right) ^{i}+%
\bar{F}_{S}F_{S}+$ \\
&  & $\frac{1}{2}D^{\prime 2}+\frac{1}{2}D_{A}D^{A}$%
\end{tabular}%
\end{equation}%
with%
\begin{equation}
\begin{tabular}{lll}
&  &  \\
$\left( \bar{F}_{S}\right) _{n-mssm}$ & $=$ & $\kappa S^{2}+\lambda
\left(
h^{i}\varepsilon _{ij}\varphi ^{j}-\mathrm{\nu }\right) $ \\
$\left( \bar{F}_{ui}\right) _{n-mssm}$ & $=$ & $+\lambda S\varphi _{i}$ \\
$\left( \bar{F}_{di}\right) _{n-mssm}$ & $=$ & $-\lambda Sh_{i}$ \\
&  &  \\
$\left( D^{\prime }\right) _{n-mssm}$ & $=$ & $\frac{g^{\prime
}}{2}\left[
\bar{h}_{i}h^{i}-\varphi _{i}\bar{\varphi}^{i}-\mathrm{r}\right] $ \\
$\left( D^{A}\right) _{n-mssm}$ & $=$ & $\frac{g}{2}\left( \tau
^{A}\right)
_{j}^{i}\left( \bar{h}_{ui}h_{u}^{j}+\bar{h}_{di}h_{d}^{j}\right) $ \\
&  &
\end{tabular}%
\end{equation}%
The field equation of $D^{A}$ of eq(\ref{es}) and the one of $\left(
D^{A}\right) _{n-mssm}$ differ from the sign in front of the terms
$\varphi ^{(i}\bar{\varphi}^{j)}$ and $\bar{h}_{d}^{(i}h_{d}^{j)}$;
this is due to the quantum charge property

\begin{eqnarray*}
&&%
\begin{tabular}{lll}
$\left[ T^{A},\varphi _{i}\right] $ & $=$ & $-\frac{1}{2}\varphi
_{i}\left(
\tau ^{A}\right) _{j}^{i}$ \\
$\left[ T^{A},h_{d}^{i}\right] $ & $=$ & $+\frac{1}{2}\left( \tau
^{A}\right) _{j}^{i}h_{d}^{j}$%
\end{tabular}
\\
&&
\end{eqnarray*}%
The vanishing conditions $D^{A}=0$ and the $\left( D^{A}\right) _{{\small n}%
\text{-}{\small MSSM}}=0$ lead to different solutions; and therefore
to different supersymmetric ground states. The same conclusion is
valid for the corresponding scalar potentials and their extrema.

\paragraph{\textbf{2)}\emph{\ explicit supersymmetry breaking contribution}%
\newline
}

\begin{figure}[tbph]
\begin{center}
\hspace{0cm} \includegraphics[width=10cm]{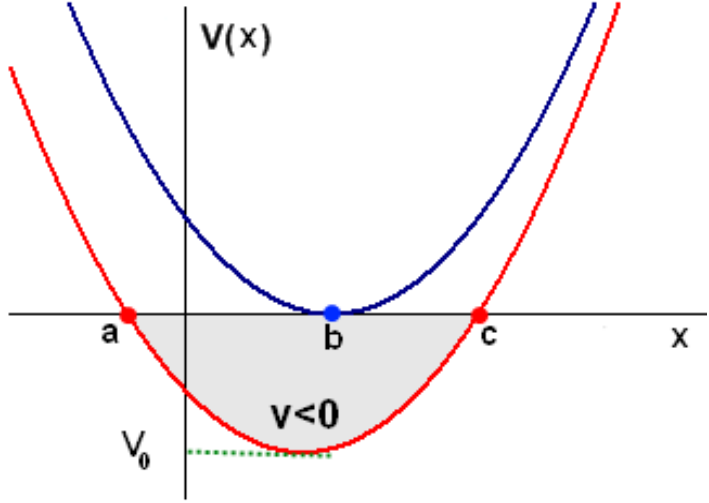}
\end{center}
\par
\vspace{-1cm}
\caption{In red, a typical non supersymmetric potential imagined as $\frac{1%
}{2}\left( x-a\right) \left( x-c\right) +V_{0}$; describing a
deformation of
the supersymmetric shape with $a\neq c$ and $V_{0}$ induced by $\mathcal{V}%
_{expl}.$ For the particular case $a=c=b$ and $V_{0}=0$ (blue) one
recovers supersymmetry. In terms of Higgs fields, the potential is
of course quartic; it has the usual shape; say with "3 extrema": 2
minima and a local maximum.} \label{1A}
\end{figure}

In dealing with the Higgs potential in \emph{n-MSSM} and in the \emph{n-MSSM}%
$^{\ast }$ \emph{conifold version} using the antidoublet $\varphi
_{i}$, one
extends the supersymmetric $\mathcal{V}_{susy}$ by an extra term $\mathcal{V}%
_{exl}$ breaking explicitly supersymmetry
\begin{equation}
\mathcal{V}_{higgs}=\mathcal{V}_{susy}+\mathcal{V}_{exl}
\end{equation}%
This term is required by low energy phenomenology. In the case of
the conifold model we are interested in here, the term
$\mathcal{V}_{exl}$ reads like
\begin{equation*}
\begin{tabular}{lll}
$\mathcal{V}_{exl}$ & $=$ & $-m_{_{u}}^{2}$
$\bar{h}_{i}h^{i}-m_{\varphi }^{2}$ $\varphi
_{i}\bar{\varphi}^{i}-m_{S}^{2}\left \vert S\right \vert
^{2} $ \\
&  & $-\left( m_{u\varphi }^{2}-\lambda A_{_{\lambda }}S\right)
\varphi
_{i}h^{i}+\frac{\kappa }{3}A_{\kappa }S^{3}+hc$%
\end{tabular}%
\end{equation*}%
The adjunction of this term to $\mathcal{V}_{susy}$ modifies the
shape of the Higgs potential which is no longer a positive function
as schematized in fig \ref{1A} . Generally, it has a remarkable
region with negative values.
\bigskip

Notice that the above $\mathcal{V}_{exl}$ can be put into the
following form
depending linearly in the auxiliary fields $D^{\prime }$\ and $\bar{F}$ like%
\begin{equation}
\begin{tabular}{lll}
$\mathcal{V}_{exl}$ & $=$ & $-\frac{m_{_{u}}^{2}}{2g^{\prime
}g^{\prime \prime }}\left( g^{\prime }\Delta ^{\prime }+g^{\prime
\prime }D^{\prime }\right) +$ $\frac{m_{\varphi }^{2}}{2g^{\prime
}g^{\prime \prime }}\left( g^{\prime \prime }D^{\prime }-g^{\prime
}\Delta ^{\prime }\right)
-m_{S}^{2}\left \vert S\right \vert ^{2}$ \\
&  &  \\
&  & $+\left( \frac{m_{u\varphi }^{2}}{\lambda }-A_{_{\lambda
}}S\right)
\bar{F}+\frac{m_{u\varphi }^{2}\kappa }{\lambda }S^{2}-\frac{2\kappa }{3}%
A_{\kappa }S^{3}+hc$ \\
&  &
\end{tabular}
\label{3V}
\end{equation}%
where $\Delta ^{\prime }$ is as in eq(\ref{exv}). \newline In the
remainder of this study, we proceed follows:

\begin{itemize}
\item switch off $\mathcal{V}_{exl}$ and study the ground state phases:%
\newline
the exact supersymmetric phase will be studied in section 5; and the
broken supersymmetric one in section 6,

\item switch on $\mathcal{V}_{exl}$ and explore the modification of the
obtained results; this is studied in section 7.
\end{itemize}

\section{Supersymmetric phase}

In quantum \emph{supersymmetric} gauge theories, the energy of the
supersymmetric ground state vanishes $\mathcal{E}_{\min }^{susy}=0$;
a remarkable property manifested at the level of the component field
lagrangian by the positivity of the scalar potential
$\mathcal{V}_{susy}$
which reads in our case like%
\begin{equation}
\begin{tabular}{lll}
$\mathcal{V}_{susy}$ & $=$ & $\bar{F}_{i}F^{i}+G_{i}\bar{G}^{i}+\bar{F}%
_{S}F_{S}+\frac{1}{2}D^{\prime 2}+\frac{1}{2}D_{A}D^{A}\geq 0$%
\end{tabular}
\label{po}
\end{equation}%
This hermitian function involves various kinds of field
representations
namely iso-singlets $F_{S}$ and $D^{\prime }$, iso-doublets $F^{i}$, $\bar{G}%
^{i}$ and corresponding anti-doublets $\bar{F}_{i},$ $G_{i}$ as well
as the iso-triplet $D_{A}$.

\subsection{Exact supersymmetry}

On supersymmetric ground state $\left \vert \Sigma _{susy}\right
\rangle $ of the conifold model (n-MSSM with the anti-doublet
$\boldsymbol{\Phi }$ instead of $\boldsymbol{H}_{d}$), described by
the superspace lagrangian density (\ref{kt}-\ref{lt}), the scalar
potential of the Higgs fields
vanishes%
\begin{equation}
\left. \mathcal{V}_{susy}\right \vert _{\Sigma _{susy}}=\mathcal{V}\left( S,%
\bar{S},h,\bar{h},\varphi ,\bar{\varphi}\right) =0
\end{equation}%
This equation should be thought of as
\begin{equation}
\sum_{I}\mathcal{V}_{I}=0
\end{equation}%
a sum of several constraint equations given by $\mathcal{V}_{I}=0$;
and
capturing data on the allowed values of the VEVs of the Higgs fields $S,$ $%
h_{i},$ $\varphi _{i}$ that preserve supersymmetry. \newline Below,
we study the set of solutions of these equations and determine the
explicit expressions of these VEVs.

\subsubsection{Supersymmetric ground state}

The set $\Sigma _{susy}$ of the Higgs field moduli $S,$ $h_{i},$
$\varphi _{i}$, solving the vanishing condition
$\mathcal{V}_{susy}=0,$ defines the exact supersymmetric phase of
the model. For a geometric interpretation, we
will refer to this set as%
\begin{equation}
\Sigma _{susy}=\left \{ \left.
\begin{array}{c}
\\
\end{array}%
\right. \left( S,h_{i},\varphi _{i}\right) \in \mathbb{C}^{5}\text{
\TEXTsymbol{\vert} \ }\mathcal{V}_{susy}=0\text{ }\left.
\begin{array}{c}
\\
\end{array}%
\right. \text{ }\right \}  \label{ho}
\end{equation}%
and, in connection with this manifold, we also need other spaces and
parameterizations; in particular the real 3-spheres
$\mathbb{S}_{h}^{3}$ and
$\mathbb{S}_{\varphi }^{3}$ respectively parameterized by the Higgs moduli $%
h^{i}$ and $\varphi _{i}$ as follows%
\begin{equation}
\begin{tabular}{lll}
$\mathbb{S}_{h}^{3}$ & $=$ & $\left \{ h_{i}\in
\mathbb{C}_{h}^{2}\text{ \  \
\TEXTsymbol{\vert} \ }\bar{h}_{i}h^{i}=\varrho ^{2}\right \} $ \\
$\mathbb{S}_{\varphi }^{3}$ & $=$ & $\left \{ \varphi _{i}\in \mathbb{C}%
_{\varphi }^{2}\text{ \  \  \TEXTsymbol{\vert} \ }\varphi _{i}\bar{\varphi}%
^{i}=R^{2}\right \} $ \\
&  &
\end{tabular}%
\end{equation}%
To deal with these spheres, we need moreover the harmonic field coordinates $%
\mathrm{\bar{f}}_{i}$ and $\mathrm{f}^{i}$ given by
\textrm{eq(\ref{arr}); see also sub-section 5.1.3}. Other related
quantities are needed as well; they will be introduced at the
appropriate time.

\  \  \  \  \newline From a geometric view, the space $\Sigma
_{susy}$ can be imagined as a
hypersurface contained in the real \emph{10} dimension space $\mathbb{R}%
^{10} $ parameterized by the \emph{10} real degrees of freedom
captured by the \emph{5} complex Higgs fields
\begin{equation}
\Sigma _{susy}\text{ \ }\mathbf{\subset }\text{ \
}\mathbb{R}^{10}\sim \mathbb{C}^{5}
\end{equation}%
Because of the algebraic structure of the scalar potential, which is
given by the sum of positive quantities, the condition
$\mathcal{V}_{susy}=0$ requires then the vanishing of all auxiliary
fields of eq(\ref{po}). \newline Moreover, seen that the auxiliary
fields are of two kinds: $\left( i\right) $
type $F$- complex auxiliary fields coming from chiral multiplets; and $%
\left( ii\right) $ type $D$ -hermitian auxiliary following from the
hermitian gauge multiplets; it is useful to split these constraint
relations into complex holomorphic constraint eqs and hermitian ones
as follows:

$\mathbf{\alpha })$ \emph{complex}%
\begin{eqnarray}
\bar{F}_{i} &=&\bar{G}^{i}=0  \notag \\
\bar{F}_{S} &=&0
\end{eqnarray}

$\mathbf{\beta })$ \emph{hermitian}
\begin{eqnarray}
D^{\prime } &=&0  \notag \\
D^{A} &=&0
\end{eqnarray}%
This splitting teaches us that the set $\Sigma _{susy}$, defined by eq(\ref%
{ho}), is therefore given by the intersection of two hypersurfaces $%
\mathfrak{C}_{\mathrm{\nu }}$ and $\mathfrak{R}_{\mathrm{r}}$ as
follows
\begin{equation}
\Sigma _{susy}\text{ \ }=\text{ \ }\mathfrak{C}_{\mathrm{\nu }}\text{ \ }%
\dbigcap \text{\  \ }\mathfrak{R}_{\mathrm{r}}
\end{equation}%
with%
\begin{equation}
\mathfrak{C}_{\mathrm{\nu }}=\left \{ \left.
\begin{array}{c}
\\
\end{array}%
\right. \left( S,h^{i},\varphi _{i}\right) \in \mathbb{C}^{5}\text{ }\mathbf{%
|}\text{ \ }\bar{F}_{i}=0,\text{ }\bar{G}^{i}=0,\text{ }\bar{F}_{S}=0\text{ }%
\left.
\begin{array}{c}
\\
\end{array}%
\right. \text{ }\right \}
\end{equation}%
and%
\begin{eqnarray}
\mathfrak{R}_{\mathrm{r}} &=&\left \{ \left.
\begin{array}{c}
\\
\end{array}%
\right. \left( S,\bar{S},h^{i},\bar{h}_{i},\varphi _{i},\bar{\varphi}%
^{i}\right) \in \mathbb{R}^{10}\text{ \TEXTsymbol{\vert} \ }D^{\prime }=0,%
\text{ }D^{A}=0\text{ }\left.
\begin{array}{c}
\\
\end{array}%
\right. \text{ }\right \} \\
&&  \notag
\end{eqnarray}%
Using the field equations of motion of the auxiliary fields
(\ref{es}), we then get the explicit expression of the constraint
relations among the Higgs fields $S,$ $h^{i}$ and $\varphi _{i}$;
they are given by the \emph{5}
algebraic complex holomorphic relations%
\begin{equation}
\begin{tabular}{lll}
&  &  \\
$\mathfrak{C}_{\mathrm{\nu }}$ & $=$ & $\left \{ \text{ }\left(
S,h_{i},\varphi _{i}\right) \text{ \ }\mathbf{|}\text{ }\left[ \text{\ }%
\begin{tabular}{lll}
$\lambda S\varphi _{i}$ & $=$ & $0$ \\
$\lambda Sh^{i}$ & $=$ & $0$ \\
$\kappa S^{2}+\lambda \left( \varphi _{i}h^{i}-\mathrm{\nu }\right)
$ & $=$
& $0$%
\end{tabular}%
\right. \text{ }\left.
\begin{array}{c}
\\
\end{array}%
\right. \right \} $ \\
&  &
\end{tabular}
\label{E11}
\end{equation}

and the \emph{4} hermitian ones%
\begin{equation}
\begin{tabular}{lll}
&  &  \\
$\mathfrak{R}_{\mathrm{r}}$ & $=$ & $\left \{ \text{ }\left(
h_{i},\varphi
_{i}\right) \text{ \ }\mathbf{|}\text{ }\left[ \text{\ }%
\begin{tabular}{lll}
$\frac{g^{\prime }}{2}\left( \bar{h}_{i}h^{i}-\varphi _{i}\bar{\varphi}^{i}-%
\mathrm{r}\right) $ & $=$ & $0$ \\
$\frac{g}{2}\left( \tau ^{A}\right) _{j}^{i}\left(
\bar{h}_{i}h^{j}-\varphi
_{i}\bar{\varphi}^{j}\right) $ & $=$ & $0$%
\end{tabular}%
\right. \text{ }\left.
\begin{array}{c}
\\
\end{array}%
\right. \right \} $ \\
&  &
\end{tabular}
\label{E2}
\end{equation}%
Clearly, these are strong constraint relations as there are much
more
equations, [\emph{5 complex }(\ref{E11})\emph{\ }plus\emph{\ 4 real }(\ref%
{E2})], than the number of variables ( \emph{5} complex field
moduli).

\  \  \  \  \  \newline To derive the solutions of these eqs, we
shall distinguish two cases depending on the value of the hermitian
FI coupling constant: $\left( i\right) $ $r=0$ and $\left( ii\right)
$ $r\neq 0.$

\begin{itemize}
\item case $r=0:$ \emph{exact supersymmetric} \emph{phase}.\newline
We will show that in this phase the supersymmetric ground state is
completely characterized by the absolute value $\left \vert \mathrm{\nu }%
\right \vert $ of the complex parameter $\mathrm{\nu }$ of the
deformed conifold geometry. \newline Denoting the VEVs of the two
Higgs fields $h^{i}$ and $\varphi _{i}$
\begin{equation}
\begin{tabular}{lllllll}
$\upsilon _{h}^{0}$ & $=$ & $\left \langle
\sqrt{\bar{h}_{i}h^{i}}\right \rangle _{{\small r=0}}$ & \  \ , \  \
\  \  & $\upsilon _{\varphi }^{0}$ & $=$
& $\left \langle \sqrt{\varphi _{i}\bar{\varphi}^{i}}\right \rangle _{%
{\small r=0}}$ \\
&  &  &  &  &  &
\end{tabular}%
\end{equation}%
we have%
\begin{equation*}
\begin{tabular}{lllllll}
$\upsilon _{h}^{0}$ & $=$ & $\sqrt{\left \vert \mathrm{\nu }\right
\vert }$ & \  \ , \  \  \  \  & $\upsilon _{\varphi }^{0}$ & $=$ &
$\sqrt{\left \vert
\mathrm{\nu }\right \vert }$%
\end{tabular}%
\end{equation*}%
and so
\begin{equation}
\frac{\upsilon _{h}^{{\small 0}}}{\upsilon _{\varphi }^{{\small
0}}}\equiv \tan \beta _{susy}=1
\end{equation}%
The supersymmetric phase of the Higgs ground state requires
therefore $\beta _{susy}=\frac{\pi }{4}$.

\item case $r\neq 0:$ \emph{broken supersymmetric} \emph{phase}.\newline
In this case, the values $\upsilon _{h}$ and $\upsilon _{\varphi }$
of the Higgs VEVs are no longer equal; and, in addition to $\left
\vert \mathrm{%
\nu }\right \vert $, they depend as well on the hermitian FI
coupling
constant r and on the gauge coupling constants $g$ and $g^{\prime }$%
\begin{equation}
\begin{tabular}{lll}
$\upsilon _{h}$ & $=$ & $\upsilon _{h}\left( \nu ;r,g,g^{\prime
}\right) $
\\
$\upsilon _{\varphi }$ & $=$ & $\upsilon _{\varphi }\left( \nu
;r,g,g^{\prime }\right) $%
\end{tabular}%
\end{equation}%
\textrm{with explicit expression as in eqs(3.92)}.\newline Because
of this deviation induced by the Kahler parameter r, the
supersymmetric value of $\tan \beta _{susy}$ gets modified into
$\tan \beta $ given \textrm{by eq(\ref{tanb})}; and which reads in
the limit of small r as
follows%
\begin{equation}
\tan \beta \simeq 1+\frac{g^{\prime 2}}{2\left( g^{2}+g^{\prime 2}\right) }%
\frac{\mathrm{r}}{\left \vert \mathrm{\nu }\right \vert }
\end{equation}%
In this view, the second term proportional to
$\frac{\mathrm{r}}{\left \vert \mathrm{\nu }\right \vert }$ in above
relation captures a breaking of supersymmetry effect. In what
follows, we give the explicit details.
\end{itemize}

\subsubsection{Switching off FI coupling constant r}

Setting the FI coupling constant parameter r to zero; but keeping
the complex $\mathrm{\nu }$ arbitrary, the supersymmetric constraint
relations
become%
\begin{equation}
\text{\ }\left \{
\begin{tabular}{lll}
$\lambda S\varphi _{i}$ & $=$ & $0$ \\
$\lambda Sh^{i}$ & $=$ & $0$ \\
$\kappa S^{2}+\lambda \left( \varphi _{i}h^{i}-\mathrm{\nu }\right)
$ & $=$
& $0$%
\end{tabular}%
\right.  \label{1S}
\end{equation}%
and
\begin{equation}
\left \{
\begin{tabular}{lll}
$\bar{h}_{i}h^{i}-\varphi _{i}\bar{\varphi}^{i}$ & $=$ & $0$ \\
&  &  \\
$\left( \tau ^{A}\right) _{j}^{i}\left( \bar{h}_{i}h^{j}-\varphi _{i}\bar{%
\varphi}^{j}\right) $ & $=$ & $0$%
\end{tabular}%
\right.  \label{2S}
\end{equation}

\  \  \  \  \newline To derive the solutions of these constraint
eqs, we proceed in \emph{3 steps} as follows: \newline First we
solve\ the complex holomorphic eqs(\ref{1S}); these solutions give
the structure of the set $\mathfrak{C}_{\mathrm{\nu }}$.\newline
Then we require to the obtained solutions; i.e ($\varphi _{i},$
$h^{i}\in
\mathfrak{C}_{\mathrm{\nu }}$), to satisfy also the iso-singlet constraint $%
D^{\prime }=0$ given by the first eq of (\ref{2S}). \newline
These solutions, that belong to a subspace of $\mathfrak{C}_{\mathrm{\nu }}$%
, are required to satisfy as well the iso-triplet constraint
$D^{A}=0$ given by the second eq of (\ref{2S}).

\  \  \

\textbf{1)} \emph{solving the complex eq(\ref{1S})}\newline A non
trivial set of solutions of the complex holomorphic constraints is
obtained into two stages: first by solving the vanishing conditions
of the auxiliary field doublets $\bar{F}_{i}=0$ and $\bar{G}^{i}=0$
ensured by
setting to zero the complex iso-singlet%
\begin{equation}
S=0
\end{equation}%
Then putting this value back into the third complex holomorphic
constraint between the two complex doublets $h^{i}$ and $\varphi
_{i}$; one ends with
\begin{equation}
\varphi _{i}h^{i}=\mathrm{\nu }  \label{3S}
\end{equation}%
But this relation is a well known equation; as it is precisely the
defining equation of the complex deformed conifold singularity of
the cotangent bundle of the real 3-sphere,
\begin{equation}
T^{\ast }S^{3}
\end{equation}%
Explicitly, by setting $\varphi _{i}=\left( \varphi _{1},\varphi
_{2}\right) $ and $h^{i}=\left( h^{1},h^{2}\right) $, which by
lowering the indices using $\varepsilon ^{ij}$ tensor reads also as
$h^{i}=\left( h_{2},-h_{1}\right) $, we have
\begin{equation}
\varphi _{1}h_{2}-\varphi _{2}h_{1}=\mathrm{\nu }
\end{equation}%
Notice also that eq(\ref{3S}) is invariant under the $U_{Y}\left(
1\right) \times SU_{L}\left( 2\right) $ gauge symmetry; this is a
remarkable feature that is helpful for working out exact explicit
solutions of this equation.

\  \  \  \newline For a physical interpretation; but also for
geometric view, it is interesting to use \textrm{the following real
4D space equivalences}
\begin{equation}
\mathbb{C}^{2}\text{ \ }\sim \text{ \ }\mathbb{R}^{4}\text{ \ }\sim
\text{ \ }\mathbb{R}^{+}\times \mathbb{S}_{h}^{3}
\end{equation}%
where $\mathbb{R}_{+}$ is the positive half line and
$\mathbb{S}_{h}^{3}$ is the real 3-sphere parameterized by the
complex $h^{i}$ and $\bar{h}_{i}$. Its radius is related to the
Higgs field moduli like
\begin{equation}
\varrho _{_{h}}=\sqrt{\sum_{i=1,2}\bar{h}_{i}h^{i}}=\sqrt{\left \vert \bar{h}%
_{1}\right \vert ^{2}+\left \vert \bar{h}_{2}\right \vert
^{2}}=\sqrt{\left \vert h_{u}^{0}\right \vert
^{2}+h_{u}^{+}\bar{h}_{u}^{-}}
\end{equation}%
it exhibits explicitly the conifold singularity at the origin of the
complex 2 space; as well as the residual $U_{em}\left( 1\right) $
symmetry of the Higgs fields.

\  \  \  \  \  \  \newline By help of the $U_{Y}\left( 1\right)
\times SU_{L}\left( 2\right) $ gauge symmetry of eq(\ref{3S}), one
can use the harmonic coordinate field variables $\left(
f^{i},\bar{f}_{i}\right) $ of the real \emph{unit} 3-sphere
\begin{equation*}
\mathbb{S}^{3}\sim SU\left( 2\right)
\end{equation*}%
obeying $\bar{f}_{i}f^{i}=1$, to decompose\footnote{%
the charges carried by $h^{+i}$ and $\varphi _{i}^{-}$ refer to the $%
U_{Y}\left( 1\right) $ hypercharges of the Higgs fields; they have
been exhibited in order to fix the ideas. Later on they will be
dropped out.} the field moduli $h^{i}\equiv h^{+i}$ and $\varphi
_{i}\equiv \varphi _{i}^{-}$ as

\begin{eqnarray}
h^{+i} &=&\chi ^{0}f^{+i}-\chi ^{++}\bar{f}^{-i}  \notag \\
\bar{h}_{i}^{-} &=&\chi ^{--}f_{i}^{+}+\bar{\chi}^{0}\bar{f}_{i}^{-}
\label{fi}
\end{eqnarray}%
and%
\begin{equation}
\begin{tabular}{lll}
$\varphi _{i}^{-}$ & $=$ & $\xi ^{--}f_{i}^{+}+\xi ^{0}\bar{f}_{i}^{-}$ \\
$\bar{\varphi}^{+i}$ & $=$ & $\bar{\xi}^{0}f^{+i}-\xi ^{++}\bar{f}^{-i}$ \\
&  &
\end{tabular}
\label{gi}
\end{equation}%
The explicit expressions of the harmonic field variables $\left( f^{+i},\bar{%
f}_{i}^{-}\right) \equiv \left( f^{i},\bar{f}_{i}\right) $ and their
basic features are given by eqs(\ref{fhar}); their physical meaning
will be given later by requiring $\left( \chi ^{0}\right) ^{\ast
}=\chi ^{0}$ and $\chi ^{++}=0$; see eqs(\ref{gfh}).

\  \  \  \newline
Notice that in the decomposition (\ref{fi}), the complex component fields $%
\chi ^{0},$ $\chi ^{++}$ and their complex conjugates $\bar{\chi}^{0},$ $%
\chi ^{--}$ are related to $h^{i}$ and $\bar{h}_{i}$ like

\begin{equation}
\begin{tabular}{lllllll}
$\chi ^{0}$ & $=$ & $\bar{f}_{i}h^{i}$ & \  \ , \  \  \  \  &
$\bar{\chi}^{0}$ &
$=$ & $f^{i}\bar{h}_{i}$ \\
$\chi ^{++}$ & $=$ & $f_{i}h^{i}$ & \  \ , & $\chi ^{++}$ & $=$ & $-\bar{f}%
^{i}\bar{h}_{i}$ \\
&  &  &  &  &  &
\end{tabular}
\label{xf}
\end{equation}%
and are nothing else the usual components Higgs fields; but in the
harmonic
coordinate basis. Moreover, being iso-singlets under $SU_{L}\left( 2\right) $%
, the hypercharges carried by the fields $\chi ^{0},$ $\chi ^{++}$
are precisely the electric charges of the residual gauge symmetry
$U_{em}\left(
1\right) $ generated by%
\begin{equation}
Q_{em}=T_{L}^{3}+\frac{Y}{2}
\end{equation}%
Notice also that $\chi ^{0},$ $\chi ^{++}$ still form an isodoublet;
but
under a dual $\widetilde{SU}_{L}\left( 2\right) $ symmetry generated by%
\begin{equation}
\begin{tabular}{lll}
$D^{++}$ & $=$ & $f^{+i}\frac{\partial }{\partial \bar{f}^{-i}}$ \\
$D^{--}$ & $=$ & $\bar{f}^{-i}\frac{\partial }{\partial f^{+i}}$ \\
$D^{0}$ & $=$ & $f^{+i}\frac{\partial }{\partial f^{+i}}-\bar{f}^{-i}\frac{%
\partial }{\partial \bar{f}^{-i}}$ \\
&  &
\end{tabular}
\label{hal}
\end{equation}%
This feature can be checked explicitly by using eq(\ref{xf}); and
computing for instance
\begin{equation*}
D^{--}\chi ^{0},\qquad D^{++}\chi ^{0},\qquad \left( D^{++}\right)
^{2}\chi ^{0}
\end{equation*}%
we have%
\begin{equation}
\begin{tabular}{lllll}
$D^{--}\chi ^{0}$ & $=$ & $0$ &  &  \\
&  &  &  &  \\
$D^{++}\chi ^{0}$ & $=$ & $D^{++}\left( -\bar{f}^{i}h_{i}\right) $ &  &  \\
& $=$ & $-f^{+i}h_{i}$ & $=$ & $\chi ^{++}$ \\
&  &  &  &  \\
$D^{++}\chi ^{++}$ & $=$ & $\left( D^{++}\right) ^{2}\chi ^{0}$ &
$=$ & $0$
\\
&  &  &  &
\end{tabular}%
\end{equation}%
this means that $\chi ^{0},$ $\chi ^{++}$ is indeed a doublet under $%
\widetilde{SU}_{L}\left( 2\right) $.

\  \  \  \  \newline
The same analysis is valid for the decomposition (\ref{gi}); there the pair $%
\xi ^{0},$ $\xi ^{--}$ and its complex conjugates $\bar{\xi}^{0},$
$\xi ^{++} $ are given by

\begin{equation}
\begin{tabular}{lllllll}
$\xi ^{0}$ & $=$ & $\varphi _{i}f^{i}$ & , \  \  \  \  &
$\bar{\xi}^{0}$ & $=$
& $\bar{\varphi}^{i}\bar{f}_{i}$ \\
&  &  &  &  &  &  \\
$\xi ^{--}$ & $=$ & $\bar{f}_{i}\varphi ^{i}$ & , & $\xi ^{++}$ & $=$ & $%
-f^{i}\bar{\varphi}_{i}$ \\
&  &  &  &  &  &
\end{tabular}%
\end{equation}%
Here also the pair $\left( \xi ^{0},\xi ^{--}\right) $ and its
complex
conjugate $\left( \bar{\xi}^{0},\xi ^{++}\right) $ form two $\widetilde{SU}%
_{L}\left( 2\right) $ doublets; the exhibited hypercharges coincide
with the
usual electric ones; seen that $\xi ^{0},$ $\xi ^{--}$ and $\bar{\xi}^{0},$ $%
\xi ^{++}$ behave as singlets under $SU_{L}\left( 2\right) $.%
\begin{equation}
\left[ T_{_{L}}^{A},\xi ^{0}\right] =\left[ T_{_{L}}^{A},\xi
^{--}\right] =0
\end{equation}

\  \  \  \

\textbf{2}) \emph{Physical meaning of }$f^{i}$\emph{\ and }$\bar{f}_{i}$%
\newline
To deal with the Higgs fields in the harmonic field coordinate basis $%
\left
\{ f^{i},\text{ }\bar{f}_{i}\right \} $, we shall think about eqs(\ref%
{fi}-\ref{gi}) in ground state as follows
\begin{equation}
\begin{tabular}{lllllll}
$h^{i}$ & $=$ & $\varrho f^{i}$ & \  \  \  \ , \  \  \  \  \  &
$\varphi _{i}$ & $=$
& $\zeta \bar{f}_{i}-\gamma f_{i}$ \\
$\bar{h}_{i}$ & $=$ & $\varrho \bar{f}_{i}$ & \  \  \  \ , &
$\bar{\varphi}^{i}$
& $=$ & $\bar{\zeta}f^{i}+\bar{\gamma}\bar{f}^{i}$%
\end{tabular}
\label{gfh}
\end{equation}%
\begin{equation*}
\end{equation*}%
This choice corresponds to fixing \emph{3} of the real \emph{4}
degrees of freedom, captured by the complex moduli $\chi ^{0},$
$\chi ^{++}$, as given
below%
\begin{equation}
\begin{tabular}{lllll}
$\left( \chi ^{0}\right) ^{\ast }=\chi ^{0}$ & $=\varrho $ & , \  \  \  & $%
\chi ^{++}$ & $=0$%
\end{tabular}%
\end{equation}%
So the quantity
\begin{equation}
\begin{tabular}{lll}
$\bar{h}_{i}h^{i}$ & $=$ & $\left \vert \chi ^{0}\right \vert ^{2}+\chi ^{++}%
\bar{\chi}^{--}=\varrho ^{2}$ \\
&  &
\end{tabular}%
\end{equation}%
With this choice, the \emph{4} real degrees of freedom of the
complex Higgs field doublets $h^{i}$ and $\bar{h}_{i}$ are then
split as $1+3$ respectively described by the real $\varrho $ and
\begin{equation}
f^{i}=\frac{1}{\varrho }h^{i},\qquad \bar{f}_{i}=\frac{1}{\varrho }\bar{h}%
_{i}
\end{equation}%
Seen $\bar{h}_{i}h^{i}=\varrho ^{2}$ is the defining equation of a
real 3- sphere $\mathbb{S}_{h}^{3}$ , one learns that $\left \vert
\varrho \right
\vert $ is nothing but the radius of the Higgs sphere $\mathbb{S}%
_{h}^{3}$. With the above coordinate basis choice, we have

\begin{equation}
\begin{tabular}{lllllll}
$\zeta $ & $=$ & $\varphi _{i}f^{i}$ & \  \  \  \ , \  \  \  \  \  &
$\bar{\zeta}$
& $=$ & $+\bar{f}_{i}\bar{\varphi}^{i}$ \\
$\gamma $ & $=$ & $\varphi _{i}\bar{f}^{i}$ & \  \  \  \ , & $\bar{\gamma}$ & $%
= $ & $-f_{i}\bar{\varphi}^{i}$ \\
&  &  &  &  &  &
\end{tabular}%
\end{equation}%
In the decomposition (\ref{gfh}), the moduli $\zeta $ and $\gamma $
are
complex iso-singlet variables with hypercharges (electric charges) as%
\begin{equation}
\begin{tabular}{lllllll}
$\left[ \frac{Y}{2},\gamma \right] $ & $=$ & $-\gamma $ & \  \  \  \
\ , \  \  \
\  \  \  \  & $\left[ T^{A},\gamma \right] $ & $=$ & $0$ \\
$\left[ \frac{Y}{2},\zeta \right] $ & $=$ & $0$ & \  \  \  \  \ , &
$\left[
T^{A},\zeta \right] $ & $=$ & $0$%
\end{tabular}%
\end{equation}%
We also have%
\begin{equation}
\begin{tabular}{lllllll}
$\left[ \frac{Y}{2},f^{i}\right] $ & $=$ & $+\frac{1}{2}f^{i}$ & \
\  \  \  \ , & $\left[ T^{A},f^{i}\right] $ & $=$ &
$\frac{1}{2}\left( \tau ^{A}\right)
_{j}^{i}f^{j}$ \\
&  &  &  &  &  &
\end{tabular}%
\end{equation}

\subsubsection{The harmonics $(f^{i},\bar{f}_{i})$ and solution of complex
equation}

We first give some details on the harmonic field variables; then we
derive the solution of the complex holomorphic equations.

\  \  \  \  \

\textbf{1}) \emph{More on harmonics} $(f^{i},\bar{f}_{i})$\newline
Recall that the defining equation of the real unit 3-sphere
$\mathbb{S}^{3}$ in terms of the harmonic coordinate variables
$f^{i},$ $\bar{f}_{i}$ reads
like \textrm{\cite{Y1}-\cite{Y3}}%
\begin{equation}
\sum_{i=1,2}\bar{f}_{i}f^{i}=1
\end{equation}%
with indices raised and lowered by the antisymmetric $\varepsilon $-
tensors. Moreover, being bosonic isodoublets, they satisfy as well
the identities
\begin{equation}
\varepsilon _{ij}f^{i}f^{j}=0,\qquad \varepsilon ^{ij}\bar{f}_{i}\bar{f}%
_{j}=0
\end{equation}%
These harmonic variables $f^{i}$ and $\bar{f}_{i}$ form a pair of
complex doublet/anti-doublet that are interchanged by complex
conjugation; i.e
\begin{equation}
\bar{f}_{i}=\overline{\left( f^{i}\right) }
\end{equation}%
and are related to the usual real \emph{3} angles $\left( \theta
,\psi ,\phi
\right) $ of the sphere $\mathbb{S}^{3}$ as follows%
\begin{equation}
\begin{tabular}{lllll}
&  &  &  &  \\
$f^{i}$ & $=\left(
\begin{array}{c}
\cos \frac{\theta }{2}e^{\frac{i}{2}\left( \psi +\phi \right) } \\
\\
\sin \frac{\theta }{2}e^{\frac{i}{2}\left( \psi -\phi \right) }%
\end{array}%
\right) $ & , & $\bar{f}_{i}$ & $=\left(
\begin{array}{c}
\cos \frac{\theta }{2}e^{-\frac{i}{2}\left( \psi +\phi \right) }\text{ } \\
\\
\text{ }\sin \frac{\theta }{2}e^{-\frac{i}{2}\left( \psi -\phi \right) }%
\end{array}%
\right) $ \\
&  &  &  &
\end{tabular}
\label{fhar}
\end{equation}%
From this solution, we learn that the generators $T_{3,\pm }$ and
$Y$ of the
gauge symmetry are realized like%
\begin{equation*}
\begin{tabular}{lll}
$\frac{Y}{2}$ & $=$ & $\frac{\partial }{i\partial \psi }$ \\
$T_{3}$ & $=$ & $\frac{\partial }{i\partial \phi }$ \\
$T_{+}$ & $=$ & $e^{i\phi }\left( \frac{\partial }{\partial \theta
}+i\cot
\frac{\theta }{2}\frac{\partial }{\partial \phi }\right) $ \\
$T_{-}$ & $=$ & $e^{-i\phi }\left( \frac{\partial }{\partial \theta
}-i\cot
\frac{\theta }{2}\frac{\partial }{\partial \phi }\right) $ \\
&  &
\end{tabular}%
\end{equation*}%
Moreover, being complex and related by complex conjugation, it is
sometimes useful to exhibit the Cartan-Weyl charge carried by these
variables; this is
achieved by using the following correspondence%
\begin{equation}
\begin{tabular}{lll}
$f^{i}$ & $\  \  \rightarrow $ \  \  \  & $f^{+i}$ \\
$\bar{f}_{i}$ & $\  \  \rightarrow $ \  \  \  & $\bar{f}_{i}^{-}$%
\end{tabular}%
\end{equation}%
with the Cartan-Weyl charge operator precisely given by the
$U_{CW}\left( 1\right) $ generator
\begin{equation*}
D^{0}=f^{+i}\frac{\partial }{\partial f^{+i}}-\bar{f}^{-i}\frac{\partial }{%
\partial \bar{f}^{-i}}
\end{equation*}%
of the $\widetilde{su}\left( 2\right) $ algebra generated by the operators (%
\ref{hal}). Indeed, we have%
\begin{equation}
\begin{tabular}{lll}
$\left[ D^{0},f^{+i}\right] $ & $=$ & $+f^{+i}$ \\
$\left[ D^{0},\bar{f}^{-i}\right] $ & $=$ & $-\bar{f}^{-i}$%
\end{tabular}%
\end{equation}%
and%
\begin{equation}
\begin{tabular}{lll}
$\left[ D^{++},f^{+i}\right] $ & $=$ & $0$ \\
$\left[ D^{++},\bar{f}^{-i}\right] $ & $=$ & $f^{+i}$%
\end{tabular}%
\end{equation}%
showing that ($f^{+i},\bar{f}^{-i}$) form a doublet under $\widetilde{su}%
\left( 2\right) $.

\  \  \  \  \

\textbf{2}) \emph{the solution} \emph{of holomorphic constraint
eqs}\newline Substituting the decomposition (\ref{gfh}) back into
the complex holomorphic constraint $\varphi _{i}h^{i}=\mathrm{\nu }$
of eq(\ref{3S}), one ends with a relation between the iso-singlets
$\varrho $ and $\zeta $; but $\gamma $ free;
\begin{equation}
\left \{
\begin{tabular}{lll}
$\varrho \zeta $ & $=$ & $\mathrm{\nu }$ \\
$\gamma $ & $=$ & {\small arbitrary C-number}%
\end{tabular}%
\right.
\end{equation}%
from which we learn
\begin{equation}
\zeta =\frac{\mathrm{\nu }}{\varrho }
\end{equation}%
For the case $\varrho \neq 0$ and the complex parameter $\mathrm{\nu }%
\rightarrow 0$, the variable $\zeta $ $\rightarrow 0$; while for the case $%
\zeta \neq 0$, the limit $\mathrm{\nu }\rightarrow 0$ requires the
vanishing of the radius $\varrho $ of the 3-spheres
$\mathbb{S}_{h}^{3}$ and so this limit is singular and describes the
shrinking of $\mathbb{S}_{h}^{3}$ to the origin of
$\mathbb{C}_{h}^{2}$.

\  \  \  \newline In conclusion, by using the harmonic coordinates
$f^{i}$ and $\bar{f}_{i}$ of the unit 3-sphere $\mathbb{S}^{3}$, the
set $\mathfrak{C}_{\mathrm{\nu }}$ of solutions of the complex
holomorphic constraint equations
\begin{equation}
\left \{
\begin{tabular}{lll}
$\lambda S\varphi _{i}$ & $=$ & $0$ \\
$\lambda Sh^{i}$ & $=$ & $0$ \\
$\kappa S^{2}+\lambda \left( \varphi _{i}h^{i}-\mathrm{\nu }\right)
$ & $=$
& $0$%
\end{tabular}%
\right.
\end{equation}%
is given by the following complex 3D hypersurface in
$\mathbb{C}^{5}$,
\begin{equation}
\begin{tabular}{lll}
$S$ & $=$ & $0$ \\
$h^{i}$ & $=$ & $\varrho f^{i}$ \\
$\varphi _{i}$ & $=$ & $\frac{\mathrm{\nu }}{\varrho
}\bar{f}_{i}-\gamma
f_{i}$%
\end{tabular}
\label{cs}
\end{equation}%
\textrm{with }$\gamma $\textrm{\ an arbitrary complex iso-singlet carrying }$%
\left( -2\right) $\textrm{\ hypercharges units. The metric of this
threefold in terms of the coordinate}s $f^{i}$ and $\gamma $ reads
as
\begin{equation}
\begin{tabular}{lll}
$ds_{\mathfrak{C}_{\nu }}^{2}$ & $=$ & $\left( 1+\frac{\mathrm{\nu \bar{\nu}}%
}{\varrho ^{4}}\right) d\varrho ^{2}+\left( \varrho
^{2}+\frac{\mathrm{\nu
\bar{\nu}}}{\varrho ^{2}}+\bar{\gamma}\gamma \right) df^{i}d\bar{f}%
_{i}+d\gamma d\bar{\gamma}$ \\
&  & $-\left( \bar{\gamma}d\gamma -\gamma d\bar{\gamma}\right) f_{i}d\bar{f}%
^{i}$ \\
&  & $+\frac{\mathrm{\nu }}{\varrho }\left( d\bar{\gamma}+\bar{\gamma}\frac{%
d\varrho }{\varrho }\right) \bar{f}_{i}d\bar{f}^{i}-\frac{\mathrm{\bar{\nu}}%
}{\varrho }\left( d\gamma +\gamma \frac{d\varrho }{\varrho }\right)
f^{i}df_{i}$ \\
&  &
\end{tabular}%
\end{equation}%
see appendix B for details and for the expression of
$ds_{\mathfrak{C}_{\nu }}^{2}$ in terms of the angles $\theta ,\psi
$ and $\phi $.

\subsection{Solving the hermitian eqs(\protect \ref{2S})}

There are two kinds of hermitian constraint equations; $\left(
i\right) $ the iso-singlet constraint relation following from the
equation of motion of
the $U_{Y}\left( 1\right) $ auxiliary field $D^{\prime }$,%
\begin{equation}
\bar{h}_{i}h^{i}-\varphi _{i}\bar{\varphi}^{i}=0  \label{G1}
\end{equation}%
and $\left( ii\right) $ the hermitian iso-triplet constraint
relation coming
from the equation of motion of the $SU_{L}\left( 2\right) $ auxiliary field $%
D^{A}$ namely
\begin{equation}
\left( \tau ^{A}\right) _{j}^{i}\left( \bar{h}_{i}h^{j}-\varphi _{i}\bar{%
\varphi}^{j}\right) =0  \label{G2}
\end{equation}%
We shall first solve eq(\ref{G1}) by using the Higgs configurations (\ref{cs}%
), obtained by solving the complex holomorphic constraints. Then, we
put the obtained solutions solving (\ref{G1}) back into (\ref{G2})
to get the final Higgs configurations in the supersymmetric ground
state. Actually this solution constitutes one of the basic results
of this paper.

\subsubsection{Solving iso-singlet constraint eq(\protect \ref{G1})}

Using the expressions the Higgs doublet $\varphi _{i}=\frac{\mathrm{\nu }}{%
\varrho }\bar{f}_{i}-\gamma f_{i}$ and its complex conjugate $\bar{\varphi}%
^{i}=\bar{\gamma}\bar{f}^{i}+\frac{\mathrm{\bar{\nu}}}{\varrho
}f^{i}$ as well as the identities
$f_{i}\bar{f}^{i}=\bar{f}_{i}f^{i}=-1$, it not difficult to check
that the real number $\varphi _{i}\bar{\varphi}^{i}$
reads as%
\begin{equation}
\varphi _{i}\bar{\varphi}^{i}=\left( \gamma
\bar{\gamma}+\frac{\mathrm{\nu \bar{\nu}}}{\varrho ^{2}}\right)
\label{fif}
\end{equation}%
and its square root $\sqrt{\varphi _{i}\bar{\varphi}^{i}}$ may be
thought of
as the radius of the real 3-sphere $\mathbb{S}_{\varphi }^{3}$ embedded in $%
\mathbb{C}_{\varphi }^{2}\sim \mathbb{R}_{+}\times \mathbb{S}_{\varphi }^{3}$%
. Clearly this 3-sphere $\mathbb{S}_{\varphi }^{3}$ is fibred over
the 3- sphere $\mathbb{S}_{h}^{3}$ with radius $\varrho $ introduced
previously and
associated with the Higgs doublet $h^{i}$. This fibration, manifested in (%
\ref{fif}) by the dependence in $\varrho $, can be seen for instance
on the example where $\mathrm{\nu }$ is a fixed complex number and
$\varrho \rightarrow 0$; there the sphere $\mathbb{S}_{h}^{3}$
shrinks to a point while $\mathbb{S}_{\varphi }^{3}$ expands to
infinity.

\  \  \newline Substituting the above expression of $\varphi
_{i}\bar{\varphi}^{i}$ back into eq(\ref{G1}) and using
$\bar{h}_{i}h^{i}=\varrho ^{2}$, the iso-singlet constraint relation
$D^{\prime }=0$ becomes
\begin{equation}
\varrho ^{2}-\left( \gamma \bar{\gamma}+\frac{\mathrm{\nu \bar{\nu}}}{%
\varrho ^{2}}\right) =0  \label{vs}
\end{equation}%
it depends on the complex deformation parameter; and gives a
constraint relation between the real $\varrho $ and the complex
$\gamma $ reducing thus the dimension of the space of solutions down
to 4 real dimensions (2
complex). As this equation may be also put into the form%
\begin{equation}
\varrho ^{4}-\gamma \bar{\gamma}\varrho ^{2}-\mathrm{\nu
\bar{\nu}}=0
\end{equation}%
the acceptable solution leading to $\varrho ^{2}>0$ and expressing
$\varrho $ in terms of $\gamma $ reads like
\begin{equation}
\varrho ^{2}=\frac{1}{2}\left( \gamma \bar{\gamma}+\sqrt{\left( \gamma \bar{%
\gamma}\right) ^{2}+4\left( \mathrm{\nu \bar{\nu}}\right) }\right)
\label{rr}
\end{equation}%
From (\ref{vs}), we can also express $\gamma \bar{\gamma}$ in terms of $%
\varrho $; we have%
\begin{equation}
\gamma \bar{\gamma}=\varrho ^{2}-\frac{\mathrm{\nu
\bar{\nu}}}{\varrho ^{2}}
\end{equation}%
Combining the solutions eqs(\ref{cs}), of the complex constraint
eqs, with the solution (\ref{rr}) of the hermitian iso-singlet
constraint, we then
have $S=0$ and%
\begin{equation}
\begin{tabular}{lll}
&  &  \\
$h^{i}$ & $=$ & $\varrho f^{i}=\left( \frac{\gamma
\bar{\gamma}+\sqrt{\left(
\gamma \bar{\gamma}\right) ^{2}+4\left( \mathrm{\nu \bar{\nu}}\right) }}{2}%
\right) ^{\frac{1}{2}}f^{i}$ \\
&  &  \\
$\varphi _{i}$ & $=$ & $-\gamma f_{i}+\frac{\mathrm{\nu }}{\varrho }\bar{f}%
_{i}$ \\
&  &
\end{tabular}
\label{3}
\end{equation}%
The Higgs field configurations solving the complex $F_{I}=0$ and the
isosinglet $D^{\prime }=0$ are expressed in terms of $\gamma
\bar{\gamma}$ and on the harmonic field variables. Moreover seen
that $\gamma \bar{\gamma}$ is gauge invariant; in particular under
the U$_{Y}\left( 1\right) $ hypercharge transformations, one may use
this arbitrariness to fix a real degree of freedom of $\gamma $, say
the phase of $\left \vert \gamma \right
\vert e^{-2i\delta }$, and ends afterwards with a real 4D manifold $%
\mathfrak{M}_{4}$ (a complex surface ) given by
\begin{equation*}
\mathfrak{M}_{4}=\left \{ \left.
\begin{array}{c}
\\
\\
\end{array}%
\right. \frac{eqs(\ref{3})}{U_{Y}\left( 1\right) }\right \} \subset
\mathfrak{C}_{\nu }
\end{equation*}%
and contained in turns into $\mathbb{C}^{4}$ $\sim $
$\mathbb{R}^{8}$.

\subsubsection{Solving the iso-triplet eq(\protect \ref{G2})}

We begin by computing the quantity $\varphi _{i}\bar{\varphi}^{j}$
as it plays a central role in the isotriplet constraint; its trace
$\delta _{j}^{i}\varphi _{i}\bar{\varphi}^{j}$ is precisely given
(\ref{fif}).
\newline
Putting the expression (\ref{cs}) of the Higgs fields $\varphi _{i}$ and $%
\bar{\varphi}^{j}$, in terms of the harmonic variables $\bar{f}_{i}$ and $%
f^{j}$ back into the constraint eq(\ref{G2}), we obtain%
\begin{equation}
\begin{tabular}{lll}
&  &  \\
$\varphi _{i}\bar{\varphi}^{j}$ & $=$ & $\frac{\mathrm{\nu \bar{\nu}}}{%
\varrho ^{2}}\bar{f}_{i}f^{j}-\gamma \bar{\gamma}f_{i}\bar{f}^{j}+$ \\
&  &  \\
&  & $\bar{\gamma}\frac{\mathrm{\nu }}{\varrho }\bar{f}_{i}\bar{f}%
^{j}-\gamma \frac{\mathrm{\bar{\nu}}}{\varrho }f_{i}f^{j}$ \\
&  &
\end{tabular}
\label{ffd}
\end{equation}%
Then, using the symmetry property of the iso-triplet manifested by
the feature $\left( \varepsilon \tau ^{A}\right) _{ij}=\left(
\varepsilon \tau ^{A}\right) _{ji}$ with matrices $\left(
\varepsilon \tau ^{A}\right) _{ij}=\varepsilon _{ik}\left( \tau
^{A}\right) _{j}^{k}$, we can put the
constraint relation $D^{A}=0$ into the form%
\begin{equation}
\left( \varepsilon \tau ^{A}\right) ^{ij}\left[ \frac{-1}{2}\left(
\varrho ^{2}-\frac{\mathrm{\nu \bar{\nu}}}{\varrho ^{2}}+\gamma
\bar{\gamma}\right)
\bar{f}_{(i}f_{j)}-\frac{\mathrm{\bar{\nu}}\gamma }{2\varrho }f_{i}f_{j}+%
\frac{\mathrm{\nu }\bar{\gamma}}{\varrho
}\bar{f}_{i}\bar{f}_{j}\right] =0
\end{equation}%
whose solution requires the vanishing of the coefficients of each of
the iso-triplets $\bar{f}_{(i}f_{j)},$ $f_{i}f_{j}$ and
$\bar{f}_{i}\bar{f}_{j}$ namely
\begin{equation}
\begin{tabular}{lll}
$\varrho ^{2}-\frac{\mathrm{\nu \bar{\nu}}}{\varrho ^{2}}+\gamma
\bar{\gamma}
$ & $=$ & $0$ \\
$\  \  \  \  \  \  \  \  \  \frac{\mathrm{\bar{\nu}}\gamma
}{2\varrho }$ & $=$ & $0$
\\
$\  \  \  \  \  \  \  \  \  \frac{\mathrm{\nu }\bar{\gamma}}{\varrho }$ & $=$ & $0$%
\end{tabular}%
\end{equation}%
Seen that the complex parameter $\mathrm{\nu }$ has been taken as an
arbitrary complex number of the deformed conifold, the two last
relations are solved by
\begin{equation}
\gamma =0
\end{equation}%
reducing further the dimension of the space of the Higgs VEVs down
to a complex curve. \newline Putting $\gamma =0$ back into the first
relation, we end with the following
non trivial solution for $\varrho $,%
\begin{equation}
\varrho ^{4}=\mathrm{\nu \bar{\nu}}  \label{XS}
\end{equation}%
and then a non trivial solution for the chargeless Higgs component of the $%
\varphi _{i}$ Higgs doublet.

\  \  \  \  \  \  \newline To conclude, the set $\Sigma _{susy}$ of
solutions the constraint equations for a supersymmetric ground state
$\left \vert \Sigma _{susy}\right \rangle $ with the complex FI
coupling constant $\mathrm{\nu }$ arbitrary; but the
real $r=0$, reads as follows%
\begin{equation}
\begin{tabular}{lll}
$S$ & $=$ & $0$ \\
$h^{i}$ & $=$ & $\left( \mathrm{\nu \bar{\nu}}\right)
^{\frac{1}{4}}f^{i}$
\\
$\varphi _{i}$ & $=$ & $\frac{\mathrm{\nu }}{\left( \mathrm{\nu \bar{\nu}}%
\right) ^{1/4}}\bar{f}_{i}$ \\
&  &
\end{tabular}
\label{los}
\end{equation}%
These Higgs configurations parameterize a real 2-sphere $\mathbb{S}%
_{susy}^{2}$ with radius $\varrho _{0}=\left( \mathrm{\nu
\bar{\nu}}\right) ^{\frac{1}{4}}$. More precisely, the sphere
$\mathbb{S}_{susy}^{2}$ may be parameterized either by $h^{i}$ or by
$\varphi _{i}$. If using the complex
coordinates $h^{i}=\left( \mathrm{\nu \bar{\nu}}\right) ^{\frac{1}{4}}f^{i}$%
; the doublet $\varphi _{i}$ is nothing but
\begin{equation}
\begin{tabular}{lll}
$\varphi _{i}$ & $=$ & $\frac{\mathrm{\nu }}{\sqrt{\mathrm{\nu \bar{\nu}}}}%
\bar{h}_{i}$%
\end{tabular}%
\end{equation}%
and so one is left with the 3 real degrees of freedom captured by
the
harmonic field variables parameterizing a 3- sphere $\mathbb{S}%
_{susy}^{3}\sim SU\left( 2\right) $. However, seen that the Higgs
configuration are symmetric under the hypercharge symmetry%
\begin{equation}
f^{i}\text{ \  \ }\rightarrow \text{ \  \ }f^{i\prime }\equiv f^{i}
\end{equation}%
the ground state reduces then to%
\begin{equation*}
\mathbb{S}_{susy}^{2}\text{ \ }\sim \text{ \ }\frac{SU\left( 2\right) }{%
U_{Y}\left( 1\right) }
\end{equation*}%
Notice that the radius $\varrho _{0}$ of the real 2-sphere $\mathbb{S}%
_{susy}^{2}$ can be computed either by using the $h^{i}$ variable or the $%
\varphi _{i}$ one; it is given by%
\begin{equation}
\begin{tabular}{lll}
&  &  \\
$\bar{h}_{i}h^{i}$ & $=$ & $\sqrt{\mathrm{\nu \bar{\nu}}}$ \\
$\varphi _{i}\bar{\varphi}^{i}$ & $=$ & $\sqrt{\mathrm{\nu \bar{\nu}}}$ \\
&  &
\end{tabular}%
\end{equation}%
leading in turns to
\begin{equation}
\tan \beta _{susy}=1
\end{equation}%
Regarding the isotriplet constraint eqs; they are as well
identically
satisfied by the solution (\ref{los}) due to the identity%
\begin{equation}
\begin{tabular}{lll}
&  &  \\
$\bar{h}_{i}h^{j}$ & $=$ & $\sqrt{\mathrm{\nu \bar{\nu}}}$
$\bar{f}_{i}f^{j}$
\\
$\varphi _{i}\bar{\varphi}^{j}$ & $=$ & $\sqrt{\mathrm{\nu \bar{\nu}}}$ $%
\bar{f}_{i}f^{j}$ \\
&  &
\end{tabular}%
\end{equation}%
For such VEVs of the complex Higgs fields; that is for Higgs fields
$h^{i}$ and $\varphi _{i}$ living on $\mathbb{S}_{susy}^{2}$,
supersymmetry is preserved; otherwise it is broken.

\section{Broken supersymmetry and ground state energy}

In previous section we have shown that for $r=0$ supersymmetry of \emph{%
n-MSSM} with anti-doublet $\Phi _{i}$ at place of $H_{d}$ is
preserved for arbitrary non zero complex parameter $\nu $. In this
section, we study the spontaneous breaking of supersymmetry by
switching on the hermitian FI term
\begin{equation*}
\mathrm{r}D^{\prime }=\mathrm{r}\int d^{4}\theta \text{ }\boldsymbol{V}_{%
{\small u}_{_{{\small 1}}}}^{\prime }
\end{equation*}%
in the lagrangian density of the conifold model (\ref{lt}).

\subsection{Switching on hermitian FI term r}

By switching on the FI coupling constant $r\neq 0$; the complex
holomorphic constraint eqs following from F-terms remain the same as
in the case $r=0$;
namely%
\begin{equation}
\left \{
\begin{tabular}{lll}
$\lambda S\varphi _{i}$ & $=$ & $0$ \\
$\lambda Sh^{i}$ & $=$ & $0$ \\
$\kappa S^{2}+\lambda \left( \varphi _{i}h^{i}-\mathrm{\nu }\right)
$ & $=$
& $0$%
\end{tabular}%
\right.  \label{W1}
\end{equation}%
while the hermitian constraint relations given by the D-terms get
modified.
More precisely, it is the isosinglet constraint which changes like%
\begin{equation}
\bar{h}_{i}h^{i}-\varphi _{i}\bar{\varphi}^{i}=\mathrm{r,\qquad
r}\neq 0 \label{W2}
\end{equation}%
but the isotriplet remains as before%
\begin{equation}
\left( \tau ^{A}\right) _{j}^{i}\left( \bar{h}_{i}h^{j}-\varphi _{i}\bar{%
\varphi}^{j}\right) =0  \label{W3}
\end{equation}%
Because of this Kahler deformation, the previous solution with $r=0$
is no longer valid in present case.

\  \  \  \  \newline To deal with these deformed constraint
relations, we proceed as follows:

\begin{itemize}
\item First, we solve the complex (\ref{W1}) in terms of the harmonic
variables $f_{i}$ and $\bar{f}_{i}$; these calculations give the
solutions of the complex holomorphic constraint relations $F_{I}=0$;
they are same as in previous section; and their explicit expressions
are as follows
\begin{equation}
\begin{tabular}{lll}
$S$ & $=$ & $0$ \\
$h^{i}$ & $=$ & $\varrho f^{i}$ \\
$\varphi _{i}$ & $=$ & $\frac{\mathrm{\nu }}{\varrho
}\bar{f}_{i}-\gamma
f_{i}$%
\end{tabular}
\label{N1}
\end{equation}%
with $\varrho =\sqrt{\bar{h}_{i}h^{i}}$. \ As in case $r=0$, these
Higgs configurations depend on the deformation parameter
$\mathrm{\nu }$ and on the arbitrary complex $\gamma $; they
parameterize a complex \emph{3D} manifold namely a deformed
conifold.

\item Then, we use the above expressions (\ref{N1}) of the Higgs $h^{i}$ and
$\varphi _{i}$ to solve the iso-singlet constraint relation
(\ref{W2}).

\item After that we check whether the obtained solutions for eq(\ref{W2})
are consistent or not with the solutions of the iso-triplet constraint eq(%
\ref{W3}).
\end{itemize}

\subsubsection{Solving iso-singlet constraint (\protect \ref{W2})}

Using the expressions (\ref{N1}) of the Higgs fields $h_{i}$ and
$\varphi
_{i}$, we have $\bar{h}_{i}h^{i}=\varrho ^{2}$ and $\varphi _{i}\bar{\varphi}%
^{i}=\gamma \bar{\gamma}+\frac{\mathrm{\nu \bar{\nu}}}{\varrho
^{2}}$; then substituting back into the iso-singlet constraint
eq(\ref{W2}) we obtain the following equation
\begin{equation}
\begin{tabular}{lll}
$\varrho ^{2}-\left( \gamma \bar{\gamma}+\frac{\mathrm{\nu \bar{\nu}}}{%
\varrho ^{2}}\right) $ & $=$ & $\mathrm{r}$%
\end{tabular}
\label{M1}
\end{equation}%
giving a relation between the real $\varrho $, the complex $\gamma $
and the FI coupling constants $\nu $ and r. For non zero $\varrho
^{2}$, the above
relation can be also put into the equivalent form%
\begin{equation}
\varrho ^{4}-\left( \mathrm{r}+\gamma \bar{\gamma}\right) \varrho ^{2}-%
\mathrm{\nu \bar{\nu}}=0
\end{equation}%
whose solution with positive $\varrho ^{2}$ is given by%
\begin{equation}
\varrho ^{2}=\frac{1}{2}\left[ \left( \mathrm{r}+\gamma \bar{\gamma}\right) +%
\sqrt{\left( \mathrm{r}+\gamma \bar{\gamma}\right) ^{2}+4\mathrm{\nu \bar{\nu%
}}}\right]  \label{or}
\end{equation}%
We can also express $\gamma $ in terms of $\varrho $ and the FI
parameters;
we have%
\begin{equation}
\begin{tabular}{lll}
$\gamma \bar{\gamma}$ & $=$ & $\left( \varrho ^{2}-\frac{\mathrm{\nu \bar{\nu%
}}}{\varrho ^{2}}\right) -\mathrm{r}$%
\end{tabular}%
\end{equation}%
Putting (\ref{or}) back into the expression (\ref{N1}), we obtain
the explicit solution of the iso-singlet $D^{\prime }=0$ in terms of
the Kahler parameter $\mathrm{r,}$ the complex deformation parameter
$\nu $, the
variable $\gamma $ and the harmonic field variables; it is given by $S=0$ and%
\begin{equation}
\begin{tabular}{lll}
&  &  \\
$h^{i}$ & $=$ & $\frac{1}{\sqrt{2}}\left[ \left( \mathrm{r}+\gamma \bar{%
\gamma}\right) +\sqrt{\left( \mathrm{r}+\gamma \bar{\gamma}\right) ^{2}+4%
\mathrm{\nu \bar{\nu}}}\right] ^{\frac{1}{2}}f^{i}$ \\
&  &  \\
$\varphi _{i}$ & $=$ & $-\gamma f_{i}+\nu \sqrt{2}\left[ \left( \mathrm{r}%
+\gamma \bar{\gamma}\right) +\sqrt{\left( \mathrm{r}+\gamma \bar{\gamma}%
\right) ^{2}+4\mathrm{\nu \bar{\nu}}}\right] ^{-\frac{1}{2}}\bar{f}_{i}$ \\
&  &
\end{tabular}
\label{N2}
\end{equation}%
In the limit $r=0$, one recovers the previous result, see eqs(\ref{3}).%
\newline
In what follows, we show that for $r\neq 0,$ the above Higgs
configurations are not solutions of the iso-triplet constraint
relations $D^{A}=0$.

\subsubsection{ Solving the isotriplet eq(\protect \ref{W3})}

Using the expression%
\begin{equation}
\begin{tabular}{lll}
$\left( \tau ^{A}\right) _{j}^{i}\varphi _{i}\bar{\varphi}^{j}$ & $=$ & $%
\left( \tau ^{A}\right) _{j}^{i}\left( \frac{\mathrm{\nu
\bar{\nu}}}{\varrho
^{2}}\bar{f}_{i}f^{j}-\gamma \bar{\gamma}f_{i}\bar{f}^{j}\right) $ \\
&  &  \\
&  & $+\left( \tau ^{A}\right) _{j}^{i}\left( \bar{\gamma}\frac{\mathrm{\nu }%
}{\varrho }\bar{f}_{i}\bar{f}^{j}-\gamma \frac{\mathrm{\bar{\nu}}}{\varrho }%
f_{i}f^{j}\right) $%
\end{tabular}%
\end{equation}%
the hermitian isotriplet constraint (\ref{W3}) gets mapped to
\begin{equation}
\begin{tabular}{lll}
&  &  \\
$\frac{1}{2}\left( \varepsilon \tau ^{A}\right) _{ij}\left[ \text{ \
}\left(
\varrho ^{2}-\gamma \bar{\gamma}-\frac{\mathrm{\nu \bar{\nu}}}{\varrho ^{2}}%
\right) \bar{f}^{(i}f^{j)}+\gamma \frac{\mathrm{\bar{\nu}}}{\varrho }%
f^{i}f^{j}-\bar{\gamma}\frac{\mathrm{\nu }}{\varrho }\bar{f}^{i}\bar{f}^{j}%
\text{ \  \ }\right] $ & $=$ & $0$ \\
&  &
\end{tabular}
\label{M2}
\end{equation}%
Substituting the relation $\varrho ^{2}-\frac{\mathrm{\nu \bar{\nu}}}{%
\varrho ^{2}}-\gamma \bar{\gamma}=\mathrm{r}$, required by the
iso-singlet constraint $D^{\prime }=0$, the isotriplet eq(\ref{M2})
can be brought to
the following form%
\begin{equation}
\left( \varepsilon \tau ^{A}\right) _{ij}\left[ \frac{\mathrm{r}}{2}\bar{f}%
^{(i}f^{j)}+\gamma \frac{\mathrm{\bar{\nu}}}{2\varrho }f^{(i}f^{j)}-\bar{%
\gamma}\frac{\mathrm{\nu }}{2\varrho
}\bar{f}^{(i}\bar{f}^{j)}\right] =0
\end{equation}%
whose solution requires the vanishing of each of the coefficients $\bar{f}%
^{(i}f^{j)}$, $f^{(i}f^{j)}$ and $\bar{f}^{(i}\bar{f}^{j)}$ namely%
\begin{equation}
\begin{tabular}{lll}
$\gamma \frac{\mathrm{\bar{\nu}}}{\varrho }$ & $=$ & $0$ \\
$\bar{\gamma}\frac{\mathrm{\nu }}{\varrho }$ & $=$ & $0$ \\
$\  \mathrm{r}$ & $=$ & $0$%
\end{tabular}%
\end{equation}%
However, the Kahler parameter should be non zero, $r\neq 0$; and so
there is no common solution for the constraint eqs $D^{\prime }=0$
and $D^{A}=0.$

\  \  \  \  \newline To conclude, for FI coupling $\mathrm{r}\neq
0$, the auxiliary field equations of motion have then no common
solution. So supersymmetry of the ground state with non zero Higgs
VEVs is spontaneously broken by the switching on the FI coupling
constant $\mathrm{r}$.

\subsection{ground state energy density}

Seen that supersymmetry of the model is spontaneously broken for
$r\neq 0$, then the ground state should have a non zero positive
energy. In this sub-section, we compute the amount of this energy in
terms of the coupling constants of the model. \newline To that
purpose, recall that in supersymmetric gauge theories the component
scalar field potential energy density $\mathcal{V}_{scalar}$ is the
sum of
two kinds of positive contributions: $\left( i\right) $ a term $\mathcal{V}%
_{ch}=\mathcal{V}\left( F,\bar{F}\right) ;$ and $\left( ii\right) $ a term $%
\mathcal{V}_{re}=\mathcal{V}\left( D\right) $;%
\begin{equation}
\mathcal{V}_{scalar}=\mathcal{V}_{ch}+\mathcal{V}_{re}
\end{equation}%
the first contribution comes from the superpotential of the chiral
sector of the superspace lagrangiant density $\boldsymbol{L}$
(\ref{tl}-\ref{lt}); it reads in the case we are considering here
like
\begin{equation}
\begin{tabular}{lll}
$\mathcal{V}_{ch}$ & $=$ & $\bar{F}_{i}F^{i}+\bar{G}^{i}G_{i}+\bar{F}%
_{S}F_{S}$ \\
&  &  \\
& $=$ & $\left \vert \bar{F}_{1}\right \vert ^{2}+\left \vert \bar{F}%
_{2}\right \vert ^{2}+\left \vert G_{1}\right \vert ^{2}+\left \vert
G_{2}\right \vert ^{2}+\left \vert \bar{F}_{S}\right \vert ^{2}$ \\
&  &
\end{tabular}%
\end{equation}%
The second contribution comes from the hermitian Kahler sector the
superspace lagrangian density $\boldsymbol{L}$; it is given by%
\begin{equation}
\mathcal{V}_{re}=\frac{1}{2}D^{\prime 2}+\frac{1}{2}D_{A}D^{A}
\end{equation}%
The objective is to determine, within the approximation of section
2, the expression of this potential energy density in terms of the
Higgs fields $S,$ $h^{i},$ $\varphi _{i}$; and then compute its
minimum.

\subsubsection{Non supersymmetric ground state}

We start from the superspace lagrangian density (\ref{tl}-\ref{lt})
with non zero FI coupling constants $\left( \mathrm{\nu
},\mathrm{r}\right) \neq \left( 0,0\right) $. Seen that Higgs field
configurations in the ground state are functions of the coupling
constants of the model, the energy density of the ground state is
also of function of these couplings,
\begin{equation}
\mathcal{E}_{\min }=\mathcal{E}\left( g,g^{\prime },r,\left \vert
\nu \right \vert \right)
\end{equation}%
with the property%
\begin{equation}
\lim_{r\rightarrow 0}\mathcal{E}_{\min }=0
\end{equation}%
because for $r=0$ both $\mathcal{V}_{ch}=0$ and $\mathcal{V}_{re}=0$
and so there is no vacuum energy.\newline For $r\neq 0$, the Higgs
VEV configurations solving the complex holomorphic constraint
relations (the equations of motion of the auxiliary fields F) have
indeed a non zero energy density seen that
\begin{equation}
\begin{tabular}{lllll}
$\mathcal{V}_{ch}=0$ & , & but &  & $\mathcal{V}_{re}>0$ \\
&  &  &  &
\end{tabular}
\label{R1}
\end{equation}%
where $\mathcal{V}_{re}$ is thought of as a perturbation with respect to $%
\mathcal{V}_{ch}$ within the approximation described in section 2.

\textbf{1) }\emph{computational method}\newline After recalling
useful tools on non supersymmetric ground state; we compute the
expression of its energy density $\mathcal{V}_{scalar}$ in terms of
the
Higgs fields. Then, we determine the expression of the Higgs VEVs%
\begin{equation}
\begin{tabular}{lll}
$\left \langle S\right \rangle _{\min }$ & $=$ & $S_{\min }\left( r,\nu ,%
\bar{\nu}\right) $ \\
$\left \langle h^{i}\right \rangle _{\min }$ & $=$ & $h_{\min
}^{i}\left(
r,\nu ,\bar{\nu}\right) $ \\
$\left \langle \varphi _{i}\right \rangle _{\min }$ & $=$ & $\varphi
_{i\min
}\left( r,\nu ,\bar{\nu}\right) $ \\
&  &
\end{tabular}%
\end{equation}%
that minimize the ground state energy density. We end this study by
computing the deviation $\tan \beta $ of the supersymmetric $\tan
\beta
_{0}=1$ by help of the formula%
\begin{equation}
\tan \beta =\sqrt{\frac{\left \langle \bar{h}_{i}h^{i}\right \rangle _{\min }%
}{\left \langle \varphi _{i}\bar{\varphi}^{i}\right \rangle _{\min
}}}
\end{equation}%
and determining the exact expression of $\mathcal{E}_{\min }$.

\  \

\textbf{2) }\emph{Energy density of non supersymmetric ground
state}\newline First recall that the complex holomorphic constraint
relations given by the equations of motion of the auxiliary fields
\begin{equation}
F^{i}=0,\qquad G^{i}=0,\qquad F_{S}=0
\end{equation}%
are independent on the real FI coupling parameter r. Their solutions
are also independent on r; they read as
\begin{eqnarray}
S &=&0  \notag \\
h^{i} &=&\varrho f^{i}\qquad ,\qquad \varphi _{i}=\frac{\mathrm{\nu }}{%
\varrho }\bar{f}_{i}-\gamma f_{i}  \notag \\
\bar{h}_{i} &=&\varrho \bar{f}_{i}\qquad ,\qquad \bar{\varphi}^{i}=\frac{%
\mathrm{\bar{\nu}}}{\varrho }f^{i}+\bar{\gamma}\bar{f}^{i}  \label{fs} \\
&&  \notag
\end{eqnarray}%
and lead to no energy density since
\begin{equation*}
\mathcal{V}_{ch}=0
\end{equation*}%
Using this property of $\mathcal{V}_{ch}$, the scalar potential
energy density $\mathcal{V}$ of the model reduces to the part
$\mathcal{V}_{re}$ coming from Kahler sector namely
\begin{equation}
\mathcal{V}=\frac{1}{2}D^{\prime 2}+\frac{1}{2}D_{A}D^{A}
\label{sca}
\end{equation}%
with the auxiliary D- fields as%
\begin{eqnarray}
D^{\prime } &=&\frac{g^{\prime }}{2}\left( \bar{h}_{i}h^{i}-\bar{\varphi}%
_{i}\varphi ^{i}-\mathrm{r}\right)  \notag \\
D^{A} &=&\frac{g}{2}\left( \tau ^{A}\right) _{j}^{i}D_{i}^{j}  \label{cas} \\
D_{i}^{j} &=&\bar{h}_{i}h^{j}-\varphi _{i}\bar{\varphi}^{j}  \notag \\
&&  \notag
\end{eqnarray}%
where $h^{i}$ and $\varphi _{i}$ are as in eqs(\ref{fs}). \newline
Substituting these relations back into (\ref{sca}), we find, after
straightforward calculations, that the explicit field expression of
the gauge invariant scalar potential energy density of the non
supersymmetric ground state reads in terms of the FI coupling
constants and the gauge invariant variables $\varrho ^{2}$ and
$\gamma \bar{\gamma}$ as follows

\begin{eqnarray}
\mathcal{V} &=&\frac{g^{\prime 2}}{8}\mathrm{r}^{2}-\frac{g^{2}}{2}\mathrm{%
\bar{\nu}\nu }-\frac{2g^{\prime 2}}{8}\mathrm{r}\left( \varrho ^{2}-\frac{%
\mathrm{\nu \bar{\nu}}}{\varrho ^{2}}-\gamma \bar{\gamma}\right) +  \notag \\
&&  \label{V0} \\
&&\frac{g^{\prime 2}}{8}\left( \varrho ^{2}-\frac{\mathrm{\nu \bar{\nu}}}{%
\varrho ^{2}}-\gamma \bar{\gamma}\right) ^{2}+\frac{g^{2}}{8}\left[
\varrho ^{2}+\frac{\mathrm{\nu \bar{\nu}}}{\varrho ^{2}}+\gamma
\bar{\gamma}\right] ^{2}  \notag
\end{eqnarray}%
This scalar energy density is a function of the hermitian gauge invariants $%
\varrho ^{2}$ and $\gamma \bar{\gamma}$
\begin{equation}
\mathcal{V}=\mathcal{V}\left( \varrho ^{2};\gamma
\bar{\gamma}\right)
\end{equation}%
and depends on the \emph{5} coupling moduli namely the \emph{2} real
gauge coupling constants $g$ and $g\prime $ and the \emph{3} the FI
couplings $r$, $\mathrm{\nu },\mathrm{\bar{\nu}}$.

\  \  \

\emph{Deriving eq(\ref{V0})}\newline Here we give the main steps of
the explicit derivation of the component
field expression (\ref{V0}) of the scalar potential energy density $\mathcal{%
V}$. First, we compute the contribution coming from the term $\frac{1}{2}%
D^{\prime 2}$; then we calculate the contribution of
$\frac{1}{2}D_{A}D^{A}$.

\begin{itemize}
\item \emph{computing} $\frac{1}{2}D^{\prime 2}$\newline
The relation between the auxiliary field $D^{\prime }$ and the Higgs fields $%
h^{i}$ and $\varphi _{i}$ is given by (\ref{cas}). Using $h^{i}\bar{h}%
_{i}=\varrho ^{2}$, one of the two basic variables of (\ref{V0})
scaling as mass$^{2}$ and manifestly gauge invariant, the
contribution to the energy of
the non supersymmetric ground state, coming from $D^{\prime },$ reads then as%
\begin{equation}
\frac{1}{2}D^{\prime 2}=\frac{g^{\prime 2}}{8}\left( \varrho
^{2}-\varphi _{i}\bar{\varphi}^{i}-\mathrm{r}\right) ^{2}
\end{equation}%
or equivalently by expanding%
\begin{equation}
\begin{tabular}{lll}
$\frac{1}{2}D^{\prime 2}$ & $=$ & $\frac{g^{\prime 2}}{8}\mathrm{r}^{2}-%
\frac{2g^{\prime 2}}{8}\mathrm{r}\left( \varrho ^{2}-\varphi _{i}\bar{\varphi%
}^{i}\right) $ \\
&  & $+\frac{g^{\prime 2}}{8}\left( \varrho ^{2}-\varphi _{i}\bar{\varphi}%
^{i}\right) ^{2}$%
\end{tabular}
\label{ag}
\end{equation}%
with the $\varphi _{i}\bar{\varphi}^{i}$ gauge invariant as
\begin{equation}
\varphi _{i}\bar{\varphi}^{i}=\frac{\mathrm{\nu \bar{\nu}}}{\varrho ^{2}}%
+\gamma \bar{\gamma}  \label{ga}
\end{equation}%
For later use notice that for $\gamma =0$, eq(\ref{ag}) reduces to%
\begin{equation}
\frac{1}{2}D^{\prime 2}=\frac{g^{\prime 2}}{8}\left( \varrho ^{2}-\frac{%
\mathrm{\nu \bar{\nu}}}{\varrho ^{2}}-\mathrm{r}\right) ^{2}
\end{equation}%
and then%
\begin{equation}
\frac{1}{2}\frac{\partial \left( D^{\prime 2}\right) }{\partial \varrho ^{2}}%
=\frac{g^{\prime 2}}{4}\left( \varrho ^{2}-\frac{\mathrm{\nu \bar{\nu}}}{%
\varrho ^{2}}-\mathrm{r}\right) \left( 1+\frac{\mathrm{\nu \bar{\nu}}}{%
\varrho ^{4}}\right)
\end{equation}

\item \emph{computing} $\frac{1}{2}D_{A}D^{A}$\newline
Expanding the isotriplet auxiliary field $D^{A}$ in terms of the 3
Pauli matrices $\tau ^{A}$ like $D^{A}=\frac{g}{2}\left( \tau
^{A}\right)
_{j}^{i}D_{i}^{j}$ and using the identity%
\begin{equation}
\left( \tau _{A}\right) _{k}^{l}\left( \tau ^{A}\right)
_{j}^{i}=2\delta _{k}^{i}\delta _{j}^{l}-\delta _{j}^{i}\delta
_{k}^{l}
\end{equation}%
we can put the term $D_{A}D^{A}$ into the form%
\begin{equation}
D_{A}D^{A}=\frac{g^{2}}{4}\left[ 2\left( D_{i}^{j}D_{j}^{i}\right)
-\left( D_{i}^{i}\right) ^{2}\right]  \label{dd}
\end{equation}%
with $\left( D_{i}^{j}D_{j}^{i}\right) $ and $\left(
D_{i}^{i}\right) ^{2}$
as follows:%
\begin{eqnarray}
2D_{i}^{j}D_{j}^{i} &=&2\varrho ^{4}+2\left(
\bar{\varphi}_{i}\varphi
^{i}\right) ^{2}-4\left( \varphi _{i}h^{i}\right) \left( \bar{h}_{j}\bar{%
\varphi}^{j}\right)  \notag \\
\left( D_{i}^{i}\right) ^{2} &=&\varrho ^{4}+\left(
\bar{\varphi}_{i}\varphi
^{i}\right) ^{2}-2\varrho ^{2}\left( \bar{\varphi}_{i}\varphi ^{i}\right) \\
&&  \notag
\end{eqnarray}%
where $\bar{\varphi}_{i}\varphi ^{i}$ is as in eq(\ref{ga}), and where $%
\varphi _{i}h^{i}$ and $\bar{h}_{j}\bar{\varphi}^{j}$ are given by%
\begin{eqnarray}
\varphi _{i}h^{i} &=&\mathrm{\nu }  \notag \\
\bar{h}_{j}\bar{\varphi}^{j} &=&\mathrm{\bar{\nu}} \\
&&  \notag
\end{eqnarray}%
Substituting these relations back into (\ref{dd}), we obtain the
contribution to the scalar energy density coming from the isotriplet
term;
it reads as%
\begin{eqnarray}
&&  \notag \\
\frac{1}{2}D_{A}D^{A} &=&\frac{g^{2}}{8}\left( \left[ \varrho
^{2}+\left(
\bar{\varphi}_{i}\varphi ^{i}\right) \right] ^{2}-4\mathrm{\bar{\nu}\nu }%
\right)  \notag \\
&& \\
&=&\frac{g^{2}}{8}\left[ \varrho ^{2}+\left(
\bar{\varphi}_{i}\varphi ^{i}\right) +2\left \vert \mathrm{\nu
}\right \vert \right] \left[ \varrho
^{2}+\left( \bar{\varphi}_{i}\varphi ^{i}\right) -2\left \vert \mathrm{\nu }%
\right \vert \right]  \notag \\
&&  \notag
\end{eqnarray}%
Explicitly, we have%
\begin{equation}
\frac{1}{2}D_{A}D^{A}=\frac{g^{2}}{8}\left[ \varrho
^{2}+\frac{\mathrm{\nu
\bar{\nu}}}{\varrho ^{2}}+\gamma \bar{\gamma}\right] ^{2}-\frac{g^{2}}{2}%
\mathrm{\bar{\nu}\nu }  \label{af}
\end{equation}%
Notice that for $\gamma =0$, we have%
\begin{equation}
\frac{1}{2}D_{A}D^{A}=\frac{g^{2}}{8}\left[ \varrho
^{2}+\frac{\mathrm{\nu \bar{\nu}}}{\varrho ^{2}}\right]
^{2}-\frac{g^{2}}{2}\mathrm{\bar{\nu}\nu }
\end{equation}%
and%
\begin{equation}
\frac{1}{2}\frac{\partial \left( D_{A}D^{A}\right) }{\partial \varrho ^{2}}=%
\frac{g^{2}}{4}\left( \varrho ^{2}+\frac{\mathrm{\nu \bar{\nu}}}{\varrho ^{2}%
}\right) \left( 1-\frac{\mathrm{\nu \bar{\nu}}}{\varrho ^{4}}\right)
\end{equation}
\end{itemize}

\  \  \  \newline Combining eqs(\ref{ag}-\ref{af}), we obtain
\begin{equation}
\mathcal{V}=\frac{g^{\prime 2}}{8}\left( \varrho ^{2}-\frac{\mathrm{\nu \bar{%
\nu}}}{\varrho ^{2}}-\gamma \bar{\gamma}-\mathrm{r}\right) ^{2}+\frac{g^{2}}{%
8}\left( \varrho ^{2}+\frac{\mathrm{\nu \bar{\nu}}}{\varrho
^{2}}+\gamma \bar{\gamma}\right)
^{2}-\frac{g^{2}}{2}\mathrm{\bar{\nu}\nu }
\end{equation}%
which can also rewritten as in (\ref{V0}).

\subsubsection{Minimizing the scalar potential}

The scalar potential $\mathcal{V}$ is function of the gauge
invariant quantities $\gamma \bar{\gamma}$ and $\varrho ^{2}$; its
extremum is then
obtained by solving the conditions%
\begin{eqnarray}
\frac{\partial \mathcal{V}}{\partial \bar{\gamma}} &=&\gamma
\frac{\partial
\mathcal{V}}{\partial \left( \gamma \bar{\gamma}\right) }=0  \notag \\
&&  \label{co} \\
\frac{\partial \mathcal{V}}{\partial \bar{h}_{i}}
&=&h^{i}\frac{\partial \mathcal{V}}{\partial \varrho ^{2}}=0  \notag
\end{eqnarray}%
The set $\Upsilon $ of solutions of these relations is then given by
the
intersection of two sets $\Omega _{1}$ and $\Omega _{2}$ like%
\begin{equation}
\Upsilon =\Omega _{1}\cap \Omega _{2}
\end{equation}%
with
\begin{eqnarray}
\Omega _{1} &=&\left \{ \gamma =0,\text{ }\frac{\partial \mathcal{V}}{%
\partial \left( \gamma \bar{\gamma}\right) }\text{ arbitrary}\right \} \cup
\left \{ \frac{\partial \mathcal{V}}{\partial \left( \gamma \bar{\gamma}%
\right) }=0\right \}  \notag \\
&& \\
\Omega _{2} &=&\left \{ h_{i}=0,\text{ }\frac{\partial
\mathcal{V}}{\partial \varrho ^{2}}\text{ arbitrary}\right \} \cup
\left \{ \frac{\partial \mathcal{V}}{\partial \varrho ^{2}}=0\right
\}  \notag
\end{eqnarray}%
\begin{equation*}
\end{equation*}%
In what follows, we determine the explicit expression of the sets
$\Omega _{1}$ and $\Omega _{2}$; but to fix the ideas, we will show
that the common
solution of (\ref{co}) corresponds to%
\begin{eqnarray}
\gamma &=&0  \notag \\
\frac{\partial \mathcal{V}}{\partial \varrho ^{2}} &=&0
\end{eqnarray}%
and reads in terms of the coupling constants and the FI parameters as follows%
\begin{eqnarray}
\gamma &=&0  \notag \\
&& \\
\varrho ^{2} &=&\frac{1}{2\left( g^{2}+g^{\prime 2}\right) }\left[
g^{\prime 2}\mathrm{r}+\sqrt{g^{\prime 4}\mathrm{r}^{2}+4\left(
g^{2}+g^{\prime 2}\right) ^{2}\mathrm{\nu \bar{\nu}}}\right]  \notag
\end{eqnarray}

\textbf{1)} \emph{computing} $\frac{\partial \mathcal{V}}{\partial
\left( \gamma \bar{\gamma}\right) }=0$\newline
Using the expression (\ref{V0}) of the scalar potential, we have%
\begin{eqnarray}
\frac{\partial \mathcal{V}^{0}}{\partial \left( \gamma
\bar{\gamma}\right) }
&=&-\frac{g^{\prime 2}}{4}\left( \varrho ^{2}-\frac{\mathrm{\nu \bar{\nu}}}{%
\varrho ^{2}}-\gamma \bar{\gamma}-\mathrm{r}\right)
+\frac{g^{2}}{4}\left(
\varrho ^{2}+\frac{\mathrm{\nu \bar{\nu}}}{\varrho ^{2}}+\gamma \bar{\gamma}%
\right) \\
&&  \notag
\end{eqnarray}%
which reads also like%
\begin{eqnarray}
&&  \notag \\
\frac{\partial \mathcal{V}}{\partial \left( \gamma \bar{\gamma}\right) } &=&%
\frac{1}{\varrho ^{2}}\left[ \frac{g^{2}-g^{\prime 2}}{4}\varrho
^{2}+\left(
\frac{2g^{\prime 2}}{8}\mathrm{r}+\frac{g^{2}+g^{\prime 2}}{4}\gamma \bar{%
\gamma}\right) \varrho ^{2}+\mathrm{\nu \bar{\nu}}\frac{g^{2}+g^{\prime 2}}{4%
}\right] \\
&&  \notag
\end{eqnarray}%
whose solution, for $g^{2}\neq g^{\prime 2}$, is obtained by solving
\begin{equation}
\varrho ^{4}-\frac{g^{\prime 2}\mathrm{r}+\left( g^{2}+g^{\prime
2}\right)
\gamma \bar{\gamma}}{g^{\prime 2}-g^{2}}\varrho ^{2}-\mathrm{\nu \bar{\nu}}%
\frac{g^{2}+g^{\prime 2}}{g^{\prime 2}-g^{2}}=0
\end{equation}%
The solution $\left \langle \varrho ^{2}\right \rangle _{_{^{\gamma
}}}=\left \langle \varrho ^{2}\right \rangle _{\frac{\partial \mathcal{V}}{%
\partial \left( \gamma \bar{\gamma}\right) }=0}$ is given by%
\begin{eqnarray*}
\left \langle \varrho ^{2}\right \rangle _{_{^{\gamma }}}
&=&\frac{g^{\prime 2}\mathrm{r}+\left( g^{2}+g^{\prime 2}\right)
\gamma \bar{\gamma}}{2\left(
g^{\prime 2}-g^{2}\right) }+\frac{1}{2}\sqrt{\left( \frac{g^{\prime 2}%
\mathrm{r}+\left( g^{2}+g^{\prime 2}\right) \gamma
\bar{\gamma}}{g^{\prime 2}-g^{2}}\right) ^{2}+\frac{4\mathrm{\nu
\bar{\nu}}\left( g^{2}+g^{\prime
2}\right) }{g^{\prime 2}-g^{2}}} \\
&&
\end{eqnarray*}

\textbf{2)} \emph{computing} $\frac{\partial \mathcal{V}}{\partial
\varrho ^{2}}=0$\newline Doing the same thing but now with respect
to the variable $\varrho ^{2}$, we
have, by using the expression (\ref{V0}), the following expression for $%
\frac{\partial \mathcal{V}}{\partial \varrho ^{2}},$%
\begin{eqnarray}
\frac{\partial \mathcal{V}}{\partial \varrho ^{2}} &=&+\frac{g^{\prime 2}}{4}%
\left[ \varrho ^{2}-\frac{\mathrm{\nu \bar{\nu}}}{\varrho ^{2}}-\gamma \bar{%
\gamma}-\mathrm{r}\right] \left( 1+\frac{\mathrm{\nu \bar{\nu}}}{\varrho ^{4}%
}\right)  \notag \\
&&  \notag \\
&&+\frac{g^{2}}{4}\left[ \varrho ^{2}+\frac{\mathrm{\nu
\bar{\nu}}}{\varrho
^{2}}+\gamma \bar{\gamma}\right] \left( 1-\frac{\mathrm{\nu \bar{\nu}}}{%
\varrho ^{4}}\right) \\
&&  \notag
\end{eqnarray}
To get the zeros of $\frac{\partial \mathcal{V}}{\partial \varrho
^{2}}$, we put the solution $\gamma =0$ solving the constraint
relation $\frac{\partial \mathcal{V}}{\partial \bar{\gamma}}=0$ back
into the above equation; this
leads to the reduced expression%
\begin{eqnarray}
\frac{\partial \mathcal{V}}{\partial \varrho ^{2}}|_{\gamma =0} &=&\left( 1+%
\frac{\mathrm{\nu \bar{\nu}}}{\varrho ^{4}}\right) \left( \frac{%
g^{2}+g^{\prime 2}}{4}\varrho ^{2}-\frac{g^{2}+g^{\prime 2}}{4}\frac{\mathrm{%
\nu \bar{\nu}}}{\varrho ^{2}}-\frac{g^{\prime 2}}{4}\mathrm{r}\right) \\
&&  \notag
\end{eqnarray}
whose zeros are obtained by solving the vanishing condition%
\begin{equation}
\frac{g^{2}+g^{\prime 2}}{4}\varrho ^{2}-\frac{g^{2}+g^{\prime 2}}{4}\frac{%
\mathrm{\nu \bar{\nu}}}{\varrho ^{2}}-\frac{g^{\prime
2}}{4}\mathrm{r}=0
\end{equation}%
Rewriting this equation like%
\begin{equation}
\varrho ^{4}-\frac{g^{\prime 2}\mathrm{r}}{g^{2}+g^{\prime 2}}\varrho ^{2}-%
\mathrm{\nu \bar{\nu}}=0  \label{acc}
\end{equation}%
it is not difficult to check that the acceptable solution is given by%
\begin{equation}
\left \langle \varrho ^{2}\right \rangle =\frac{g^{\prime 2}\mathrm{r}+\sqrt{%
g^{\prime 4}\mathrm{r}^{2}+4\left( g^{2}+g^{\prime 2}\right)
^{2}\mathrm{\nu \bar{\nu}}}}{2\left( g^{2}+g^{\prime 2}\right) }
\end{equation}%
Notice that for the limit $r=0$, one has%
\begin{equation}
\left \langle \varrho _{0}^{2}\right \rangle =\sqrt{\mathrm{\nu
\bar{\nu}}}
\end{equation}%
which should compared with eq(\ref{XS}) obtained previously when the
Kahler parameter r was switched off.

\  \  \  \newline
To conclude, the minimum of the scalar potential $\mathcal{V}=\frac{1}{2}%
D^{\prime 2}+\frac{1}{2}D_{A}D^{A}$ with the auxiliary fields
related to the
Higgs fields as follows%
\begin{eqnarray}
D^{\prime } &=&\bar{h}_{i}h^{i}-\varphi
_{i}\bar{\varphi}^{i}-\mathrm{r}
\notag \\
&&  \notag \\
D^{A} &=&\left( \tau ^{A}\right) _{j}^{i}\left( \bar{h}_{i}h^{j}-\varphi _{i}%
\bar{\varphi}^{j}\right)  \label{r} \\
&&  \notag
\end{eqnarray}%
is obtained by first solving the conditions $\mathcal{V}_{ch}=0$
leading to
\begin{equation}
\left \{
\begin{tabular}{lll}
$\lambda S\varphi _{i}$ & $=$ & $0$ \\
$\lambda Sh^{i}$ & $=$ & $0$ \\
$\kappa S^{2}+\lambda \left( \varphi _{i}h^{i}-\mathrm{\nu }\right)
$ & $=$
& $0$%
\end{tabular}%
\right.
\end{equation}%
ensured by taking $S=0$ and the Higgs fields $h_{i}$ and $\varphi
_{i}$\ as
in eq(\ref{fs}). Putting these expressions back into $\mathcal{V}=\frac{1}{2}%
D^{\prime 2}+\frac{1}{2}D_{A}D^{A}$, one gets an explicit expression
of the potential in terms of the Higgs doublets given by (\ref{V0})
with minimum given by the following non trivial VEVs completely
characterized by the
gauge coupling constants and the FI coupling parameters,%
\begin{eqnarray}
\left \langle h^{i}\right \rangle &=&\left( \frac{g^{\prime 2}\mathrm{r}+%
\sqrt{g^{\prime 4}\mathrm{r}^{2}+4\left( g^{2}+g^{\prime 2}\right) ^{2}%
\mathrm{\nu \bar{\nu}}}}{2\left( g^{2}+g^{\prime 2}\right) }\right) ^{\frac{1%
}{2}}\cdot \mathrm{f}^{i}  \notag \\
&& \\
\left \langle \bar{h}_{i}\right \rangle &=&\left( \frac{g^{\prime 2}\mathrm{r%
}+\sqrt{g^{\prime 4}\mathrm{r}^{2}+4\left( g^{2}+g^{\prime 2}\right) ^{2}%
\mathrm{\nu \bar{\nu}}}}{2\left( g^{2}+g^{\prime 2}\right) }\right) ^{\frac{1%
}{2}}\cdot \mathrm{\bar{f}}_{i}  \notag \\
&&  \notag
\end{eqnarray}

and%
\begin{eqnarray}
\varphi _{i} &=&\mathrm{\nu }\left( \frac{2\left( g^{2}+g^{\prime
2}\right) }{g^{\prime 2}\mathrm{r}+\sqrt{g^{\prime
4}\mathrm{r}^{2}+4\left(
g^{2}+g^{\prime 2}\right) ^{2}\mathrm{\nu \bar{\nu}}}}\right) ^{\frac{1}{2}%
}\cdot \mathrm{\bar{f}}_{i}  \notag \\
&& \\
\bar{\varphi}^{i} &=&\mathrm{\bar{\nu}}\left( \frac{2\left(
g^{2}+g^{\prime 2}\right) }{g^{\prime 2}\mathrm{r}+\sqrt{g^{\prime
4}\mathrm{r}^{2}+4\left(
g^{2}+g^{\prime 2}\right) ^{2}\mathrm{\nu \bar{\nu}}}}\right) ^{\frac{1}{2}%
}\cdot \mathrm{f}^{i}  \notag \\
&&  \notag
\end{eqnarray}%
From this solution, we determine the numbers%
\begin{eqnarray}
\left \langle \bar{h}_{i}h^{i}\right \rangle &=&\frac{g^{\prime 2}\mathrm{r}+%
\sqrt{g^{\prime 4}\mathrm{r}^{2}+4\left( g^{2}+g^{\prime 2}\right) ^{2}%
\mathrm{\nu \bar{\nu}}}}{2\left( g^{2}+g^{\prime 2}\right) }  \notag \\
&& \\
\left \langle \varphi _{i}\bar{\varphi}^{i}\right \rangle
&=&\mathrm{\nu
\bar{\nu}}\left( \frac{2\left( g^{2}+g^{\prime 2}\right) }{g^{\prime 2}%
\mathrm{r}+\sqrt{g^{\prime 4}\mathrm{r}^{2}+4\left( g^{2}+g^{\prime
2}\right) ^{2}\mathrm{\nu \bar{\nu}}}}\right)  \notag \\
&&  \notag
\end{eqnarray}%
whose ratio reads as%
\begin{eqnarray}
\frac{\left \langle \bar{h}_{i}h^{i}\right \rangle }{\left \langle
\varphi
_{i}\bar{\varphi}^{i}\right \rangle } &=&\frac{1}{\mathrm{\nu \bar{\nu}}}%
\left( \frac{g^{\prime 2}\mathrm{r}+\sqrt{g^{\prime
4}\mathrm{r}^{2}+4\left( g^{2}+g^{\prime 2}\right) ^{2}\mathrm{\nu
\bar{\nu}}}}{2\left(
g^{2}+g^{\prime 2}\right) }\right) ^{2} \\
&&  \notag
\end{eqnarray}%
From this quantity, we obtain $\tan \beta $ which can be put into
the form
\begin{eqnarray}
&&  \notag \\
\tan \beta &=&\frac{g^{\prime 2}\mathrm{r}}{2\left( g^{2}+g^{\prime
2}\right) \sqrt{\mathrm{\nu \bar{\nu}}}}+\sqrt{1+\left( \frac{g^{\prime 2}%
\mathrm{r}}{2\left( g^{2}+g^{\prime 2}\right) \sqrt{\mathrm{\nu \bar{\nu}}}}%
\right) ^{2}} \\
&&  \notag
\end{eqnarray}%
Notice that for $r=0$, we have $\tan \beta $ $\rightarrow $ $\tan
\beta _{susy}=1$. For non zero Kahler parameter \textrm{r}, this
quantity is
clearly different from unity. Moreover, for small values of $\frac{\mathrm{r}%
}{\left \vert \mathrm{\nu }\right \vert }$ such as,%
\begin{equation}
\frac{g^{\prime 2}}{2\left( g^{2}+g^{\prime 2}\right) }\frac{\mathrm{r}}{%
\left \vert \mathrm{\nu }\right \vert }<<1
\end{equation}%
we have%
\begin{eqnarray}
\tan \beta &\simeq &1+\frac{g^{\prime 2}\mathrm{r}}{2\left(
g^{2}+g^{\prime
2}\right) \left \vert \mathrm{\nu }\right \vert }+\frac{1}{2}\left( \frac{%
g^{\prime 2}\mathrm{r}}{2\left( g^{2}+g^{\prime 2}\right) \left
\vert
\mathrm{\nu }\right \vert }\right) ^{2} \\
&&  \notag
\end{eqnarray}%
The energy $\mathcal{E}_{\min }$ of the ground state is given by%
\begin{eqnarray}
&&  \notag \\
\mathcal{E}_{\min } &=&\frac{g^{\prime 2}}{8}\left( \varrho ^{2}-\frac{%
\mathrm{\nu \bar{\nu}}}{\varrho ^{2}}-\mathrm{r}\right) ^{2}+\frac{g^{2}}{8}%
\left( \varrho ^{2}+\frac{\mathrm{\nu \bar{\nu}}}{\varrho ^{2}}\right) ^{2}-%
\frac{g^{2}}{2}\mathrm{\bar{\nu}\nu } \\
&&  \notag
\end{eqnarray}%
which reads also like%
\begin{eqnarray}
&&  \notag \\
\mathcal{E}_{\min } &=&\frac{g^{\prime 2}}{8}\left( \varrho ^{2}-\frac{%
\mathrm{\nu \bar{\nu}}}{\varrho ^{2}}-\mathrm{r}\right) ^{2}+\frac{g^{2}}{8}%
\left[ 2\varrho ^{2}-\left( \varrho ^{2}-\frac{\mathrm{\nu \bar{\nu}}}{%
\varrho ^{2}}\right) \right] ^{2}-\frac{g^{2}}{2}\mathrm{\bar{\nu}\nu } \\
&&  \notag
\end{eqnarray}%
But using the relations%
\begin{eqnarray}
\varrho ^{2}-\frac{\mathrm{\nu \bar{\nu}}}{\varrho ^{2}}
&=&\frac{g^{\prime
2}}{g^{2}+g^{\prime 2}}\mathrm{r}  \notag \\
&& \\
2\varrho ^{2} &=&\frac{g^{\prime 2}\mathrm{r}+\sqrt{g^{\prime 4}\mathrm{r}%
^{2}+4\left( g^{2}+g^{\prime 2}\right) ^{2}\mathrm{\nu
\bar{\nu}}}}{\left(
g^{2}+g^{\prime 2}\right) }  \notag \\
&&  \notag
\end{eqnarray}%
we can put $\mathcal{E}_{\min }$ into the form%
\begin{equation}
\mathcal{E}_{\min }=\frac{g^{\prime 2}g^{2}}{8\left( g^{2}+g^{\prime
2}\right) }\mathrm{r}^{2}
\end{equation}

\section{Explicit supersymmetry breaking}

Explicit supersymmetry breaking in \emph{n-MSSM} is achieved by
adding to the supersymmetric scalar potential $\mathcal{V}_{susy}$
the following extra term
\begin{equation}
\begin{tabular}{lll}
$\mathcal{V}_{exl}$ & $=$ & $-m_{_{_{u}}}^{2}$ $\bar{h}%
_{_{ui}}h_{_{u}}^{i}-m_{_{d}}^{2}$ $\bar{h}_{_{di}}h_{_{d}}^{i}-m_{s}^{2}%
\left \vert S\right \vert ^{2}$ \\
&  & $-\left( m_{ud}^{2}-\lambda _{_{s}}A_{_{\lambda }}S\right)
\varepsilon
_{ij}h_{_{u}}^{i}h_{_{d}}^{j}+\frac{\kappa _{_{s}}}{3}A_{\kappa }S^{3}+hc$%
\end{tabular}%
\end{equation}%
\begin{equation*}
\end{equation*}%
involving 5 new coupling parameters in addition to the existing
ones: 3 real
masses $m_{_{_{u}}},$ $m_{_{d}}$, $m_{s}$; and the 2 complex $m_{ud}$ and $%
A_{_{\lambda }}$\textrm{; see also and appendix A.}

\subsection{Explicit potential with anti-doublet $\protect \varphi _{i}$}

By replacing the doublet $h_{_{d}}^{i}$ by the anti-doublet $\varphi
_{i}$,
the above explicit potential gets mapped to%
\begin{equation}
\begin{tabular}{lll}
$\mathcal{V}_{exl}$ & $=$ & $-m_{_{u}}^{2}$
$\bar{h}_{i}h^{i}-m_{\varphi }^{2}$ $\varphi
_{i}\bar{\varphi}^{i}-m_{S}^{2}\left \vert S\right \vert
^{2} $ \\
&  & $-\left( m_{u\varphi }^{2}-\lambda A_{_{\lambda }}S\right)
\varphi
_{i}h^{i}+\frac{\kappa }{3}A_{\kappa }S^{3}+hc$%
\end{tabular}
\label{et}
\end{equation}%
with the two following remarkable features.

\begin{itemize}
\item \emph{gauge invariant composites}.\newline
Besides the complex singlet $S$, the potential $\mathcal{V}_{exl}$
involves four other gauge invariant composites of the Higgs field
doublets and anti-doublets; these are:
\begin{equation}
\begin{tabular}{lllll}
{\small hermitian} & : \  \  \  \  \  & $\bar{h}_{i}h^{i}$ & , & $\varphi _{i}%
\bar{\varphi}^{i}$ \\
{\small complex} & : & $\varphi _{i}h^{i}$ & , & $\bar{h}_{i}\bar{\varphi}%
^{i}$%
\end{tabular}
\label{bl}
\end{equation}%
Using the harmonic field coordinate basis,
\begin{equation}
\begin{tabular}{lll}
$h^{i}$ & $=$ & $\varrho f^{i}$ \\
$\varphi _{i}$ & $=$ & $\xi $ $\bar{f}_{i}+\eta \varepsilon _{ij}f^{j}$%
\end{tabular}%
\end{equation}%
with $\xi $\ and $\eta $\ two complex variables; and the relations%
\begin{equation}
\begin{tabular}{lll}
$\bar{h}_{i}h^{i}$ & $=$ & $\varrho ^{2}$ \\
$\varphi _{i}\bar{\varphi}^{i}$ & $=$ & $\xi \bar{\xi}+\eta \bar{\eta}$ \\
$\varphi _{i}h^{i}$ & $=$ & $\varrho \xi $%
\end{tabular}%
\end{equation}%
the above potential $\mathcal{V}_{exl}$ reads as%
\begin{equation}
\begin{tabular}{lll}
$\mathcal{V}_{exl}$ & $=$ & $-m_{_{u}}^{2}$ $\varrho ^{2}-m_{\varphi }^{2}$ $%
\left( \xi \bar{\xi}+\eta \bar{\eta}\right) -m_{S}^{2}\left \vert
S\right
\vert ^{2}$ \\
&  & $-\left( m_{u\varphi }^{2}-\lambda A_{_{\lambda }}S\right) \varrho \xi +%
\frac{\kappa }{3}A_{\kappa }S^{3}+hc$%
\end{tabular}%
\end{equation}

\item $\mathcal{V}_{exl}$ \emph{in terms of auxiliary fields}\newline
The above gauge invariant bilinear (\ref{bl}) appear as well into
the
equations of motion of the auxiliary fields $F_{S}$ and $D^{\prime }$\textrm{%
\ }%
\begin{equation}
\begin{tabular}{lll}
$\bar{F}$ & $=$ & $-\lambda \varphi _{i}h^{i}-\kappa S^{2}$ \\
$D^{\prime }$ & $=$ & $g^{\prime }\left( \bar{h}_{i}h^{i}-\varphi _{i}\bar{%
\varphi}^{i}\right) $ \\
&  &
\end{tabular}
\label{RR1}
\end{equation}%
For the particular case $m_{_{u}}^{2}$ $=-m_{\varphi }^{2}=m^{2}$, the term $%
m_{_{u}}^{2}$ $\left( \bar{h}_{i}h^{i}-\varphi
_{i}\bar{\varphi}^{i}\right) $ coincides exactly with
$\frac{m^{2}}{g^{\prime }}D^{\prime }$, and the explicit breaking
potential is expressed in terms of the auxiliary fields
like%
\begin{equation}
\begin{tabular}{lll}
$\mathcal{V}_{exl}$ & $=$ & $-\frac{m^{2}}{g^{\prime }}D^{\prime
}-m_{S}^{2}\left \vert S\right \vert ^{2}$ \\
&  & $+\left( \frac{m_{u\varphi }^{2}}{\lambda }-A_{_{\lambda
}}S\right)
\bar{F}+\frac{m_{u\varphi }^{2}\kappa }{\lambda }S^{2}-\frac{2\kappa }{3}%
A_{\kappa }S^{3}+hc$ \\
&  &
\end{tabular}
\label{RR2}
\end{equation}%
In the general case where $m_{_{u}}^{2}$ and $m_{\varphi }^{2}$ are
arbitrary real numbers, one needs to introduce the following exotic
auxiliary field%
\begin{equation}
\begin{tabular}{lll}
$\Delta ^{\prime }$ & $=$ & $g^{\prime \prime }\left( \bar{h}%
_{i}h^{i}+\varphi _{i}\bar{\varphi}^{i}\right) $%
\end{tabular}
\label{RR3}
\end{equation}%
This fields may be thought of as related to an extra $U^{\prime
\prime }\left( 1\right) $ gauge symmetry with gauge coupling
constant $g^{\prime \prime }$ and under which the Higgs fields
$h^{i}$ and $\varphi _{i}$\ have
the same charge. By inverting the relations (\ref{RR1}-\ref{RR2}), we obtain%
\begin{equation}
\begin{tabular}{lll}
$\varphi _{i}h^{i}$ & $=$ & $-\frac{1}{\lambda }\left(
\bar{F}+\kappa
S^{2}\right) $ \\
&  &  \\
$\bar{h}_{i}\bar{\varphi}^{i}$ & $=$ & $-\frac{1}{\bar{\lambda}}\left( F+%
\bar{\kappa}\bar{S}^{2}\right) $ \\
&  &  \\
$\bar{h}_{i}h^{i}$ & $=$ & $\frac{g^{\prime }\Delta ^{\prime
}+g^{\prime
\prime }D^{\prime }}{2g^{\prime }g^{\prime \prime }}$ \\
&  &  \\
$\varphi _{i}\bar{\varphi}^{i}$ & $=$ & $\frac{g^{\prime }\Delta
^{\prime
}-g^{\prime \prime }D^{\prime }}{2g^{\prime }g^{\prime \prime }}$ \\
&  &
\end{tabular}%
\end{equation}%
and ends with the following expression%
\begin{equation}
\begin{tabular}{lll}
$\mathcal{V}_{exl}$ & $=$ & $-\frac{m_{_{u}}^{2}-m_{\varphi }^{2}}{%
2g^{\prime }}D^{\prime }-$ $\frac{m_{_{u}}^{2}+m_{\varphi
}^{2}}{2g^{\prime
\prime }}\Delta ^{\prime }-m_{S}^{2}\left \vert S\right \vert ^{2}$ \\
&  &  \\
&  & $+\left( \frac{m_{u\varphi }^{2}}{\lambda }-A_{_{\lambda
}}S\right)
\bar{F}+\frac{m_{u\varphi }^{2}\kappa }{\lambda }S^{2}-\frac{2\kappa }{3}%
A_{\kappa }S^{3}+hc$ \\
&  &
\end{tabular}
\label{exv}
\end{equation}%
depending linearly on the auxiliary fields $\Delta ^{\prime },$
$D^{\prime }$ and $\bar{F}$; and having an indefinite sign
responsible for the negative region in fig \ref{1A}.
\end{itemize}

\subsection{Potential $\mathcal{V}_{exl}$ in superspace}

Here, we consider a particular $\mathcal{V}_{exl}$ resulting from
giving
special non zero VEVs to the two hyperchargeless iso-singlet superfields $%
\boldsymbol{S}_{u_{1}}$ and $\boldsymbol{V}_{u_{1}}$. By special non
zero VEVs we mean those superfield vacuum expectation values that
break supersymmetry explicitly in the sense they depend on the
Grassmann variables
as follows%
\begin{equation}
\begin{tabular}{lll}
$\left \langle \boldsymbol{V}_{u_{1}}\right \rangle $ & $=$ & $\theta ^{2}%
\bar{\theta}^{2}m_{x}^{2}$ \\
$\left \langle \boldsymbol{S}_{u_{1}}\right \rangle $ & $=$ &
$-\theta
^{2}A_{\kappa }$%
\end{tabular}
\label{vss}
\end{equation}%
with $m_{x}$ and $A_{\kappa }$ constant coefficients as
above.\newline
Substituting these expressions back into the lagrangian density $\boldsymbol{%
L}_{higgs}$ given by eq(\ref{lt}); but with $\boldsymbol{V}^{\prime }=%
\boldsymbol{V}_{0}\frac{Y}{2}$ and $\boldsymbol{S}$ thought of as
describing supersymmetric fluctuations like
\begin{equation}
\begin{tabular}{lllll}
$\boldsymbol{V}_{0}$ & $\rightarrow $ \  \  & $\mathbb{V}_{0}$ & $=$
& $\left
\langle \boldsymbol{V}_{u_{1}}\right \rangle +\boldsymbol{V}_{0}$ \\
$\boldsymbol{S}$ & $\rightarrow $ & $\mathbb{S}_{0}$ & $=$ & $\left
\langle
\boldsymbol{S}_{u_{1}}\right \rangle +\boldsymbol{S}$%
\end{tabular}%
\end{equation}%
with $\left \langle \boldsymbol{V}_{u_{1}}\right \rangle $ and $%
\left \langle \boldsymbol{S}_{u_{1}}\right \rangle $ as in
eqs(\ref{vss}), we obtain
\begin{equation}
\begin{tabular}{lll}
&  &  \\
$\mathcal{L}_{higgs}$ & $=$ & $\dint d^{4}\theta $ $\mathbb{S}_{0}^{\dagger }%
\mathbb{S}_{0}\boldsymbol{+}\dint d^{4}\theta \boldsymbol{H}_{i}^{\dagger }%
\left[ e^{-g\boldsymbol{V}}\right] _{j}^{i}e^{\boldsymbol{-g}^{\prime }%
\mathbb{V}^{\prime }}\boldsymbol{H}^{j}$ \\
&  & $\mathrm{+}\dint d^{4}\theta $ $\boldsymbol{\Phi }_{i}$ $\left[ e^{+g%
\boldsymbol{V}}\right] _{j}^{i}$ $e^{+\boldsymbol{g}^{\prime }\mathbb{V}%
^{\prime }}\boldsymbol{\Phi }^{\dagger j}$ \\
&  & $-\dint d^{2}\theta \left( \lambda \mathbb{S}_{0}\left[ \Phi _{i}%
\boldsymbol{H}^{i}\right] +\frac{\kappa
}{3}\mathbb{S}_{0}^{3}+hc\right) $
\\
&  & $+\left( \frac{g^{\prime }}{2}\mathrm{r}\dint d^{4}\theta \mathbb{V}%
_{0}\right) +\left( \lambda \mathrm{\bar{\nu}}\dint d^{2}\theta \mathbb{S}%
_{0}+hc\right) $ \\
&  &
\end{tabular}%
\end{equation}%
Using the relations
\begin{equation}
\begin{tabular}{lll}
$e^{\boldsymbol{-g}^{\prime }\left( \boldsymbol{V}^{\prime }+\theta ^{2}\bar{%
\theta}^{2}m_{x}^{2}\right) }$ & $=$ & $e^{\boldsymbol{-g}^{\prime }%
\boldsymbol{V}^{\prime }}\left[ 1-\boldsymbol{g}^{\prime }\theta ^{2}\bar{%
\theta}^{2}m_{x}^{2}\right] $ \\
&  &  \\
$\left( \boldsymbol{S}-\theta ^{2}A_{\kappa }\right) ^{3}$ & $=$ & $%
\boldsymbol{S}^{3}-3\theta ^{2}A_{\kappa }S^{2}$ \\
&  &
\end{tabular}%
\end{equation}%
and integrating with respect to the Grassmann variables $\theta $ and $\bar{%
\theta}$, we get%
\begin{equation}
\begin{tabular}{lll}
$\mathcal{L}_{higgs}$ & $=$ & $\boldsymbol{L}_{higgs}+L_{0}$ \\
&  & $-g^{\prime }m_{x}^{2}\left( \bar{h}_{i}^{\prime }h^{i\prime
}-\varphi
_{i}^{\prime }\bar{\varphi}^{i\prime }\right) $ \\
&  & $-\left( \lambda A_{\kappa }\varphi _{i}h^{i}+\kappa A_{\kappa
}S^{2}\right) +hc$ \\
&  &
\end{tabular}
\label{fl}
\end{equation}%
with $L_{0}=\left \vert A_{\kappa }\right \vert ^{2}-\nu A_{\kappa }-\bar{\nu%
}\bar{A}_{\kappa }$ a constant term scaling as mass$^{4}$ and
interpreted in
terms of vacuum energy density; it will be dropped out in what follows.%
\newline
The field doublets $h^{i\prime }$ and $\varphi _{i}^{\prime }$ of
the second line of eq(\ref{fl}) are related to the usual doublets
$h^{i}$ and $\varphi
_{i}$ like%
\begin{eqnarray}
h^{i\prime } &\equiv &e^{\boldsymbol{-}\frac{\boldsymbol{g}^{\prime }}{4}%
\boldsymbol{\upsilon }_{0}}\left( e^{-\frac{g}{4}\boldsymbol{\upsilon }%
_{A}\tau ^{A}}\right) _{j}^{i}h^{j}  \notag \\
&&  \notag \\
\varphi _{i}^{\prime } &\equiv &\varphi _{j}\left( e^{+\frac{g}{4}%
\boldsymbol{\upsilon }_{A}\tau ^{A}}\right) _{i}^{j}e^{\boldsymbol{+}\frac{%
\boldsymbol{g}^{\prime }}{4}\boldsymbol{\upsilon }_{0}}  \label{vv} \\
&&  \notag
\end{eqnarray}%
with $\boldsymbol{\upsilon }_{A}$ and $\boldsymbol{\upsilon }_{0}$
standing respectively for the leading component fields of the
$\theta $- expansion of the $SU_{L}\left( 2\right) $ and the
$U_{Y}\left( 1\right) $ gauge
multiplets $\boldsymbol{V}_{A}$ and $\boldsymbol{V}_{0}$. They read as%
\begin{eqnarray}
\boldsymbol{\upsilon }_{A} &=&\boldsymbol{V}_{A}\boldsymbol{|}_{\theta =\bar{%
\theta}=0}  \notag \\
\boldsymbol{\upsilon }_{0} &=&\left( \boldsymbol{V}_{0}\right) \boldsymbol{|}%
_{\theta =\bar{\theta}=0}
\end{eqnarray}%
and are pure gauge degrees of freedom. However, because of $\mathbb{C}%
_{Y}^{\ast }\times SL\left( 2,C\right) $ property, the doublet
$h^{i\prime }$ and anti-doublet $\varphi _{i}^{\prime }$ can be
identified with $h^{i}$ and $\varphi _{i}$ respectively due to the
residual symmetry transformations leaving in the coset space,
\begin{equation*}
\frac{\mathbb{C}_{Y}^{\ast }\times SL\left( 2,C\right) }{U_{Y}\left(
1\right) \times SU_{L}\left( 2\right) }
\end{equation*}%
Notice also that in the Wess- Zumino gauge vector multiplets, the
quantities
$\boldsymbol{\upsilon }_{A}$ and $\boldsymbol{\upsilon }_{0}$ appearing in (%
\ref{vv}) are simply set to zero.

\subsection{Energy of the ground state}

In this subsection, we use the potential approximation%
\begin{equation*}
\mathcal{V}=\mathcal{V}_{ch}+\mathcal{V}_{pert}
\end{equation*}%
with%
\begin{equation*}
\mathcal{V}_{pert}=\left( \frac{1}{2}D^{\prime 2}+\frac{1}{2}%
D_{A}D^{A}\right) +\mathcal{V}_{exl}
\end{equation*}%
to study the effect induced by the $\mathcal{V}_{exl}$ potential on
the predictions of previous \textrm{sections. }In this
approximation, the
minimum of $\mathcal{V}_{ch}$ is given by the Higgs fields configuration%
\begin{equation}
\left \{
\begin{tabular}{lll}
$\lambda S\varphi _{i}$ & $=$ & $0$ \\
$\lambda Sh^{i}$ & $=$ & $0$ \\
$\kappa S^{2}+\lambda \left( \varphi _{i}h^{i}-\mathrm{\nu }\right)
$ & $=$
& $0$%
\end{tabular}%
\right.  \label{F1}
\end{equation}%
solved as%
\begin{eqnarray}
S &=&0  \notag \\
\varphi _{i} &=&\frac{\mathrm{\nu }}{\varrho ^{2}}\bar{h}_{i}-\frac{\gamma }{%
\varrho }h_{i} \\
h^{i} &=&\varrho f^{i}  \notag
\end{eqnarray}%
Notice that there is no contribution of the Higgs iso-singlet Higgs
field $S$ nor in $\varphi _{i}h^{i}$; the main role of $S$ is to fix
of the shape of the Higgs VEVS as
\begin{equation}
\varphi _{i}h^{i}=\mathrm{\nu }  \label{F3}
\end{equation}%
Then, because of $\varphi _{i}h^{i}=\mathrm{\nu }$ and $S=0$, the
contribution of $\mathcal{V}_{exl}$ coming from the chiral sector
drops out; no the complex term plus its complex conjugate in
$\mathcal{V}_{exl}$; so the explicit potential reduces to
\begin{equation*}
\mathcal{V}_{exl}=\mathcal{V}_{0}+g^{\prime }m_{x}^{2}\left( \bar{h}%
_{i}h^{i}-\varphi _{i}\bar{\varphi}^{i}\right)
\end{equation*}%
with $\mathcal{V}_{0}=cte$. Therefore, the resulting Higgs potential
reads,
up to a constant, like%
\begin{equation}
\mathcal{V}=\frac{1}{2}D^{\prime 2}+\frac{1}{2}D_{A}D^{A}+g^{\prime
}m_{x}^{2}\left( \bar{h}_{i}h^{i}-\varphi
_{i}\bar{\varphi}^{i}\right) \label{vh}
\end{equation}%
with an indefinite sign and a field variation completely originating
from the Kahler sector of supersymmetry. The last feature can be
explicitly
exhibited by expressing $\left( \bar{h}_{i}h^{i}-\varphi _{i}\bar{\varphi}%
^{i}\right) $ in term of the auxiliary field $D^{\prime }$. Using
\begin{equation}
\begin{tabular}{lll}
$D^{\prime }$ & $=$ & $\frac{g^{\prime }}{2}\left[ \left( \bar{h}%
_{i}h^{i}-\varphi _{i}\bar{\varphi}^{i}\right) -2\mathrm{r}\right] $ \\
& $=$ & $\frac{g^{\prime }}{2}\left( \varrho ^{2}-\frac{\mathrm{\nu \bar{\nu}%
}}{\varrho ^{2}}-\gamma \bar{\gamma}-2\mathrm{r}\right) $%
\end{tabular}%
\end{equation}%
we have
\begin{equation*}
\left( \bar{h}_{i}h^{i}-\varphi _{i}\bar{\varphi}^{i}\right) =\frac{2}{%
g^{\prime }}\left( D^{\prime }+2\mathrm{r}\right)
\end{equation*}%
and then
\begin{equation}
\mathcal{V}=\frac{1}{2}D_{A}D^{A}+\frac{1}{2}D^{\prime 2}+\frac{2m_{x}^{2}}{%
g^{\prime }}\left( D^{\prime }+\mathrm{r}\right)
\end{equation}%
In addition to the usual quadratic $\frac{1}{2}D^{\prime 2}$, the
Higgs potential $\mathcal{V}$ has also a linear dependence on
$D^{\prime }$. Moreover, seen that this potential may be also
rewritten as
\begin{equation}
\mathcal{V}=\frac{1}{2}D_{A}D^{A}+\frac{1}{2}\left( D^{\prime }+\frac{%
m_{x}^{2}}{g^{\prime }}\right) ^{2}+\frac{2m_{x}^{2}}{g^{\prime
}}\left( \mathrm{r}-\frac{m_{x}^{2}}{4g^{\prime }}\right)
\end{equation}%
or more explicitly like%
\begin{equation}
\mathcal{V}=\frac{g^{\prime 2}}{8}\left( \varrho ^{2}-\frac{\mathrm{\nu \bar{%
\nu}}}{\varrho ^{2}}-\gamma \bar{\gamma}-\mathrm{r}+\frac{m_{x}^{2}}{%
g^{\prime }}\right) ^{2}+\frac{g^{2}}{8}\left( \varrho ^{2}+\frac{\mathrm{%
\nu \bar{\nu}}}{\varrho ^{2}}+\gamma \bar{\gamma}\right) ^{2}-\frac{g^{2}}{2}%
\mathrm{\bar{\nu}\nu }
\end{equation}%
its minimum is obtained by following the same steps as in case
$m_{x}^{2}=0$
studied in previous section. We have%
\begin{equation}
\begin{tabular}{lllll}
$\frac{\partial \mathcal{V}}{\partial \bar{\gamma}}$ & $=$ & $\gamma \frac{%
\partial \mathcal{V}}{\partial \left( \gamma \bar{\gamma}\right) }$ & $=$ & $%
0$ \\
$\frac{\partial \mathcal{V}}{\partial h^{i}}$ & $=$ & $\bar{h}_{i}\frac{%
\partial \mathcal{V}}{\partial \varrho ^{2}}$ & $=$ & $0$%
\end{tabular}%
\end{equation}%
and is solved by taking $\gamma =0$ and $\frac{\partial \mathcal{V}}{%
\partial \varrho ^{2}}=0$. This leads to%
\begin{equation}
\frac{\partial \mathcal{V}}{\partial \varrho ^{2}}|_{\gamma =0}=\left( 1+%
\frac{\mathrm{\nu \bar{\nu}}}{\varrho ^{4}}\right) \left( \frac{%
g^{2}+g^{\prime 2}}{4}\varrho ^{2}-\frac{g^{2}+g^{\prime 2}}{4}\frac{\mathrm{%
\nu \bar{\nu}}}{\varrho ^{2}}-\frac{g^{\prime
2}}{4}\mathrm{R}\right)
\end{equation}%
where we have set
\begin{equation*}
R=\mathrm{r}-\frac{m_{x}^{2}}{g^{\prime }}
\end{equation*}%
Following the analysis of section 5, the zeros of the above relation
is given by the solution of
\begin{equation}
\varrho ^{4}-\frac{g^{\prime 2}\mathrm{R}}{g^{2}+g^{\prime 2}}\varrho ^{2}-%
\mathrm{\nu \bar{\nu}}=0  \label{ca}
\end{equation}%
whose acceptable solution reads as follows%
\begin{equation}
\left \langle \varrho ^{2}\right \rangle _{\min }=\frac{g^{\prime 2}\mathrm{R%
}+\sqrt{g^{\prime 4}\mathrm{R}^{2}+4\left( g^{2}+g^{\prime 2}\right) ^{2}%
\mathrm{\nu \bar{\nu}}}}{2\left( g^{2}+g^{\prime 2}\right) }
\end{equation}%
This expression leads in turns to the following Higgs configurations%
\begin{eqnarray}
\left \langle h^{i}\right \rangle _{\min } &=&\left( \frac{g^{\prime 2}%
\mathrm{R}+\sqrt{g^{\prime 4}\mathrm{R}^{2}+4\left( g^{2}+g^{\prime
2}\right) ^{2}\mathrm{\nu \bar{\nu}}}}{2\left( g^{2}+g^{\prime 2}\right) }%
\right) ^{\frac{1}{2}}\mathrm{f}^{i}  \notag \\
&& \\
\left \langle \varphi _{i}\right \rangle _{\min } &=&\mathrm{\nu
}\left( \frac{g^{\prime 2}\mathrm{R}+\sqrt{g^{\prime
4}\mathrm{R}^{2}+4\left( g^{2}+g^{\prime 2}\right) ^{2}\mathrm{\nu
\bar{\nu}}}}{2\left( g^{2}+g^{\prime 2}\right) }\right)
^{-\frac{1}{2}}\mathrm{\bar{f}}_{i}
\notag \\
&&  \notag
\end{eqnarray}%
From these relations, we determine $\tan \beta $
\begin{eqnarray}
\tan \beta &=&\frac{g^{\prime 2}\mathrm{R}}{2\left( g^{2}+g^{\prime
2}\right) \sqrt{\mathrm{\nu \bar{\nu}}}}+\sqrt{1+\left( \frac{g^{\prime 2}%
\mathrm{R}}{2\left( g^{2}+g^{\prime 2}\right) \sqrt{\mathrm{\nu \bar{\nu}}}}%
\right) ^{2}} \\
&&  \notag
\end{eqnarray}%
and the energy $\mathcal{E}_{\min }$ of the ground state%
\begin{equation}
\mathcal{E}_{\min }=cte+\frac{g^{\prime 2}g^{2}}{8\left(
g^{2}+g^{\prime 2}\right) }\left(
\mathrm{r}-\frac{m_{x}^{2}}{g^{\prime }}\right) ^{2}
\end{equation}%
Therefore the effect of $\mathcal{V}_{exl}$ is manifested mainly by
a shift
of the energy of the ground state; the term $\mathrm{R}=\mathrm{r}%
-m_{x}^{2}/g^{\prime }$ has the same geometric interpretation as the
FI coupling constant $\mathrm{r}$.

\section{Conclusion}

Assuming that the Higgs potential \ $\mathcal{V}_{higgs}$ of
\emph{n-MSSM} as dominated by the contribution $\mathcal{V}_{ch}$
coming from the chiral sector of supersymmetry; and replacing the
Higgs superfield doublet $\left( \boldsymbol{H}_{d}\right) ^{i}$ by
an anti-doublet $\boldsymbol{\Phi }_{i}$, we have derived in this
paper the explicit geometry of the Higgs ground state $\left \vert
\Sigma _{higgs}\right \rangle $ in terms of the coupling constants
of the model. To achieve this goal, we have used tools on
supersymmetric gauge theory, conifold geometry, harmonic field
coordinates of the real 3-sphere as well as special features of
$SU_{L}\left( 2\right) \times U_{Y}\left( 1\right) $ gauge symmetry
representations.\newline The basic idea behind these solutions is as
follows:

\  \  \  \  \

\emph{an} \emph{anti-doublet} $\boldsymbol{\Phi }_{d}$ \emph{at
place of} \emph{the doublet} $\boldsymbol{H}_{d}$\newline Instead of
describing the usual $\boldsymbol{H}_{u}$ and $\boldsymbol{H}_{d}$
chiral Higgs superfields of \emph{n-MSSM} in terms of the two
doublets with opposite hypercharges and same charge under
$SU_{L}\left( 2\right) $, we
have modified the quantum numbers of down Higgs by replacing the doublet $%
\left( H_{d}\right) ^{i}$ by the anti-doublet $\Phi _{i}$. With this
change,
$\boldsymbol{H}_{u}$ and $\Phi $ have now opposite charges under $%
SU_{L}\left( 2\right) \times U_{Y}\left( 1\right) $ gauge symmetry;
i.e opposite hypercharge $y_{u}=-y_{d}$ and opposite $SU_{L}\left(
2\right) $ charge leading to change of sign of the $SU_{L}\left(
2\right) $ gauge
coupling constant g of certain superfield interactions such as%
\begin{eqnarray}
&&\left( -g\right) \boldsymbol{V}_{A}^{\left( su_{2}\right) }\boldsymbol{H}%
_{x}^{\dagger }T^{A}\boldsymbol{H}_{x}\qquad ,\qquad \left(
+g\right)
\boldsymbol{V}_{A}^{\left( su_{2}\right) }\boldsymbol{\Phi }T^{A}\boldsymbol{%
\Phi }^{\dagger }  \label{p} \\
&&  \notag
\end{eqnarray}%
with $x=u,$ $d$. The obtained model, refereed in this study as \emph{n-MSSM}$%
^{\ast }$, has an interpretation in terms of intersecting conifold
geometries. Indeed, the equation of motion of the auxiliary fields
of the
supersymmetric \emph{n-MSSM}$^{\ast }$ with non zero FI terms lead to%
\begin{eqnarray}
\varphi _{i}h^{i} &=&\nu  \notag \\
\bar{h}_{i}h^{i}-\varphi _{i}\bar{\varphi}^{i} &=&r \\
\left( \tau ^{A}\right) _{j}^{i}\left( \bar{h}_{i}h^{j}-\varphi _{i}\bar{%
\varphi}^{j}\right) &=&0  \notag
\end{eqnarray}

\  \  \  \newline and turn out to have non trivial solutions given
by the intersection of two
conifolds. Notice that in case of using the usual $H_{d}^{i}$ of \emph{n-MSSM%
}, the iso-triplet relation of above eqs should be modified like in (\ref{2D}%
) as required by (\ref{p}).

\  \  \  \  \  \  \  \

\emph{supersymmetric phase}\newline In the case where the
contributions of the explicit supersymmetric breaking potential is
switched off ($\mathcal{V}_{_{exl}}=0$), a common solution of
the above auxiliary field equations requires $r=0$, and is given by%
\begin{eqnarray}
h^{i} &=&\varrho f^{i}  \notag \\
\varphi _{i} &=&\frac{\nu }{\varrho }\bar{f}_{i}
\end{eqnarray}%
with
\begin{equation*}
\varrho ^{4}=\nu \bar{\nu}
\end{equation*}%
and%
\begin{eqnarray}
f^{i} &=&\left(
\begin{array}{c}
\cos \frac{\theta }{2}e^{\frac{i}{2}\left( \psi +\phi \right) } \\
\sin \frac{\theta }{2}e^{\frac{i}{2}\left( \psi -\phi \right) }%
\end{array}%
\right) \\
&&  \notag
\end{eqnarray}%
as well as the identification%
\begin{equation*}
f^{i\prime }\equiv e^{i\gamma }f^{i}
\end{equation*}%
under the hypercharge gauge symmetry; which can used to set $\psi
-\phi =0$. This feature shows that the Higgs configurations in the
supersymmetric ground state parameterize a real 2-sphere
\begin{equation*}
\mathbb{S}^{2}=\frac{SU\left( 2\right) }{U\left( 1\right) }
\end{equation*}%
In this case the ratio $\frac{\upsilon _{h}}{\upsilon _{\varphi }}$
of the Higgs VEVs and the energy $\mathcal{E}_{\min }^{\left(
r=0\right) }$ of the
ground state are as follows%
\begin{eqnarray}
\tan \beta _{susy} &=&1  \notag \\
\left. \mathcal{E}_{\min }^{\left( r=0\right) }\right \vert _{\mathcal{V}%
_{_{exl}}} &=&0
\end{eqnarray}

\emph{non supersymmetric ground state}\newline For $r\neq 0$ and
$\mathcal{V}_{_{exl}}=0$, the equations of motion of the auxiliary
fields; in particular the D-fields, have no common solution and so
supersymmetry is broken. \newline We have computed the scalar
potential energy density inducing supersymmetry breaking; and showed
the energy $\mathcal{E}_{\min }^{\left( r\neq 0\right)
} $ of the non supersymmetric ground state is exactly given by%
\begin{equation}
\left. \mathcal{E}_{\min }^{\left( r\neq 0\right) }\right \vert _{\mathcal{V}%
_{_{exl}}}=\frac{g^{\prime 2}g^{2}}{8\left( g^{2}+g^{\prime 2}\right) }%
\mathrm{r}^{2}
\end{equation}%
Moreover, because of supersymmetry breaking, the ratio of the Higgs
VEVs gets a deviation from unity ($\tan \beta _{susy}=1$) and reads
as
\begin{eqnarray}
\tan \beta &=&\frac{g^{\prime 2}\mathrm{r}}{2\left( g^{2}+g^{\prime
2}\right) \sqrt{\mathrm{\nu \bar{\nu}}}}+\sqrt{1+\left( \frac{g^{\prime 2}%
\mathrm{r}}{2\left( g^{2}+g^{\prime 2}\right) \sqrt{\mathrm{\nu \bar{\nu}}}}%
\right) ^{2}} \\
&&  \notag
\end{eqnarray}%
We have also explored the effect of the explicit supersymmetry
breaking terms; but for the special case $m_{_{u}}^{2}$
$=-m_{\varphi }^{2}=m^{2}$;
leading to a shift of the Kahler parameter $\mathrm{r}$ like $\mathrm{r}-%
\frac{m^{2}}{g^{\prime }}$; this issue needs further exploration.

\section{Appendix A: general on MSSM and extensions}

In this section, we give some useful tools on \emph{MSSM} and \emph{%
next-to-MSSM }in superspace.

\subsection{MSSM in superspace}

In superspace with local graded coordinates $\left( x^{\mu },\theta
^{\alpha },\bar{\theta}_{\dot{\alpha}}\right) $, the superfield
spectrum of MSSM with $SU_{L}\left( 2\right) \times U_{Y}\left(
1\right) $ gauge symmetry contains \emph{29} superfields arranged
into two subsets: \emph{4} hermitian superfields and \emph{25}
complex chiral ones as follows:

\begin{description}
\item[1)] \emph{4 gauge superfields}\newline
These are the hermitian superfields $\boldsymbol{V}_{A},$ $\boldsymbol{V}%
_{0} $ in one to one with the generators of $SU_{L}\left( 2\right)
\times U_{Y}\left( 1\right) $ gauge symmetry; they can be combined
together within a unique hermitian superfield $\boldsymbol{V}$ ; but
valued in the Lie algebra of the $SU_{L}\left( 2\right) \times
U_{Y}\left( 1\right) $ gauge symmetry like
\begin{equation}
\boldsymbol{V}=\boldsymbol{V}_{A}T^{A}+\boldsymbol{V}_{0}\frac{Y}{2}
\label{va}
\end{equation}%
where the $T^{A}$'s are the 3 generators of $SU_{L}\left( 2\right) $ and $%
\frac{Y}{2}$ the generator of $U_{Y}\left( 1\right) $. From these
\emph{4} dimensionless superfields, we build two basic spinor
superfields namely the chiral $\mathcal{W}_{\alpha }$ and antichiral
$\mathcal{W}_{\dot{\alpha}}$ superfield strengths given by
\begin{equation}
\begin{tabular}{llllllll}
\multicolumn{3}{l}{$\  \  \  \  \  \  \  \ SU_{L}\left( 2\right) $}
& : &  &
\multicolumn{3}{l}{$\  \  \  \  \  \  \  \  \ U_{Y}\left( 1\right) $} \\
&  &  &  &  &  &  &  \\
$\mathcal{W}_{\alpha }$ & $=$ & $\frac{1}{4}\mathcal{\bar{D}}^{2}\text{ }%
e^{2gV}\mathcal{D}_{\alpha }e^{-2gV}$ & , &  & $\mathcal{W}_{\alpha
}^{\prime }$ & $=$ &
$\frac{1}{4}\mathcal{\bar{D}}^{2}\mathcal{D}_{\alpha
}V^{\prime }$ \\
$\mathcal{\bar{W}}_{\dot{\alpha}}$ & $=$ &
$\frac{1}{4}\mathcal{D}^{2}\text{
}e^{-2gV}\mathcal{\bar{D}}_{\dot{\alpha}}e^{2gV}$ & , &  & $\mathcal{\bar{W}}%
_{\dot{\alpha}}^{\prime }$ & $=$ & $\frac{1}{4}\mathcal{D}^{2}\mathcal{\bar{D%
}}_{\dot{\alpha}}V^{\prime }$ \\
&  &  &  &  &  &  &
\end{tabular}%
\end{equation}%
These are gauge covariant superfields appearing in the superspace
lagrangian density of the $SU_{L}\left( 2\right) \times U_{Y}\left(
1\right) $
supersymmetric gauge theory%
\begin{equation}
\boldsymbol{L}_{su_{2}\times u_{1}}=\int d^{2}\theta \left[ Tr\left( \frac{1%
}{16g^{2}}\mathcal{W}^{\alpha }\mathcal{W}_{\alpha }\right) +\frac{1}{4}%
\mathcal{W}^{\prime \alpha }\mathcal{W}_{\alpha }^{\prime }\right]
+hc
\end{equation}%
with $g$ the gauge coupling of $SU_{L}\left( 2\right) $. By
integration with respect to the Grassmann variables $\theta $ and
$\bar{\theta}$, we obtain the component field lagrangian density
\begin{equation}
\begin{tabular}{lll}
$\boldsymbol{L}_{su_{2}\times u_{1}}$ & $=$ & $-\frac{1}{4}W_{\mu
\nu
}^{A}W_{A}^{\mu \nu }-i\lambda ^{A}\sigma ^{\mu }\nabla _{\mu }\bar{\lambda}%
_{A}+\frac{1}{2}D^{A}D_{A}$ \\
&  & $-\frac{1}{4}B_{\mu \nu }B^{\mu \nu }-i\lambda ^{\prime }\sigma
^{\mu
}\partial _{\mu }\bar{\lambda}^{\prime }+\frac{1}{2}D^{\prime 2}$ \\
&  &
\end{tabular}%
\end{equation}%
where $W_{\mu \nu }^{A}$, $B_{\mu \nu }$ are respectively the field
strengths of the bosonic vector gauge fields $W_{\mu }^{A}$, $B_{\mu
}$; the fermionic gauginos $\lambda ^{A},$ $\lambda $ are the
supersymmetric partners; and the $D^{A}$, $D^{\prime }$ scalars the
usual D-auxiliary fields

\item[2)] \emph{25 chiral superfields} \newline
These are complex chiral superfields describing supersymmetric
matter and
Higgs multiplets; they belong to special representations of the $%
SU_{L}\left( 2\right) \times U_{Y}\left( 1\right) $ symmetry as given below%
\begin{eqnarray*}
&&%
\begin{tabular}{llllll}
&  &  &  &  &  \\
\ {\small sectors} & {\small :} & {\small chiral superfields } & $\
\  \  \  \ \  \  \  \ {\small G}$ & \multicolumn{2}{l}{\  \ {\small
number}} \\ \hline \hline \multicolumn{1}{|l}{} &  &  &
\multicolumn{1}{l|}{} &  & \multicolumn{1}{l|}{
} \\
\multicolumn{1}{|l}{\ {\small Leptons}} & : & $\left.
\begin{array}{ccc}
{\small L}_{i} & {\small =} & \left( {\small N}_{iL}{\small ,E}%
_{iL}^{-}\right) \\
{\small \bar{R}}^{i} & {\small =} & {\small \bar{E}}_{R}^{i}%
\end{array}%
\right. $ & \multicolumn{1}{l|}{\  \ $\  \  \left.
\begin{array}{c}
\left( {\small 1,2,-1}\right) \\
\left( {\small 1,1,+2}\right)%
\end{array}%
\right. $} & $\left.
\begin{array}{c}
{\small 3\times 2}\  \\
{\small 3\times 1}%
\end{array}%
\right. $ & \multicolumn{1}{l|}{${\small 9}$} \\
\multicolumn{1}{|l}{} &  &  & \multicolumn{1}{l|}{} &  &
\multicolumn{1}{l|}{ } \\ \hline \multicolumn{1}{|l}{} &  &  &
\multicolumn{1}{l|}{} &  & \multicolumn{1}{l|}{
} \\
\multicolumn{1}{|l}{\ {\small Quarks}} & : & $\left.
\begin{array}{ccc}
{\small Q}_{i} & {\small =} & \left( {\small U}_{iL}{\small
,D}_{iL}\right)
\\
{\small \bar{U}}^{i} & {\small =} & {\small \bar{U}}_{R}^{i} \\
{\small \bar{D}}^{i} & {\small =} & {\small \bar{D}}_{R}^{i}%
\end{array}%
\right. $ & \multicolumn{1}{l|}{\  \ $\  \  \left.
\begin{array}{c}
\left( {\small 3,2,+}\frac{{\small 1}}{{\small 3}}\right) \\
\left( {\small \bar{3},1,-}\frac{{\small 4}}{{\small 3}}\right) \\
\left( {\small \bar{3},1,+}\frac{{\small 2}}{{\small 3}}\right)%
\end{array}%
\right. $} & $\left.
\begin{array}{c}
{\small 3\times 2} \\
{\small 3\times 1} \\
{\small 3\times 1}%
\end{array}%
\right. $ & \multicolumn{1}{l|}{${\small 12}$} \\
\multicolumn{1}{|l}{} &  &  & \multicolumn{1}{l|}{} &  &
\multicolumn{1}{l|}{ } \\ \hline \multicolumn{1}{|l}{} &  &  &
\multicolumn{1}{l|}{} &  & \multicolumn{1}{l|}{
} \\
\multicolumn{1}{|l}{\  \ {\small Higgs}} & : & $\left.
\begin{array}{ccc}
{\small H}_{u} & {\small =} & \left( {\small H}_{u}^{+}{\small ,H}_{u}^{%
{\small 0}}\right) \\
{\small H}_{d} & {\small =} & \left( {\small H}_{d}^{{\small 0}}{\small ,H}%
_{d}^{-}\right)%
\end{array}%
\right. $ & \multicolumn{1}{l|}{\  \ $\  \  \left.
\begin{array}{c}
\left( {\small 1,2,+1}\right) \\
\left( {\small 1,2,-1}\right)%
\end{array}%
\right. $} & $\left.
\begin{array}{c}
{\small 2} \\
{\small 2}%
\end{array}%
\right. $ & \multicolumn{1}{l|}{${\small 4}$} \\
\multicolumn{1}{|l}{} &  &  & \multicolumn{1}{l|}{} &  &
\multicolumn{1}{l|}{ } \\ \hline
\end{tabular}
\\
&&
\end{eqnarray*}%
with $G=SU_{{\small C}}\left( {\small 3}\right) \times SU_{L}\left(
2\right)
\times U_{{\small Y}}\left( {\small 1}\right) $; the extra factor $SU_{%
{\small C}}\left( {\small 3}\right) $ stands for color symmetry
which is understood in present analysis. \newline The superspace
lagrangian density $\boldsymbol{L}_{_{MSSM}}$ describing the
dynamics of these chiral superfields and the interactions among
themselves
and with gauge superfields is given by%
\begin{eqnarray}
\boldsymbol{L}_{_{MSSM}} &=&\int d^{4}\theta \left( \mathcal{K}_{_{\text{%
\textit{L-sector}}}}+\mathcal{K}_{_{\text{\textit{Q-sector}}}}+\mathcal{K}%
_{_{\text{\textit{H-sector}}}}\right)  \notag \\
&&+\int d^{2}\theta \left( W_{_{\text{\textit{L-H}}}}+W_{_{\text{\textit{Q-H}%
}}}+W_{_{\text{\textit{H-H}}}}\right) +hc \\
&&+\boldsymbol{L}_{soft}  \notag
\end{eqnarray}%
where the Kahler terms $\mathcal{K}_{i},$ the chiral
super-potentials $W_{i}$ and the space time lagrangian densities
$\boldsymbol{L}_{soft}$ as collected below:

\item[a)] \emph{Kahler type terms}%
\begin{eqnarray}
&&%
\begin{tabular}{lll}
$\mathcal{K}_{_{\text{\textit{H-sector}}}}$ & $=$ & $\boldsymbol{H}%
_{u}^{\dagger }.\left( e^{-g\boldsymbol{V}}e^{-g^{\prime }\boldsymbol{V}%
^{\prime }}\right) .\boldsymbol{H}_{u}+\boldsymbol{H}_{d}^{\dagger
}.\left(
e^{-g\boldsymbol{V}}e^{-g^{\prime }\boldsymbol{V}^{\prime }}\right) .%
\boldsymbol{H}_{d}$ \\
&  &  \\
$\mathcal{K}_{_{\text{\textit{L-sector}}}}$ & $=$ & $\sum \limits_{{\small L}%
\text{{\small - }}{\small superfields}}\boldsymbol{\Phi }^{\dagger
}.\left(
e^{-g\boldsymbol{V}}e^{-g^{\prime }\boldsymbol{V}^{\prime }}\right) .%
\boldsymbol{\Phi }$ \\
&  &  \\
$\mathcal{K}_{_{\text{\textit{Q-sector}}}}$ & $=$ & $\sum \limits_{{\small Q}%
\text{{\small -}}{\small superfields}}\boldsymbol{\Phi
}_{i}^{\dagger }.\left( e^{-g\boldsymbol{V}}e^{-g^{\prime
}\boldsymbol{V}^{\prime }}\right)
.\boldsymbol{\Phi }_{i}$%
\end{tabular}
\\
&&  \notag
\end{eqnarray}%
To get the component field expression corresponding to these Kahler
terms, we have to expand the superfields in $\theta $-series and
integrate with respect to the Grassmann variables.\newline
For example, expanding the chiral superfield Higgs doublets $\boldsymbol{H}%
_{x}$ in $\theta $-series like
\begin{equation}
\boldsymbol{H}_{x}=\boldsymbol{h}_{x}+\sqrt{2}\theta .\boldsymbol{\tilde{h}}%
_{x}+\theta ^{2}\boldsymbol{F}_{x}
\end{equation}%
the integration with respect to the Grassmann variables leads to the
following gauge covariant quadratic terms%
\begin{equation}
\begin{tabular}{lll}
$\boldsymbol{L}_{\text{\textit{H-sector}}}$ & $=$ & $\left( \nabla _{\mu }%
\boldsymbol{h}_{u}\right) ^{\dagger }.\left( \nabla ^{\mu }\boldsymbol{h}%
_{u}\right) +i\boldsymbol{\tilde{h}}_{u}^{\dagger
}.\bar{\sigma}^{\mu
}\nabla _{\mu }\boldsymbol{\tilde{h}}_{u}+\boldsymbol{F}_{u}^{\dagger }.%
\boldsymbol{F}_{u}$ \\
&  &  \\
&  & $+\left( \nabla _{\mu }\boldsymbol{h}_{d}\right) ^{\dagger
}.\nabla
^{\mu }\boldsymbol{h}_{d}+i\boldsymbol{\tilde{h}}_{d}^{\dagger }.\bar{\sigma}%
^{\mu }\nabla _{\mu
}\boldsymbol{\tilde{h}}_{d}+\boldsymbol{F}_{d}^{\dagger
}.\boldsymbol{F}_{d}$ \\
&  &
\end{tabular}%
\end{equation}%
with
\begin{equation}
\mathcal{\nabla }_{\mu }=\partial _{\mu }-igW_{\mu }-ig^{\prime
}B_{\mu }
\end{equation}%
the space time gauge covariant derivative valued in the Lie algebra
of the
gauge symmetry; and where the $\boldsymbol{F}_{x}$ doublets%
\begin{equation}
\boldsymbol{F}_{u}=\left(
\begin{array}{c}
F_{u}^{+} \\
F_{u}^{0}%
\end{array}%
\right) ,\qquad \boldsymbol{F}_{d}=\left(
\begin{array}{c}
F_{d}^{0} \\
F_{d}^{-}%
\end{array}%
\right)
\end{equation}%
are auxiliary fields.

\item[b)] \emph{the superpotential terms}\newline
they are given by%
\begin{equation}
\begin{tabular}{lll}
$W_{{\small MSSM}}$ & $=$ & $\mu \text{ }\boldsymbol{H}_{u}.\boldsymbol{H}%
_{d}+Y_{lu}^{ij}\boldsymbol{\bar{R}}_{i}.\boldsymbol{L}_{j}.\boldsymbol{H}%
_{d}$ \\
&  & $Y_{qu}^{ij}\text{ }\boldsymbol{\bar{U}}_{Ri}.\boldsymbol{Q}_{j}.%
\boldsymbol{H}_{u}+Y_{qd}^{ij}\text{ }\boldsymbol{\bar{D}}_{Ri}.\boldsymbol{Q%
}_{j}.\boldsymbol{H}_{d}$%
\end{tabular}%
\end{equation}%
with the dimensionless $3\times 3$ matrices $Y_{rs}^{ij}$ standing
for the coupling constants. Explicitly, we have
\begin{equation}
\begin{tabular}{lll}
$W_{{\small MSSM}}$ & $=$ & $\mathrm{\mu }\left(
H_{u}^{0}H_{d}^{0}-H_{u}^{+}H_{d}^{-}\right) +\left(
\mathrm{Y}_{ld}\right)
_{ij}\bar{R}^{i}\left( E_{L}^{-j}H_{d}^{0}-N_{L}^{j}H_{d}^{-}\right) +$ \\
&  & $(\mathrm{Y}_{qu})_{ij}\bar{U}^{i}\left(
D_{L}^{j}H_{u}^{+}-U_{L}^{j}H_{u}^{0}\right) +(\mathrm{Y}_{qd})_{ij}\bar{D}%
^{i}\left( D_{L}^{j}H_{d}^{0}-U_{L}^{j}H_{d}^{-}\right) $%
\end{tabular}%
\end{equation}%
The contribution of $W_{{\small MSSM}}$ to the scalar Higgs
potential comes
therefore from%
\begin{equation}
W_{\text{\textit{H-H}}}=\mu \text{
}\boldsymbol{H}_{u}.\boldsymbol{H}_{d} \label{ud}
\end{equation}%
with massive coupling constant $\mu $.

\item[c)] \emph{soft supersymmetry breaking term}\newline
This is a scalar potential needed to break supersymmetry explicitly
before the breaking of gauge symmetry happens. It reads as
\begin{equation}
\begin{tabular}{lll}
$\boldsymbol{L}_{_{{\small soft}}}^{{\small MSSM}}$ & $=$ & $-\mathrm{m}%
_{_{d}}^{2}$ $h_{_{d}}^{\dagger }.h_{_{d}}-\mathrm{m}_{_{_{u}}}^{2}$ $%
h_{_{u}}^{\dagger }.h_{_{u}}$ \\
&  & $+\mathrm{m}_{ud}^{2}h_{_{u}}.h_{_{d}}+\mathrm{\bar{m}}%
_{ud}^{2}h_{_{u}}^{\dagger }.h_{_{d}}^{\dagger }$%
\end{tabular}
\label{ls}
\end{equation}%
and may be thought as following from the superspace density%
\begin{eqnarray}
\boldsymbol{L}_{_{{\small soft}}}^{{\small MSSM}} &=&\int
d^{4}\theta \text{
}\left( \boldsymbol{H}_{_{u}}^{\dagger }.\mathfrak{N}_{_{u}}.\boldsymbol{H}%
_{_{u}}+\boldsymbol{H}_{_{d}}^{\dagger }.\mathfrak{N}_{_{d}}.\boldsymbol{H}%
_{_{d}}\right)  \notag \\
&&-\int d^{2}\theta \text{ }\boldsymbol{H}_{_{u}}.\mathfrak{N}_{_{ud}}.%
\boldsymbol{H}_{_{d}}+hc  \label{sl}
\end{eqnarray}%
with
\begin{equation}
\begin{tabular}{lll}
$\mathfrak{N}_{_{i}}=-\mathrm{m}_{_{i}}^{2}\theta
^{2}\bar{\theta}^{2}$ & $,$
& $\mathfrak{N}_{_{ud}}=-\mathrm{m}_{_{ud}}^{2}\theta ^{2}$%
\end{tabular}
\label{tu}
\end{equation}%
breaking explicitly supersymmetry.
\end{description}

\  \  \  \  \  \  \newline
The scalar potential, involving the Higgs field doublets $\boldsymbol{h}%
_{_{u}},$ $\boldsymbol{h}_{_{d}}$, the corresponding auxiliary doublets $%
\boldsymbol{F}_{u},$ $\boldsymbol{F}_{d}$ and the hermitian
auxiliary
triplet $D_{A}$ and singlet $D^{\prime }$ reads in general like%
\begin{equation}
\begin{tabular}{lll}
$\boldsymbol{L}_{_{scalar}}$ & $=$ & $\frac{1}{2}D^{A}D_{A}+\frac{1}{2}%
D^{\prime 2}+\boldsymbol{F}_{u}^{\dagger }.\boldsymbol{F}_{u}+\boldsymbol{F}%
_{d}^{\dagger }.\boldsymbol{F}_{d}$ \\
&  & $-gD_{A}\left( \boldsymbol{h}_{u}^{\dagger }.\frac{\tau ^{A}}{2}.%
\boldsymbol{h}_{u}+\boldsymbol{h}_{d}^{\dagger }.\frac{\tau ^{A}}{2}.%
\boldsymbol{h}_{d}\right) $ \\
&  & $-g^{\prime }D^{\prime }\left( \boldsymbol{h}_{u}^{\dagger }.\frac{Y}{2}%
.\boldsymbol{h}_{u}+\boldsymbol{h}_{d}^{\dagger }.\frac{Y}{2}.\boldsymbol{h}%
_{d}\right) $ \\
&  & $-\mathrm{m}_{_{d}}^{2}$ $\boldsymbol{h}_{_{d}}^{\dagger }.\boldsymbol{h%
}_{_{d}}-\mathrm{m}_{_{_{u}}}^{2}$ $\boldsymbol{h}_{_{u}}^{\dagger }.%
\boldsymbol{h}_{_{u}}$ \\
&  & $+\mu \left( \boldsymbol{h}_{d}.\boldsymbol{F}_{u}+\boldsymbol{h}_{u}.%
\boldsymbol{F}_{d}\right) +\mathrm{m}_{ud}^{2}\boldsymbol{h}_{_{u}}.%
\boldsymbol{h}_{_{d}}+hc$ \\
&  &
\end{tabular}
\label{ld}
\end{equation}%
by eliminating the auxiliary fields, we get the standard relation
\begin{equation}
\begin{tabular}{lll}
$\mathcal{V}_{_{scalar}}$ & $=$ & $\boldsymbol{F}_{u}^{\dagger }.\boldsymbol{%
F}_{u}+\boldsymbol{F}_{d}^{\dagger }.\boldsymbol{F}_{d}+\frac{1}{2}%
\boldsymbol{D}^{A}\boldsymbol{D}_{A}+\frac{1}{2}\boldsymbol{D}^{\prime
2}$
\\
&  & $+\boldsymbol{L}_{_{{\small soft}}}^{{\small MSSM}}$%
\end{tabular}%
\end{equation}%
with
\begin{equation}
\begin{tabular}{lll}
$\boldsymbol{F}_{u}^{\dagger }$ & $=$ & $-\mu \boldsymbol{h}_{d}$ \\
$\boldsymbol{F}_{d}^{\dagger }$ & $=$ & $-\mu \boldsymbol{h}_{u}$ \\
$\boldsymbol{D}^{A}$ & $=$ & $g\boldsymbol{D}_{A}\left( \boldsymbol{h}%
_{u}^{\dagger }.\frac{\tau ^{A}}{2}.\boldsymbol{h}_{u}+\boldsymbol{h}%
_{d}^{\dagger }.\frac{\tau ^{A}}{2}.\boldsymbol{h}_{d}\right) $ \\
$\boldsymbol{D}^{\prime }$ & $=$ & $g^{\prime }\left( \boldsymbol{h}%
_{u}^{\dagger }.\frac{Y}{2}.\boldsymbol{h}_{u}+\boldsymbol{h}_{d}^{\dagger }.%
\frac{Y}{2}.\boldsymbol{h}_{d}\right) $%
\end{tabular}%
\end{equation}

\subsection{Extended Higgs sector}

Phenomenological studies on MSSM predictions; in particular the one
regarding the lowest bound on the mass of the lightest Higgs
particle, suggest amongst other scenarios, that quite realistic
models might be described by a lagrangian density with more than
\emph{4} complex Higgs
superfields. In this view, and because of properties of tensor products of $%
SU_{L}\left( 2\right) \times U_{Y}\left( 1\right) $ representations;
consistent extensions of the MSSM Higgs sector are obtained by adding $%
SU_{L}\left( 2\right) $ singlet superfields $\boldsymbol{S}_{i}$ and / or $%
SU_{L}\left( 2\right) $ triplets $\vec{\Delta}_{i}$. Indeed, using
the fact that products of two complex $SU_{L}\left( 2\right) $
doublets carrying hypercharges $y$ and $y^{\prime }$ decompose like
\begin{equation}
2_{y}\otimes 2_{y^{\prime }}=1_{y+y^{\prime }}\oplus 3_{y+y^{\prime
}}
\end{equation}%
we learn that gauge invariant cubic couplings of chiral superfields
allow
extensions of the MSSM Higgs sector by the above mentioned superfields $%
\boldsymbol{S}_{i}$ and $\vec{\Delta}_{i}$ with hypercharges as given below%
\begin{eqnarray}
&&%
\begin{tabular}{lllllll}
{\small singlets} & : & {\small hypercharge} & \  \  \  \  \  \  \  \  \  & {\small %
triplets} &  & {\small hypercharge} \\ \hline
&  &  &  &  &  &  \\
${\small S}_{ud}$ &  & ${\small 0}$ &  & ${\small \vec{\Delta}}_{ud}$ & : & $%
{\small 0}$ \\
${\small S}_{uu}$ &  & ${\small -2}$ &  & ${\small
\vec{\Delta}}_{uu}$ & : &
${\small -2}$ \\
${\small S}_{dd}$ &  & ${\small +2}$ &  & ${\small
\vec{\Delta}}_{dd}$ & : &
${\small +2}$ \\
&  &  &  &  &  &  \\ \hline
\end{tabular}
\\
&&  \notag
\end{eqnarray}%
These quantum numbers lead, amongst others, to the following extra
chiral tri-superfield couplings
\begin{equation}
\begin{tabular}{lll}
$\boldsymbol{S}_{ud}\left(
\boldsymbol{H}_{u}.\boldsymbol{H}_{d}\right) $ &
, \  \  \  \  \  & $\boldsymbol{\vec{\Delta}}_{ud}.\left( \boldsymbol{H}_{u}%
\frac{\boldsymbol{\vec{\tau}}}{2}\boldsymbol{H}_{d}\right) $ \\
$\boldsymbol{S}_{uu}\left(
\boldsymbol{H}_{u}.\boldsymbol{H}_{u}\right) $ & ,
& $\boldsymbol{\vec{\Delta}}_{uu}.\left( \boldsymbol{H}_{u}\frac{\boldsymbol{%
\vec{\tau}}}{2}\boldsymbol{H}_{u}\right) $ \\
$\boldsymbol{S}_{dd}\left(
\boldsymbol{H}_{d}.\boldsymbol{H}_{d}\right) $ & ,
& $\boldsymbol{\vec{\Delta}}_{dd}.\left( \boldsymbol{H}_{d}\frac{\boldsymbol{%
\vec{\tau}}}{2}\boldsymbol{H}_{d}\right) $ \\
&  &
\end{tabular}%
\end{equation}%
In the next -to- MSSM we have been interested in this paper, one
extends the Higgs sector of MSSM by adding the hyperchargeless
singlet $S_{ud}\equiv
\boldsymbol{S}$ to the superfield spectrum; so the chiral superpotential $%
W_{_{MSSM}}$ of MSSM gets modified to
\begin{equation}
\boldsymbol{W}_{_{N-MSSM}}=\boldsymbol{W}_{_{MSSM}}+\lambda \boldsymbol{S}%
\left( \boldsymbol{H}_{u}.\boldsymbol{H}_{d}\right) +\frac{\kappa }{3}%
\boldsymbol{S}^{3}
\end{equation}%
with $\lambda $ and $\kappa $ new dimensionless parameters. The soft
supersymmetry breaking terms extending (\ref{ld}) have moreover $S$-
coupling terms and are given by \textrm{\cite{x}}%
\begin{equation}
\begin{tabular}{lll}
$\boldsymbol{L}_{_{{\small soft}}}^{{\small N-MSSM}}$ & $=$ &
$-m_{_{d}}^{2}$ $\boldsymbol{h}_{_{d}}^{\dagger }\cdot
\boldsymbol{h}_{_{d}}-m_{_{_{u}}}^{2}$
$\boldsymbol{h}_{_{u}}^{\dagger }\cdot \boldsymbol{h}_{_{u}}-m_{s}^{2}\left%
\vert S\right \vert ^{2}$ \\
&  & $-\left( m_{ud}^{2}-\lambda _{_{s}}A_{_{\lambda }}S\right) \boldsymbol{h%
}_{_{u}}\cdot \boldsymbol{h}_{_{d}}+\frac{\kappa
_{_{s}}}{3}A_{\kappa
}S_{ud}^{3}+hc$%
\end{tabular}%
\end{equation}%
with $m_{_{d}},$ $m_{_{_{u}}},$\ $m_{s}$, $m_{ud},$\ $\lambda _{_{s}}$, $%
A_{_{\lambda }}$\ and $\kappa _{_{s}}$ new dimensionful
constants.\newline Combining the scalar terms and eliminating the
auxiliary fields, one gets
the scalar potential of \emph{n-MSSM} namely%
\begin{equation}
\begin{tabular}{lll}
$\mathcal{V}_{_{scalar}}$ & $=$ & $\frac{1}{2}\boldsymbol{D}^{A}\boldsymbol{D%
}_{A}+\frac{1}{2}\boldsymbol{D}^{\prime
2}+\boldsymbol{F}_{u}^{\dagger
}\cdot \boldsymbol{F}_{u}+\boldsymbol{F}_{d}^{\dagger }\cdot \boldsymbol{F}%
_{d}$ \\
&  & $+\boldsymbol{F}_{S}^{\dagger }\cdot \boldsymbol{F}_{S}+\boldsymbol{L}%
_{_{{\small soft}}}^{n{\small -MSSM}}$%
\end{tabular}%
\end{equation}%
with%
\begin{equation}
\begin{tabular}{lll}
$\boldsymbol{F}_{s}^{\dagger }$ & $=$ & $-\left( \lambda \boldsymbol{h}%
_{_{u}}\cdot \boldsymbol{h}_{_{d}}+\kappa S^{2}\right) $ \\
$\boldsymbol{F}_{u}^{\dagger }$ & $=$ & $-\left( \mu +\lambda
S\right)
\boldsymbol{h}_{d}$ \\
$\boldsymbol{F}_{d}^{\dagger }$ & $=$ & $-\left( \mu +\lambda
S\right)
\boldsymbol{h}_{u}$ \\
$D^{A}$ & $=$ & $g\left( \boldsymbol{h}_{u}^{\dagger }.\frac{\tau ^{A}}{2}.%
\boldsymbol{h}_{u}+\boldsymbol{h}_{d}^{\dagger }.\frac{\tau ^{A}}{2}.%
\boldsymbol{h}_{d}\right) $ \\
$D^{\prime }$ & $=$ & $g^{\prime }\left( \boldsymbol{h}_{u}^{\dagger }.\frac{%
Y}{2}.\boldsymbol{h}_{u}+\boldsymbol{h}_{d}^{\dagger }.\frac{Y}{2}.%
\boldsymbol{h}_{d}\right) $%
\end{tabular}%
\end{equation}

\section{Appendix B: deriving the metric of Higgs curve}

In this appendix, we build the metric of the complex 3D hypersurface $%
\mathfrak{C}_{\nu }$ and the induced one on the Higgs manifold
$\Sigma $ parameterized by the Higgs moduli in the ground state.
First, we determine the metric of\ the complex \emph{3D}
hypersurface $\mathfrak{C}_{\nu }$ sitting the complex \emph{4D}
space $\mathbb{C}^{4}$; then we consider the two cases $\Sigma
_{susy}$ and $\Sigma _{non-susy}$ respectively associated with the
supersymmetric and non supersymmetric ground states.

\subsection{Metric of the complex 3D hypersurface $\mathfrak{C}_{\protect \nu %
}$}

We start from the \textrm{solution }of the holomorphic constraint relation $%
\varphi _{i}h^{i}=\mathrm{\nu }$ giving the field moduli $\left(
\varphi
_{i},\bar{\varphi}^{i}\right) $ in term of the fields $\left( h^{i},\bar{h}%
_{i}\right) $ and the complex singlets $\left( \eta
,\bar{\eta}\right) $
namely%
\begin{equation}
\begin{tabular}{lll}
$\varphi _{i}$ & $=$ & $\frac{\mathrm{\nu }}{\varrho ^{2}}\bar{h}_{i}+\frac{%
\eta }{\varrho }h_{i}$ \\
$\bar{\varphi}^{i}$ & $=$ & $\frac{\mathrm{\bar{\nu}}}{\varrho ^{2}}h^{i}-%
\frac{\bar{\eta}}{\varrho }\bar{h}^{i}$%
\end{tabular}
\label{AB0}
\end{equation}%
where $\bar{h}_{i}h^{i}=\varrho ^{2}$ and where $\eta $ is scaling
in same manner as the Higgs fields $h^{i}$ and $\varphi _{i}$.
\newline The hypercharges of these field moduli are as follows
\begin{equation}
\begin{tabular}{lll}
$\left[ Y,h^{i}\right] =h^{i}$ & , & $\left[ Y,\varphi _{i}\right]
=-\varphi
_{i}$ \\
$\left[ Y,\bar{h}_{i}\right] =-\bar{h}_{i}$ & , & $\left[ Y,\bar{\varphi}^{i}%
\right] =+\bar{\varphi}^{i}$%
\end{tabular}
\label{AB4}
\end{equation}%
\newline
and%
\begin{equation*}
\begin{tabular}{lll}
$\left[ Y,\eta \right] =-2\eta $ & , & $\left[ Y,\bar{\eta}\right] =+2\bar{%
\eta}$%
\end{tabular}%
\end{equation*}%
By using the harmonic variables $f^{i}$ and $\bar{f}_{i}$ satisfying $f^{i}%
\bar{f}_{i}=1$ and $f^{i}f_{i}=0=\bar{f}^{i}\bar{f}_{i},$ we can
express the
Higgs configuration (\ref{AB0}) like%
\begin{equation}
\begin{tabular}{lll}
$h^{i}$ & $=$ & $\varrho f^{i}$ \\
$\bar{h}_{i}$ & $=$ & $\varrho \bar{f}_{i}$%
\end{tabular}
\label{AB1}
\end{equation}%
and%
\begin{equation}
\begin{tabular}{lll}
$\varphi _{i}$ & $=$ & $\frac{\mathrm{\nu }}{\varrho
}\bar{f}_{i}+\eta f_{i}$
\\
$\bar{\varphi}^{i}$ & $=$ & $\frac{\mathrm{\bar{\nu}}}{\varrho }f^{i}-\bar{%
\eta}\bar{f}^{i}$%
\end{tabular}
\label{AB2}
\end{equation}%
with hypercharges like%
\begin{equation}
\begin{tabular}{lll}
$\left[ Y,f^{i}\right] =+f^{i}$ & , & $\left[ Y,\bar{f}_{i}\right] =-\bar{f}%
_{i}$ \\
$\left[ Y,\varrho \right] =0$ &  &
\end{tabular}
\label{AB5}
\end{equation}%
Then, using $d\mathrm{\nu }=0$, we compute the differentials%
\begin{equation}
\begin{tabular}{lll}
$dh^{i}$ & , & $d\bar{h}_{i}$ \\
$d\varphi _{i}$ & , & $d\bar{\varphi}^{i}$%
\end{tabular}%
\end{equation}%
in terms of
\begin{equation}
\begin{tabular}{lllll}
$df^{i}$ & , & $d\bar{f}_{i}$ & , & $d\varrho $ \\
$d\eta $ & , & $d\bar{\eta}$ &  &
\end{tabular}%
\end{equation}%
These differentials have same quantum numbers under the $U_{Y}\left(
1\right) \times SU_{L}\left( 2\right) $ gauge symmetry group as the
corresponding variables. We have:

\begin{description}
\item[$1)$] \emph{the} \emph{differentials} $dh^{i}$ and $d\bar{h}_{i}$
\newline
Using eq(\ref{AB1}), the complex differentials $dh^{i}$ and
$d\bar{h}_{i}$
are given by%
\begin{equation}
\begin{tabular}{lll}
$dh^{i}$ & $=$ & $f^{i}d\varrho +\varrho df^{i}$ \\
$d\bar{h}_{i}$ & $=$ & $\bar{f}_{i}d\varrho +\varrho d\bar{f}_{i}$%
\end{tabular}%
\end{equation}%
leading to%
\begin{eqnarray*}
d\bar{h}_{i}dh^{i} &=&d\varrho ^{2}+\varrho ^{2}df^{i}d\bar{f}_{i} \\
&&
\end{eqnarray*}%
Moreover, by using the explicit expression of the harmonic field
variables
in terms of the angles $\theta ,$ $\psi ,$ $\phi $ namely%
\begin{equation*}
\begin{tabular}{lllll}
&  &  &  &  \\
$f^{i}$ & $=\left(
\begin{array}{c}
\cos \frac{\theta }{2}e^{\frac{i}{2}\left( \psi +\phi \right) } \\
\\
\sin \frac{\theta }{2}e^{\frac{i}{2}\left( \psi -\phi \right) }%
\end{array}%
\right) $ & , & $\bar{f}_{i}$ & $=\left(
\begin{array}{c}
\cos \frac{\theta }{2}e^{-\frac{i}{2}\left( \psi +\phi \right) }\text{ } \\
\\
\text{ }\sin \frac{\theta }{2}e^{-\frac{i}{2}\left( \psi -\phi \right) }%
\end{array}%
\right) $ \\
&  &  &  &
\end{tabular}%
\end{equation*}%
it is not difficult to check that we have%
\begin{equation}
\begin{tabular}{lll}
&  &  \\
$d\bar{f}_{i}df^{i}$ & $=$ & $\frac{1}{4}\left( d\theta ^{2}+\cos ^{2}\frac{%
\theta }{2}d\left( \psi +\phi \right) ^{2}+\sin ^{2}\frac{\theta
}{2}d\left(
\psi -\phi \right) ^{2}\right) $ \\
&  &  \\
& $=$ & $\frac{1}{4}\left( d\theta ^{2}+d\psi ^{2}+d\phi ^{2}+2\cos
\theta
d\psi d\phi \right) $ \\
&  &
\end{tabular}
\label{ae}
\end{equation}

\item[$2)$] \emph{the} \emph{differentials} $d\varphi _{i}$ and $d\bar{%
\varphi}^{i}$ \newline Similarly using eq(\ref{AB2}), these
differentials read in terms of the harmonic field variables like
\begin{equation}
\begin{tabular}{lll}
$d\varphi _{i}$ & $=$ & $\frac{\mathrm{\nu }}{\varrho }\left( d\bar{f}_{i}-%
\bar{f}_{i}\frac{d\varrho }{\varrho }\right) +\left( f_{i}d\eta
+\eta
df_{i}\right) $ \\
$d\bar{\varphi}^{i}$ & $=$ & $\frac{\mathrm{\bar{\nu}}}{\varrho
}\left(
df^{i}-f^{i}\frac{d\varrho }{\varrho }\right) -\left( \bar{f}^{i}d\bar{\eta}+%
\bar{\eta}d\bar{f}^{i}\right) $ \\
&  &
\end{tabular}%
\end{equation}%
To get the expression of the length element $d\varphi _{i}d\bar{\varphi}^{i}$%
, it is helpful to decompose it like
\begin{equation}
d\varphi _{i}d\bar{\varphi}^{i}=A+B+C+D
\end{equation}%
with%
\begin{equation}
\begin{tabular}{lll}
$A$ & $=$ & $\frac{\mathrm{\nu \bar{\nu}}}{\varrho ^{2}}\left( df^{i}-f^{i}%
\frac{d\varrho }{\varrho }\right) \left( d\bar{f}_{i}-\bar{f}_{i}\frac{%
d\varrho }{\varrho }\right) $ \\
$B$ & $=$ & $-\frac{\mathrm{\nu }}{\varrho }\left( \bar{f}^{i}d\bar{\eta}+%
\bar{\eta}d\bar{f}^{i}\right) \left(
d\bar{f}_{i}-\bar{f}_{i}\frac{d\varrho
}{\varrho }\right) $ \\
$C$ & $=$ & $\frac{\mathrm{\bar{\nu}}}{\varrho }\left( df^{i}-f^{i}\frac{%
d\varrho }{\varrho }\right) \left( f_{i}d\eta +\eta df_{i}\right) $ \\
$D$ & $=$ & $-\left(
\bar{f}^{i}d\bar{\eta}+\bar{\eta}d\bar{f}^{i}\right)
\left( f_{i}d\eta +\eta df_{i}\right) $ \\
&  &
\end{tabular}%
\end{equation}%
then compute the 4 terms of the product $d\varphi
_{i}d\bar{\varphi}^{i}$ separately:

\begin{itemize}
\item \emph{term }$A$\newline
This term is hermitian and is proportional to the absolute value of
the complex deformation parameter $\left \vert \mathrm{\nu }\right
\vert ^{2}$:
\begin{equation}
A=\frac{\mathrm{\nu \bar{\nu}}}{\varrho ^{4}}d\varrho
^{2}+\frac{\mathrm{\nu \bar{\nu}}}{\varrho ^{2}}df^{i}d\bar{f}_{i}
\end{equation}%
it reads also like%
\begin{equation}
A=\frac{\mathrm{\nu \bar{\nu}}}{\varrho ^{4}}\left( d\varrho
^{2}+\varrho ^{2}df^{i}d\bar{f}_{i}\right)
\end{equation}%
with $df^{i}d\bar{f}_{i}$ as in eq(\ref{ae}). This term vanishes at
the conifold singulary $\nu =0.$

\item \emph{terms B and C}\newline
These are complex contributions; the term $B$ is proportional to $\mathrm{%
\nu }$ and $C$ is its complex conjugate; they read as%
\begin{eqnarray}
B &=&\frac{\mathrm{\nu }}{\varrho }\left( d\bar{\eta}+\bar{\eta}\frac{%
d\varrho }{\varrho }\right) \bar{f}_{i}d\bar{f}^{i}  \notag \\
C &=&-\frac{\mathrm{\bar{\nu}}}{\varrho }\left( d\eta +\eta \frac{d\varrho }{%
\varrho }\right) f^{i}df_{i}
\end{eqnarray}%
and also vanish as well for $\mathrm{\nu }=0.$

\item \emph{term D}\newline
this is a hermitian term with non dependence on $\varrho $ nor on $\mathrm{%
\nu }$%
\begin{equation}
D=d\eta d\bar{\eta}+\bar{\eta}\eta df^{i}d\bar{f}_{i}-\left(
\bar{\eta}d\eta -\eta d\bar{\eta}\right) f_{i}d\bar{f}^{i}
\end{equation}%
it vanishes for $\eta =0.$
\end{itemize}
\end{description}

\  \  \  \  \  \newline
Summing these 4 terms, we get%
\begin{equation}
\begin{tabular}{lll}
$d\varphi _{i}d\bar{\varphi}^{i}$ & $=$ & $\frac{\mathrm{\nu \bar{\nu}}}{%
\varrho ^{4}}d\varrho ^{2}+\left( \frac{\mathrm{\nu \bar{\nu}}}{\varrho ^{2}}%
+\bar{\eta}\eta \right) df^{i}d\bar{f}_{i}+d\eta d\bar{\eta}$ \\
&  & $\frac{\mathrm{\nu }}{\varrho }\left( d\bar{\eta}+\bar{\eta}\frac{%
d\varrho }{\varrho }\right) \bar{f}_{i}d\bar{f}^{i}-\frac{\mathrm{\bar{\nu}}%
}{\varrho }\left( d\eta +\eta \frac{d\varrho }{\varrho }\right)
f^{i}df_{i}$
\\
&  & $-\left( \bar{\eta}d\eta -\eta d\bar{\eta}\right) f_{i}d\bar{f}^{i}$ \\
&  &
\end{tabular}
\label{BA1}
\end{equation}%
The metric of the hypersurface $\mathfrak{C}_{\nu }$\ is given by by
the
metric of the space $\mathbb{C}^{4}$ with $\bar{h}_{i},$ $h^{i}$ and $%
d\varphi _{i}$, $d\bar{\varphi}^{i}$ as in eqs(\ref{AB1}-\ref{AB2}).
In
other words%
\begin{equation}
ds^{2}=d\bar{h}_{i}dh^{i}+d\varphi _{i}d\bar{\varphi}^{i}
\end{equation}%
with $d\varphi _{i}d\bar{\varphi}^{i}$ as in (\ref{BA1}). So, the
explicit
form of the metric reads like%
\begin{equation}
\begin{tabular}{lll}
$ds_{\mathfrak{C}_{\nu }}^{2}$ & $=$ & $\left( 1+\frac{\mathrm{\nu \bar{\nu}}%
}{\varrho ^{4}}\right) d\varrho ^{2}+\left( \varrho
^{2}+\frac{\mathrm{\nu
\bar{\nu}}}{\varrho ^{2}}+\bar{\eta}\eta \right) df^{i}d\bar{f}_{i}+d\eta d%
\bar{\eta}$ \\
&  & $-\left( \bar{\eta}d\eta -\eta d\bar{\eta}\right) f_{i}d\bar{f}^{i}$ \\
&  & $+\frac{\mathrm{\nu }}{\varrho }\left( d\bar{\eta}+\bar{\eta}\frac{%
d\varrho }{\varrho }\right) \bar{f}_{i}d\bar{f}^{i}-\frac{\mathrm{\bar{\nu}}%
}{\varrho }\left( d\eta +\eta \frac{d\varrho }{\varrho }\right)
f^{i}df_{i}$
\\
&  &
\end{tabular}%
\end{equation}%
By substituting $f^{i}$ and $\bar{f}_{i}$ by their expressions
(\ref{fhar}),
one can also rewrite the above metric in terms of the angles $\theta ,$ $%
\psi $ and $\phi $.

\subsection{Metric of the ground state $\Sigma $}

Here, we study the metric of the two possible phases of the ground
state namely the supersymmetric phase $r=0$ and the non
supersymmetric one $r\neq 0 $.

\subsubsection{Supersymmetric phase}

The supersymmetric phase of the ground state $\Sigma _{susy}$ of \emph{n-MSSM%
} with $\boldsymbol{H}_{d}$ replaced by $\Phi $, the Higgs fields
are given
by $S=0$ and doublet $h^{i}$ and anti-doublet $\varphi _{i}$ as%
\begin{equation}
h^{i}=\sqrt{\left \vert \mathrm{\nu }\right \vert }f^{i},\, \,
\qquad \varphi _{i}=\sqrt{\left \vert \mathrm{\nu }\right \vert
}\bar{f}_{i}
\end{equation}%
This solution corresponds to a constant radius $\varrho $ given by
the square root of the absolute value of the complex parameter
$\mathrm{\nu }$, and to $\eta =0$;
\begin{equation}
\varrho =\sqrt{\left \vert \mathrm{\nu }\right \vert }\text{\qquad
},\qquad \eta =0
\end{equation}%
Substituting these values back into the metric
$ds_{\mathfrak{C}_{\nu }}^{2}$ of the complex 3D hypersurface
$\mathfrak{C}_{\nu }$ obtained above, we get the metric
$ds_{_{\mathfrak{\Sigma }_{\text{susy}}}}^{2}$ of the
supersymmetric ground state%
\begin{eqnarray}
ds_{_{\Sigma _{\text{susy}}}}^{2} &=&2\left \vert \mathrm{\nu
}\right \vert
df^{i}d\bar{f}_{i}  \notag \\
&=&\frac{1}{2}\left \vert \mathrm{\nu }\right \vert \left( d\theta
^{2}+d\psi ^{2}+d\phi ^{2}+2\cos \theta d\psi d\phi \right)
\end{eqnarray}%
This metric is proportional to $\left \vert \mathrm{\nu }\right
\vert $
and, as expected it is singular for $\mathrm{\nu }=0$. Notice also that $%
ds_{_{\Sigma _{\text{susy}}}}^{2}$ is invariant under the phase
change
\begin{equation*}
f^{i}\rightarrow f^{i\prime }=e^{i\alpha }f^{i}
\end{equation*}%
a symmetry that allows to fix one degree of freedom; say $\psi
-\varphi =0$; this leads to the metric of a real 2-sphere as given
below
\begin{equation}
ds_{_{\Sigma _{\text{susy}}}}^{2}=\frac{1}{2}\left \vert \mathrm{\nu
}\right \vert \left( d\theta ^{2}+4\cos ^{2}\frac{\theta }{2}d\phi
^{2}\right)
\end{equation}

\subsubsection{Non supersymmetric phase}

In this case, the supersymmetric phase of the ground state $\Sigma _{\text{%
n-susy}}$ is given by the Higgs configurations%
\begin{equation}
h^{i}=\varrho \text{ }f^{i},\qquad \varphi _{i}=\frac{\mathrm{\nu
}}{\varrho }\text{ }\bar{f}_{i}
\end{equation}%
with
\begin{equation}
\varrho ^{2}=\frac{g^{\prime 2}\mathrm{r}+\sqrt{g^{\prime 4}\mathrm{r}%
^{2}+4\left( g^{2}+g^{\prime 2}\right) ^{2}\mathrm{\nu
\bar{\nu}}}}{2\left( g^{2}+g^{\prime 2}\right) }
\end{equation}%
By using the Weinberg angle $\vartheta _{_{W}}$ allowing to express
the
gauge coupling constants like%
\begin{equation}
\sin \vartheta _{_{W}}=\frac{g^{\prime }}{\sqrt{g^{2}+g^{\prime
2}}},\qquad \cos \vartheta _{_{W}}=\frac{g}{\sqrt{g^{2}+g^{\prime
2}}}
\end{equation}%
we can put $\varrho $ into the form
\begin{equation}
\varrho ^{2}=\frac{\sin ^{2}\vartheta _{_{W}}\ }{2\ }\left( \mathrm{r}+\sqrt{%
\mathrm{r}^{2}+\frac{4\left \vert \mathrm{\nu }\right \vert
^{2}}{\sin ^{4}\vartheta _{_{W}}}}\right)
\end{equation}%
Substituting these values back into the metric
$ds_{\mathfrak{C}_{\nu }}^{2}$
of the complex 3D hypersurface $\mathfrak{C}_{\nu }$, we obtain the metric $%
ds_{_{\mathfrak{\Sigma }_{\text{susy}}}}^{2}$ of the non
supersymmetric ground state. By using $\eta =0$ and $d\varrho
^{2}=0$, the metric reads as
follows%
\begin{equation}
ds_{\mathfrak{\Sigma }_{\text{n-susy}}}^{2}=\left( \varrho
^{2}+\frac{\left \vert \mathrm{\nu }\right \vert ^{2}}{\varrho
^{2}}\right) df^{i}d\bar{f}_{i}
\end{equation}%
with $\varrho ^{2}$ as above. Moreover, using the relation
\begin{equation*}
\varrho ^{2}-\frac{\mathrm{\nu \bar{\nu}}}{\varrho ^{2}}=\mathrm{r}
\end{equation*}%
the metric can be also rewritten like%
\begin{equation*}
ds_{\mathfrak{\Sigma }_{\text{n-susy}}}^{2}=\left(
\mathrm{r}+\frac{2\left \vert \mathrm{\nu }\right \vert
^{2}}{\varrho ^{2}}\right) df^{i}d\bar{f}_{i}
\end{equation*}%
In terms of Weinberg angle%
\begin{equation}
ds_{\mathfrak{\Sigma }_{\text{n-susy}}}^{2}=\left(
\mathrm{r}+\frac{2\left
\vert \mathrm{\nu }\right \vert ^{2}}{\mathrm{r}\sin ^{2}\vartheta _{_{W}}+%
\sqrt{4\left \vert \mathrm{\nu }\right \vert ^{2}+\mathrm{r}^{2}\sin
^{4}\vartheta _{_{W}}}}\right) df^{i}d\bar{f}_{i}
\end{equation}%
with%
\begin{equation*}
df^{i}d\bar{f}_{i}=\frac{1}{4}\left( d\theta ^{2}+4\cos ^{2}\frac{\theta }{2}%
d\phi ^{2}\right)
\end{equation*}%
\begin{equation*}
\end{equation*}

\acknowledgments I would like to thank the senor associateship
programm of ICTP and the the international centre for theoretical
physics, Trieste, Italy where this works has been refined. I also
thank the program\emph{\ URAC09/CNRS} for support.

\end{document}